\documentclass[]{emulateapj}

\def \sext {{\sc Sextractor}}
\def \HST{{\emph{HST}}}
\def \tf {{$24 \mu m$}}

\begin{document}
\slugcomment{12/14/07}

\title{Triggered or Self-Regulated Star Formation within Intermediate Redshift Luminous Infrared Galaxies (I). Morphologies and Spectral Energy Distributions}

\author{J. Melbourne  \altaffilmark{1},
M. Ammons \altaffilmark{2},
S. A. Wright \altaffilmark{3},
A. Metevier \altaffilmark{2},
E. Steinbring  \altaffilmark{4},
C. Max  \altaffilmark{2},
D. C. Koo  \altaffilmark{2},
J. E. Larkin \altaffilmark{3},
M. Barczys \altaffilmark{3}
}
\altaffiltext{1}{Division of Physics, Mathematics and Astronomy, Mail Stop 320-47, California Institute of Technology, Pasadena, CA 91125. jmel@caltech.edu}

\altaffiltext{2} {University of California Observatories/Lick Observatory, Department of Astronomy and Astrophysics, University of California at Santa Cruz, 1156 High Street, Santa Cruz, CA 95064.  ammons, anne, max, koo@ucolick.org}
\altaffiltext{3} {Department of Physics and Astronomy, University of California, P.O. Box 951562, Los Angeles, CA 90095-1562. saw, larkin, barczysm@astro.ucla.edu}
\altaffiltext{4} {Herzberg Institute of Astrophysics, National Research Council Canada, Victoria, BC V9E 2E7, Canada}

\begin{abstract}
As part of the Center for Adaptive Optics  (AO) Treasury Survey (CATS) we imaged a set of 15 intermediate redshift ($z\sim0.8$) luminous infrared galaxies (LIRGs) with the Keck Laser Guide Star (LGS) AO facility.  These galaxies were selected from the Great Observatories Origins Deep Survey (GOODS) southern field, allowing us to combine the high spatial resolution \HST\ optical ($B$, $V$, $i$, and $z$-bands) images with our near-infrared ($K'$-band) images to study the LIRG morphologies and spatially resolved spectral energy distributions (SEDs).  Two thirds of the LIRGs are disk galaxies, with only one third showing some evidence for interactions, minor, or major mergers. In contrast with local LIRG disks (which are primarily barred systems), only 10\% of the LIRG disks in our sample contain a prominent bar. While the optical bands tend to show significant point-like substructure, indicating distributed star formation, the AO $K$-band images tend to be smooth.  They lack point-like structures to a $K \sim 23.5$ limit  This places an upper bound on the number of red super giants per blue-knot at less than 4000.  The SEDs of the LIRGs are consistent with distributed dusty star formation, as exhibited by optical to IR colors redder than allowed by old stellar populations alone.   This effect is most pronounced in the galaxy cores, possibly indicating central star formation.  We also observed a set of 11 intermediate redshift comparison galaxies, selected to be non-ellipticals with apparent $K$-band magnitudes comparable to the LIRGs. The ``normal'' (non-LIRG) systems tended to have lower optical luminosity, lower stellar mass, and more irregular morphology than the LIRGs.   Half of the ``normal'' galaxies have SEDs consistent with intermediate aged stellar populations and minimal dust.  The other half show evidence for some dusty star formation, usually concentrated in their cores. Our work suggests that the LIRG disk galaxies are similar to large disk systems today, undergoing self regulated star formation, only at 10 - 20 times higher rates.       
\end{abstract}


\section{Introduction}
\begin{deluxetable*}{ccccccc}
\tabletypesize{\small}
\tablecaption{Keck Observation Summary \label{table:ao_obs}}
\tablehead{\colhead{object} & \colhead{R.A. } & \colhead{Dec.} & \colhead{obs. date} & \colhead{exptime } & \colhead{FWHM}& \colhead{Strehl} \\ & \colhead{(J2000)}& & & \colhead{(m)} & \colhead{of PSF ($\arcsec$)} &\colhead {estimate}}
\startdata
                   LIRG 1&              03:32:13.86&              -27:42:48.9&         Sept. 29,  2005 &      44&    0.10&    0.30\\
                   LIRG 2&              03:32:13.24&              -27:42:40.9&         Sept. 29,  2005 &      44&    0.10&    0.30\\
                   LIRG 3&              03:32:12.81&              -27:42:01.5&          Nov. 25,  2005 &      55&    0.08&    0.35\\
                   LIRG 4&              03:32:32.99&              -27:41:17.0&          Dec. 11,  2006 &      50&    0.11&    0.25\\
                   LIRG 5&              03:32:32.90&              -27:41:23.9&          Dec. 11,  2006 &      50&    0.14&    0.18\\
                   LIRG 6&              03:32:31.48&              -27:41:18.1&          Dec. 11,  2006 &      50&    0.11&    0.25\\
                   LIRG 7&              03:32:31.05&              -27:40:50.1&          Dec. 11,  2006 &      50&    0.11&    0.20\\
                   LIRG 8&              03:32:32.46&              -27:40:56.6&          Dec. 11,  2006 &      50&    0.11&    0.25\\
                   LIRG 9&              03:32:52.88&              -27:51:19.8&          Dec. 11,  2006 &      53&    0.12&    0.18\\
                  LIRG 10&              03:32:52.87&              -27:51:14.7&          Dec. 11,  2006 &      53&    0.12&    0.18\\
                  LIRG 11&              03:32:53.02&              -27:51:04.7&          Dec. 11,  2006 &      53&    0.11&    0.20\\
                  LIRG 12&              03:32:07.98&              -27:42:39.4&         Sept. 28,  2005 &      60&    0.14&    0.14\\
                  LIRG 13&              03:32:55.88&              -27:53:27.9&          Nov. 12,  2005 &      30&    0.12&    0.15\\
                  LIRG 14&              03:32:38.12&              -27:39:44.8&          Dec. 11,  2006 &      42&    0.10&    0.24\\
                  LIRG 15&              03:32:55.69&              -27:53:45.9&          Nov. 12,  2005 &      30&    0.09&    0.30\\
                COMPGAL 1&              03:32:13.06&              -27:42:04.9&          Nov. 25,  2005 &      55&    0.08&    0.35\\
                COMPGAL 2&              03:32:30.93&             -27:40:58.43&          Dec. 11,  2006 &      50&    0.11&    0.20\\
                COMPGAL 3&             03:32:53.428&             -27:50:51.06&          Dec. 11,  2006 &      53&    0.10&    0.24\\
                COMPGAL 4&              03:32:52.33&              -27:50:51.8&          Dec. 11,  2006 &      53&    0.11&    0.20\\
                COMPGAL 5&              03:32:09.40&              -27:42:36.1&         Sept. 28,  2005 &      60&    0.12&    0.15\\
                COMPGAL 6&              03:32:08.43&              -27:42:43.5&         Sept. 28,  2005 &      60&    0.12&    0.15\\
                COMPGAL 7&              03:32:55.95&              -27:53:57.1&          Nov. 12,  2005 &      30&    0.10&    0.25\\
                COMPGAL 8&             03:32:13.458&             -27:41:49.89&          Nov. 25,  2005 &      55&    0.08&    0.35\\
                COMPGAL 9&              03:32:36.17&              -27:39:55.7&          Dec. 11,  2006 &      42&    0.12&    0.20\\
               COMPGAL 10&              03:32:36.68&              -27:39:54.6&          Dec. 11,  2006 &      42&    0.10&    0.30\\
               COMPGAL 11&             03:32:55.003&             -27:50:51.63&          Dec. 11,  2006 &      53&    0.11&    0.20\\
\enddata
\end{deluxetable*}

Luminous infrared galaxies \citep[LIRGs; for review see ][]{SandersMirabel96} are defined to have total IR luminosities ($8 - 1000 \mu m$) in excess of $10^{11}L_{\odot}$.  At low redshift, LIRGs are rare, comprising less than 5\% of the total IR energy density of local galaxies \citep{Soifer91}.  However, by $z=1$, LIRGs account for $\sim70\%$ of the IR background \citep{LeFloch05}.  The high IR luminosities of LIRGs result from thermal dust heating in starbursts, or active galactic nuclei (AGN).  Dust absorbs optical and UV radiation, and reemits that energy in a broad thermal peak in the mid to far IR.  In order for a star forming galaxy to emit at a LIRG level it must have a  very high star formation rate, in excess of $\sim17 M_{\odot} \; yr^{-1}$.  In the local universe, these high star formation rates are primarily triggered by galaxy-galaxy interactions or mergers \citep{Ishida04}.  However, the rise in the number density of LIRGs with redshift appears to be unrelated to any change in the merger rate.  For instance \citet{Bell05,MKL05, Lotz07} all show that roughly 50\% of intermediate redshift LIRGs are disk galaxies with little sign of recent merger activity.  

While morphology rules out major mergers as triggers for the majority of intermediate redshift LIRGs, interactions with neighbors, minor mergers, or bar instabilities may contribute.  In addition, intermediate redshift disks may contain higher gas fractions than their local counterparts, allowing otherwise normal galaxies to reach higher levels of star formation.  Some star formation triggers have observable signatures.  For instance interactions, minor mergers and bar instabilities will tend to drive central star formation.  Centrally concentrated star formation is a feature of local LIRGs \citep{SandersMirabel96}.  Minor mergers may also show elevated star star formation in a dwarf companion which may appear as a single bright blue knot at these redshifts \citep{Ferreiro2004}.  Minor mergers and interactions may also produce tidal features \citep{Knierman05} or drive bar instabilities.

Low redshift LIRG disks are found to uniformly contain bars \citep{Wang06}.  Interestingly, \citet{Zheng05} found that intermediate redshift LIRGs are relatively free of bars.  Their study, however, was done with optical (rest-frame UV) imaging and therefore may not be sensitive to bars, which in the presence of dust will be more easily observed at rest frame red to near-IR wavelengths.

While centrally concentrated star formation is a feature of low redshift LIRGs \citep{SandersMirabel96}, it may be less important for intermediate redshift LIRGs.  If the intermediate redshift LIRG disks are experiencing smoothly distributed star formation, rather than centrally concentrated star formation, then there may be no need to invoke a ``trigger'' for the event.  These galaxies may tend to high star formation rates simply because of higher gas fractions.  If that is the case, then higher star formation rates may actually be normal for the large disks of that epoch.  

It is currently an open question, whether the intermediate redshift LIRGs are ``triggered'' or experiencing ``normal'', if elevated, star formation.  In a series of 3 papers we discuss the modes of star formation in a set of 15 intermediate redshift LIRGs and 11 comparison galaxies.  These galaxies are located in the Great Observatories Origins Deep Survey  \citep[GOODS]{giavalisco04}, in fields for which we have obtained high spatial resolution IR imaging from the Keck laser guide star (LGS) adaptive optics (AO) facility .    This paper will present the \HST\ (from GOODS) and AO images, basic photometric measurements, and will discuss the morphologies of the LIRG and comparison galaxy samples.  It will seek to identify ``triggers'' (e.g. mergers, interactions, bars, or AGN) for the LIRG events, and  will present the spectral energy distributions of subcomponents of galaxies in an attempt to identify sites of dusty star formation.     We compare the results from the LIRG sample with the set of ``normal'' (i.e. non-LIRG) galaxies drawn from the same fields.

Paper 2 will attempt to ``fit'' the SEDs of the galaxy sub-components with specific stellar population synthesis models.  Paper 2 will compare the stellar populations of the central and outer regions of each galaxy in an effort to identify where the bulk of the star formation is occurring.  It will also attempt to locate the bulk of the dust and compare to the global dust measurements from \emph{Spitzer}\ \tf\ images.

Paper 3 will expand our work on the morphologies of the LIRGs.  It will present analysis of the Sersic profiles of each galaxy.  It will then look for substructures hidden beneath the Sersic profiles, specifically looking for small central bars, or tidal features not immediately obvious in the images. 

While the sample size is small, 15 LIRGs and 11 ``normal'' galaxies, it is a unique data set.  It is the only such set of intermediate redshift galaxies that has been observed in 4 optical bands with \HST\ and with equally high spatial resolution in the near-IR.  The point-spread-function of our $K$-band AO images has a  full-width half-max of $0.1\arcsec$ comparable to \HST\ in the optical, and a factor of 3 improvement over \HST\ NICMOS.  The high spatial resolution is necessary to study galaxy subcomponents individually.  The Keck IR band is necessary to distinguish between old and young-dusty stellar populations, which are degenerate cases given the optical \HST\ data alone \citep{Melbourneetal05}.  

\begin{figure*}
\includegraphics{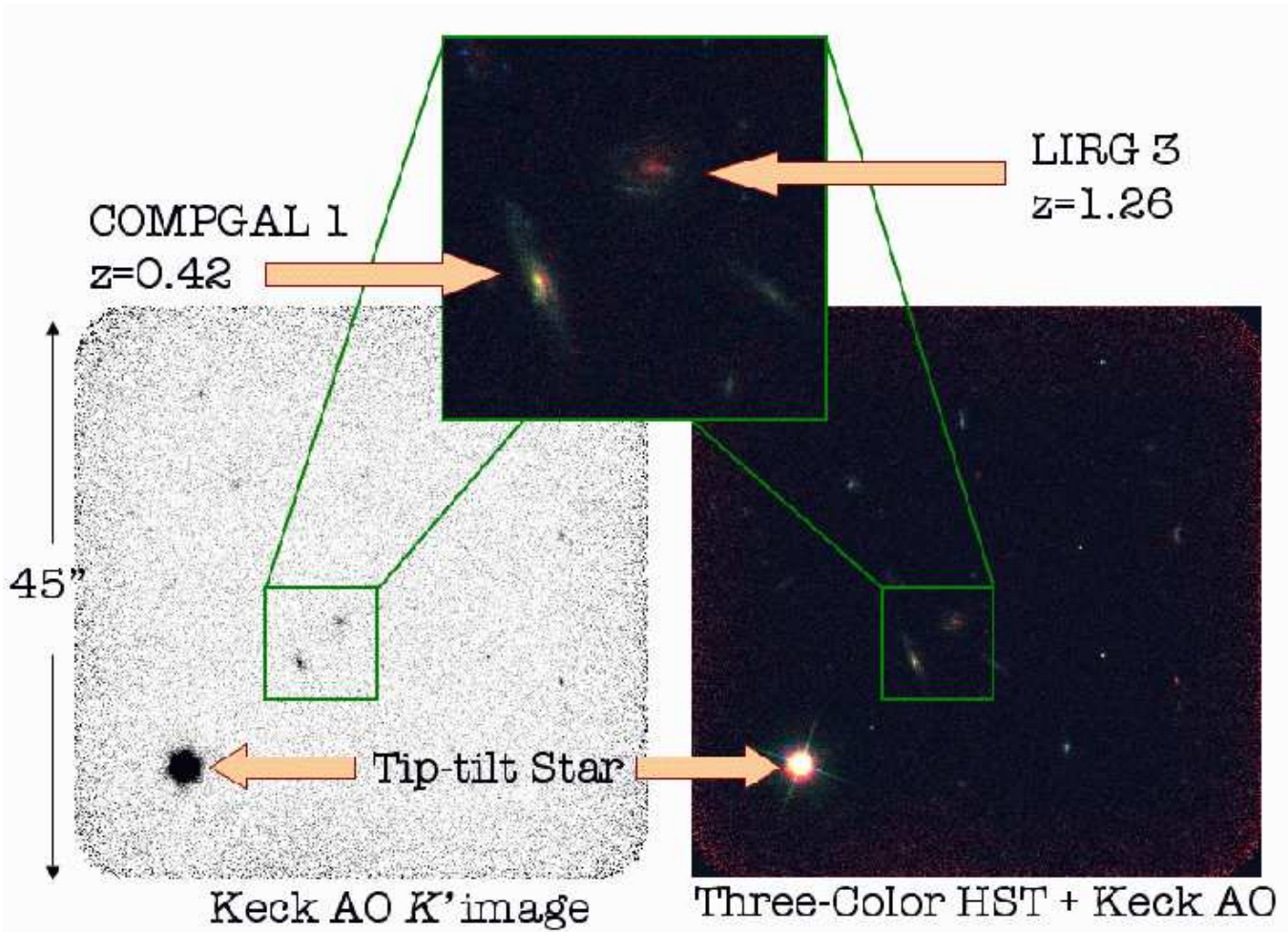}
\caption{\label{fig:ao_field} A typical CATS pointing in GOODS-S.  The left panel shows the Keck AO pointing.  The tip-tilt star is in the bottom, left-hand corner of the image.  North is up, east is to the left.  During each observation the laser was positioned at the center of the frame.  As the telescope dithered between frames the laser moved with the telescope.  The right hand panel is the same as the left only it is a three color image. combining the Keck AO $K$-band(red) and HST $i$ and $B$-band (green and blue) images.  The blow up shows LIRG 3 which contains a prominent red bulge and blue star forming regions.  It also shows COMPGAL 1, an  inclined spiral.
}
\end{figure*}

While the AO images are vital for this project, they are difficult to obtain.  Adaptive optics systems require guide stars in order to track and correct the turbulent layers in the atmosphere.  The first generation adaptive optics systems required very bright ($R<13$) natural guide stars (NGS) nearby the science target ($d<40\arcsec$).  NGS stars are rare, especially at the high galactic latitudes of the extragalactic treasury fields.  GOODS-S contains no suitable NGS guide stars.  Laser guide stars improve the situation.  The Keck sodium LGS AO system produces a bright ($R\sim10$) artificial star in the sodium layer of the Earth's atmosphere.  The LGS AO system uses the laser spot to correct the high-order aberrations of the wavefront.  LGS AO systems still require a natural guide star to correct for image motion (``tip-tilt'' correction).  These tip-tilt stars, however, can be as faint as $R\sim18$ making $\sim 30\%$ of the GOODS field accessible to the Keck AO facility.  

The images used in our study were obtained as part of the Center for Adaptive Optics Treasury Survey \citep[CATS;][; Larkin et al. in preparation]{Melbourneetal05}.   During Spring 2008, the images will be made available for download to the public through the CATS web interface \footnote{http://www.ucolick.org/$\sim$jmel/cats\_database/cats\_search.php}. CATS NGS images in the COSMOS \citep[Cosmological Evolution Survey]{Scoville06} and GEMS \citep[Galaxy Evolution From Morphology And SEDs, ]{Rix04}  regions are already available for download as a service to the community.

This paper is divided into the following sections.  Section two describes the basic data used in this study, including a discussion of the AO observing strategy, data reduction, and PSF estimation. Section 3 presents the 5-band high resolution images, and a discussion of galaxy morphologies.  Section 4 gives global photometric measurements and derived galaxy properties, such as absolute magnitude, rest-frame color, stellar mass, and IR luminosity.  This section compares the broad properties of the LIRG and ``normal'' galaxy samples.  Section 5 provides photometry and SED's of galaxy sub-components and compares these with single burst stellar population synthesis models.  Section 6 discusses the findings of the paper and puts them into the context of previous results from other groups.  Section 7 summarizes.

We assume a flat cosmology with $H_0= 70$ km/s/Mpc, $\Omega_m=0.3$, and $\Omega_{\lambda}=0.7$.     


\section{The Data}
The galaxies in this study are drawn from the GOODS-S \HST\ treasury field.  The GOODS team obtained very deep ACS imaging in  4 optical bands F435W, F606W, F775W, and F850LP ($B,V,i$, and $z$, respectively).  The field has also been imaged in the infrared with the \emph{Spitzer Space Telescope}\ \citep{Rieke04}, and in x-ray \citep{Rosati02} with the Chandra X-ray Telescope .   We identify LIRGs in GOODS-S using the MIPS $24 \mu m$ measurements and  IR luminosity estimates from \citet{LeFloch05}.  We use the x-ray images to identify x-ray luminous LIRGs which are likely to harbor AGN.  We use the optical ACS images to measure magnitudes, colors, and morphology.  The high spatial resolution (typically $0.06 \arcsec$) of these images allow us to measure photometry of galaxy subcomponents, such as bulges, star forming knots, and bars.  

In addition to the imaging data, several large redshift surveys have targeted the GOODS-S field.  The largest, COMBO-17 \citep{Wolf04}, used narrow-band imaging to match SEDs to photometric redshifts with high precision.  \citet{Vanzella06} targeted an additional 500 galaxies with the FORS2 spectrograph on the Very Large Telescope.  \citet{Szokoly04} obtained FORS/FORS2 redshifts for an additional 161  Chandra sources in the field.  Finally the DEEP2 \citep{Davis03} collaboration obtained spectra of 91 galaxies in GOODS-S with the DEIMOS spectrograph on Keck.  We make use of these redshifts for our project.

\subsection{Keck Adaptive Optics Data}
We obtained Keck LGS AO observations of seven, $\sim1$ hour pointings in the GOODS-S field over a two year period from 2005-2006.  These observations were made with the NIRC2 camera in the wide field of view ($40x40\arcsec$), in the $K'$-band.  Figure \ref{fig:ao_field} shows an example AO pointing in the GOODS-S field.  The centers of pointings where restricted to be within $20\arcsec$ of suitable tip-tilt stars (typical magnitude $R\sim15$).  

There are $\sim30$ suitable tip-tilt stars in the GOODS-S region.   If you place the tip-tilt star in the corner of the detector there are four possible pointings for each star. This allowed for a large degree of flexibility in choosing pointings.  We specifically selected pointings with a high density of bright disk galaxies, or x-ray sources.  We did not specifically target fields with LIRGs.  After the AO observations were made we identified the IR luminous galaxies from the \citet{LeFloch05} catalogue.  As we will see in the following section, the LIRGs tended to be at intermediate redshift, $0.5 < z < 1.0$.  

In addition to LIRGs, we selected a comparison sample from the same AO pointings.  Our goal in constructing this sample was to compare star formation triggers of non-LIRGs with star formation triggers of LIRGs.  We also wanted to select galaxies that had strong detections in the AO images so that signal-to-noise ratio was similar for both the LIRGs and non-LIRGs.  Therefore the criteria for selection into the comparison sample was 1) $K$-band fluxes comparable to the LIRGs; 2) intermediate redshift; 3) non-elliptical morphology.   Eleven galaxies matched these criteria.   While there were a handful of ellipticals that would have fit the $K$-band criteria we did not include these because they were not actively star forming.  In addition, none of the LIRGs exhibited elliptical-like morphology.   

\subsection{\label{sec:ao_obs} Adaptive Optics Observing Strategy}

AO observing is challenging because of the high thermal background of the atmosphere and instruments, and because the profile of the AO point-spread-function (PSF) fluctuates with time.  Exposure times should be short enough so that the detector will not saturate objects (typical saturation time is 5 min for the wide field camera in the $K$-band), but long enough to detect objects in a single sky subtracted exposure (to improve image alignment on dithered data). 

Another exposure time consideration results from the time varying PSF.  The atmospheric turbulence profile can change on the time scale of several seconds.  Whenever light passes though a particularly turbulent patch, the AO correction and image quality will suffer.  If an intermediate redshift galaxy contains high contrast point-light regions (i.e. bulges, bars, and star forming knots) they will be hard to detect above the background unless they are AO corrected.  Therefore better measurements of high contrast structures can be achieved by culling poor AO performance images.  In order to take advantage of culling, multiple shorter exposures can be an improvement over fewer long images.  It is also important to track the AO PSF, in order to measure the system performance.  The Strehl ratio, the ratio of peak brightness in a PSF to its theoretical maximum, gives a measure of AO performance.  

Over the course of two years we tried several different observing strategies.  Because we were using the wide field camera, we generally tried to keep the tip-tilt star on the detector to track the AO PSF.  At each dither position we took a short exposure of the tip-tilt star to get a PSF estimate, followed by a long science exposure.  For the first three observing nights our science exposures were 30 seconds with 4 coadds giving a total exposure time of 2 minutes per dither position.   Our short PSF exposures were generally 2-5 seconds.  Unfortunately we found that those short exposures were too short to give an accurate measure of the PSF.  They were taken immediately after dithering and the AO system did not have enough time to settle down before the exposure was over.  As a result, these were of limited value.  If the short PSF exposures were taken after the long science exposures, immediately before the dither, they would give a better representation of the PSF in that part of the field.  

For our final two observing nights we changed our strategy.  We decided to make our long exposures 30 seconds with two coadds, for one minute each, but we only dithered after two science frames.   Thus each dither position was still 2 minutes of science exposure.  We did this so that we could improve our ability to cull low Strehl images.  It was a reasonable strategy because we could detect objects for alignment in the one minute exposures.  For the short PSF exposures, we increased the coadds so that the total exposure time was $\sim20 s$.  This allowed for a better PSF measurement than the very short exposures used in the first nights of observation.  In cases where the tip-tilt star did not saturate in 30s or there were additional PSF stars in the field, we did away with the short exposures all together.

By placing the tip-tilt star on our camera, we had a real-time PSF estimate at one place in the field.  However, this PSF may not be a good match to the PSF of a galaxy at a different location in the field.  In additional to temporal variations, the PSF changes with respect to separation from the laser spot and the tip-tilt star.  In order to track the spatial variations in the PSF we observed star fields each night.  Star fields were generally observed at the beginning or the end of the night.  Because most of the observing was done with half-night runs, we only observed one star field per night.  While the GOODS-S fields were observed at typical airmasses of 1.5, the airmasses of the star fields were typically 1.2.  When the seeing was good, the spatial variations in the PSF were relatively constant between successive star field images even when the overall Strehl was changing.   The AO PSFs are discussed further in Section \ref{sec:ao_psf}.

For all of the observing runs we adopted the strategy of keeping the laser spot fixed at the center of the NIRC2 camera.  This meant that as the telescope dithered, the laser dithered as well and the location of the laser spot changed with respect to the background galaxies and stars.  Steinbring et al. (submitted) demonstrated that this strategy allows for the most uniform AO performance across the field.

\subsection{Adaptive Optics Data Reduction}
The AO images were reduced in the same manner as \citet{Melbourneetal05}.  The procedure was to create sky and flatfield frames from the science images after masking out sources.  After flatfielding and sky subtraction, frames were corrected for known NIRC2 camera distortion.  Image alignment was performed by centroiding on objects in the field.  Frames were then average combined with a clipped mean algorithm.  

Each night UKIRT IR standard stars were observed to set the photometric zeropoint.   These images were reduced in the same way as the science frames.  

\begin{figure}
\includegraphics[scale=0.6]{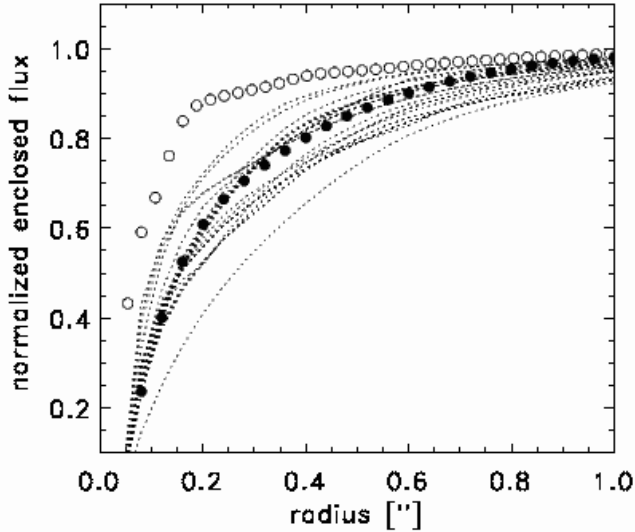}
\caption{\label{fig:psf_ee} Enclosed energy fractions as a function of radius, for the AO PSFs used in this study (dotted lines).  Also shown is the enclosed energy fraction of the \HST\ $z$-band PSF \citep[from Tiny Tim,][]{Krist95} before and after convolution with the AO PSF (open and filled circles, respectively).  This figure demonstrates the wide range of PSF shapes that occur in our sample.  It also shows that the \HST\ PSFs tend to have steeper enclosed energy distributions compared with the AO PSF.  In order to bring the data to a common PSF, we convolve each \HST\ image with the AO PSF.
} 
\end{figure}

\subsection{\label{sec:ao_psf} Adaptive Optics PSF Estimation}

In Section 5 we will present photometry of galaxy subcomponents, some of which have typical sizes of $\sim0.2\arcsec$, smaller than the seeing limit.  In general, knowledge of the AO PSF will be the limiting factor in the accuracy of the small aperture photometry.  The AO PSF is comprised of a diffraction limited core, and a seeing limited halo.  The seeing limited halo represents the light that the AO system was unable to correct, while the core represents the corrected light. In order to do accurate photometry of regions smaller than the seeing limited halo, we will need to know the fraction of light in each component.  This can be derived from the Strehl ratio which is the ratio of the peak of the observed PSF to the theoretical maximum peak of a perfect diffraction limited PSF.  We will then have to bring the \HST\ and AO images to a common PSF, so that the photometry is measured in the same way across all 5 bands.  In this section we describe how we model the AO PSF for each galaxy.


Because the AO PSF varies both temporally and spatially, it is very challenging to make an accurate estimate of the PSF of a given galaxy in the field.  However, we designed our observing strategies (see Sec. \ref{sec:ao_obs}) in order to obtain the best possible PSF estimates.  We can generate PSF estimates from a combination of the following data: 1) unsaturated images of the tip-tilt star; 2) images of star fields taken during the night; and 3) serendipitous PSF stars in the field.   

Ideally we would combine the real time Strehl estimates from the tip-tilt stars and serendipitous stars with the spatial variations tracked by the star fields to model the AO PSF anywhere in the field at any given time.  In practice, because most of our tip-tilt PSFs needed longer exposure time to be accurate representations of the PSF, we decided to base our PSF estimates entirely on the star fields.  

For each galaxy, we first identify the star in the star field that is located closest to the field position (with respect to the tip-tilt star and laser positions) of a study galaxy.  We measure the radial profile of this PSF into the noise of the sky. We use this profile to model the core of the PSF.   The halo of the PSF is more difficult to model, because, for most stars in the star field, the halo is observed with low signal-to-noise ratio.   Instead, we use observations of bright standard stars (originally obtained to set the photometric zeropoints for each night).  Because the standard stars have much higher signal-to-noise ratio (SNR) in the wings of their profiles, we use them to model the halo of the PSF. The standard stars were observed with tip-tilt AO correction on, but high-order AO correction off.  The outer profile shapes of these stars should be similar to the outer profile shapes of the AO corrected PSF, because the outer shape is set by the seeing and tip-tilt correction.

We generate a 2D model of the AO PSF by combining the core + halo profiles.  We splice the two profiles together at the point where the core PSF drops to only 9 sigma above the noise.  At this radius we scale the halo PSF to the same flux level as the core PSF, and follow the halo profile for the remainder of the AO PSF.  This method gives us a high SNR estimate of the PSF to very low flux levels at large radii from the core.  We chose 9 sigma as the splice point based on the few unsaturated, yet high SNR, stars in our science fields.

Figure \ref{fig:psf_ee} shows the AO PSF enclosed energy fractions as a function of radius (dotted lines).  This figure demonstrates the range of PSF shapes that occur in our sample.  Also plotted is the enclosed energy fraction for the ACS $z$-band PSF \citep[from Tiny Tim,][]{Krist95} before and after convolving with the AO PSF (open and filled circles, respectively).  Before convolution, the \HST\ PSFs are steeper than the AO PSFs.  After convolution the \HST\ PSFs are a good match to the AO PSFs.  For this reason, we convolve the \HST\ images with the appropriate AO PSF, before measuring photometry of galaxy subcomponents.  This is done by first resampling the \HST\ images and the model PSFs to a $0.01\arcsec/pix$ pixel scale.   After convolving, we resample the images to the pixel scale of the original AO images, $0.04 \arcsec/pix$.   

\begin{figure}
\includegraphics[scale=0.6]{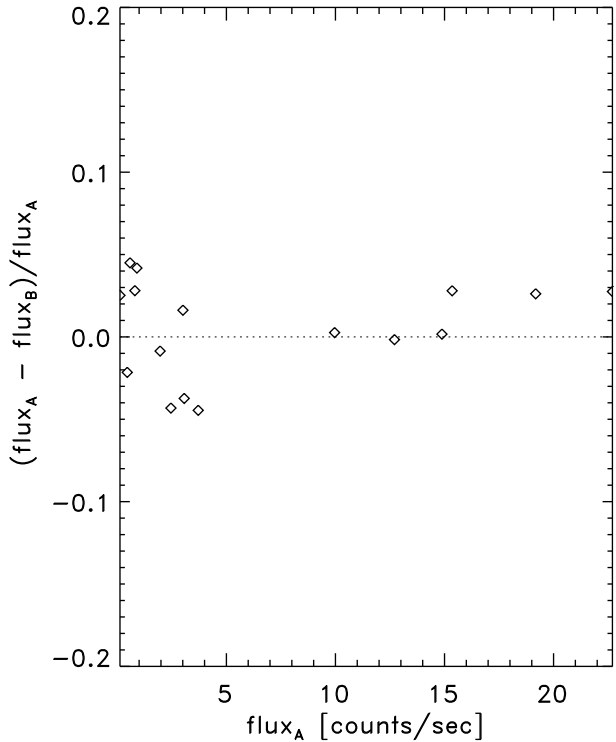}
\caption{\label{fig:psf_test} A study of the effect of AO PSF models on galaxy subcomponent photometry.  The 4-band \HST\ images of a sample galaxy where convolved with two different PSFs.  PSF$_A$ was our best guess model PSF as derived in Section \ref{sec:ao_psf}.  PSF$_B$ was an actual star in the science frame nearby ($10\arcsec$ separation) the galaxy. PSF$_B$ should be similar to the true PSF for this galaxy.  We see that after convolution, the photometry of the galaxy subcomponents is robust to minor differences in PSF.  In this case the photometry matches to within better than 10\%.} 
\end{figure}

In order to test the validity of our model PSFs, we study their effect on the  photometry of galaxy subcomponents.  Figure \ref{fig:psf_test} shows the photometry (from the \HST\ images) of subcomponents (typically $0.2 - 0.5 \arcsec$ radii regions) within one galaxy in our sample.  Flux$_A$ is the photometry after convolving with the model PSF.  Flux$_B$  is the photometry after convolving with an  AO image of a high SNR star located near the galaxy ($10\arcsec$ separation).   The star image should be a close approximation to the actual PSF at the galaxy position.  Figure \ref{fig:psf_test} shows that the photometry based on the model PSFs is robust. It matches, to within better than 10\%, the photometry based on the ``true'' PSF for the galaxy subcomponents in all four \HST\ bands. 

A summary of the AO observations, including coordinates, exposure times, and data quality, as given by the PSF full-width-half-max (FWHM), and Strehl ratio,  is given in Table \ref{table:ao_obs}.  

\section{Morphologies of LIRGs and Comparison Sample}
\begin{figure*}
\center
\includegraphics[scale=1.4]{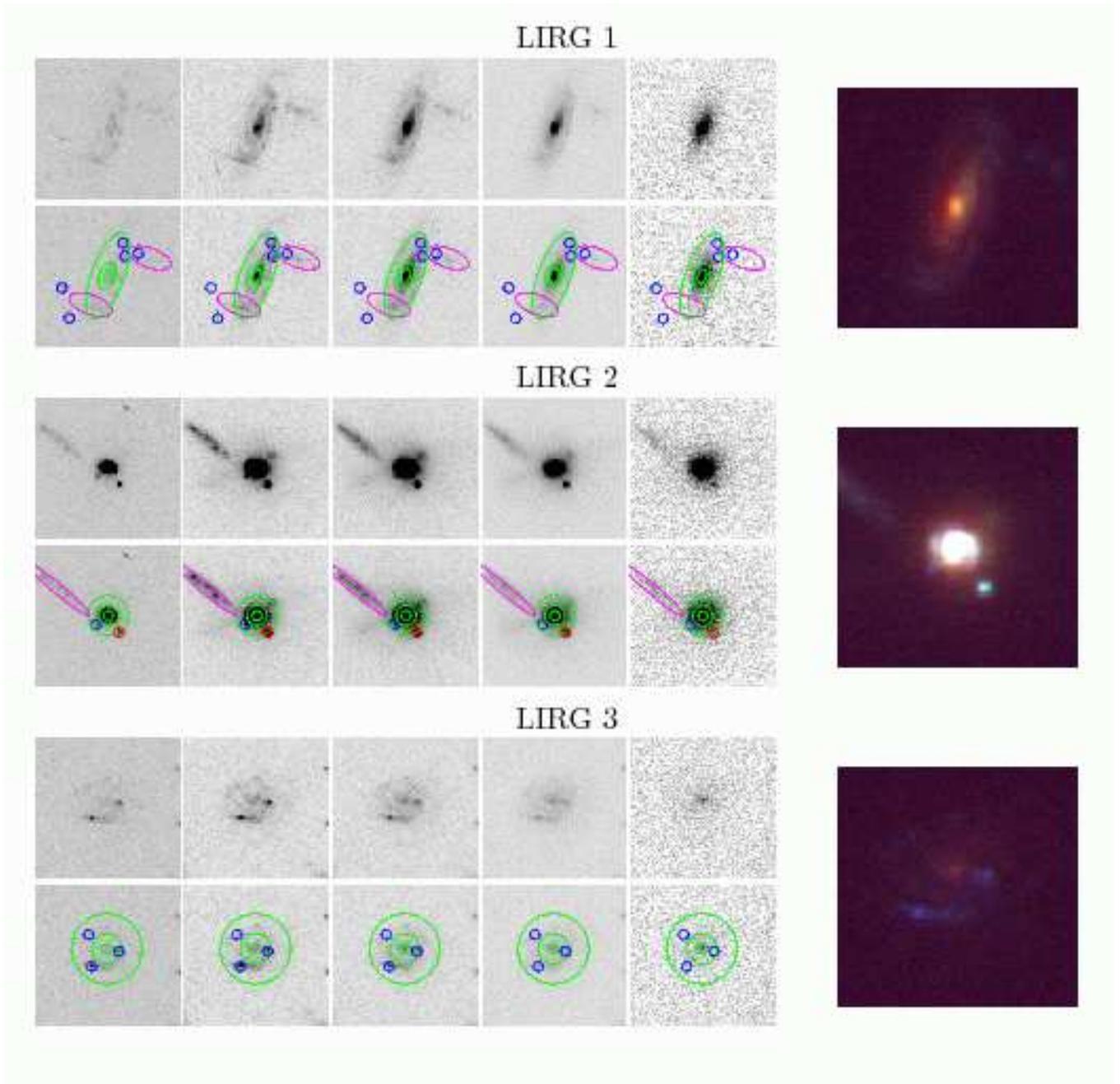}
\caption{\label{fig:LIRGsA} $B,V,i,z,$ (\HST) and $K$-band (Keck AO) images of LIRGs 1 - 3. North is up, and East is to the left.  \HST\ images are shown at the same pixel scale as the AO image, and are shown before convolution with the AO PSF. The images are $\sim 2.8 \arcsec$ on a side. The bottom row of each set of 10 images is identical to the top row, but shows the apertures used for photometry of the subcomponents.  LIRG 1 is a large spiral galaxy, possibly undergoing a minor merger.  LIRG 2 is a major merger.  LIRG 3 is at higher redshift ($z=1.08$) and looks like a late type spiral. Color images (created with the $B$, $i$, and $K$-band images) are also shown. }
\end{figure*}  

\begin{figure*}
\center
\includegraphics[scale=1.4]{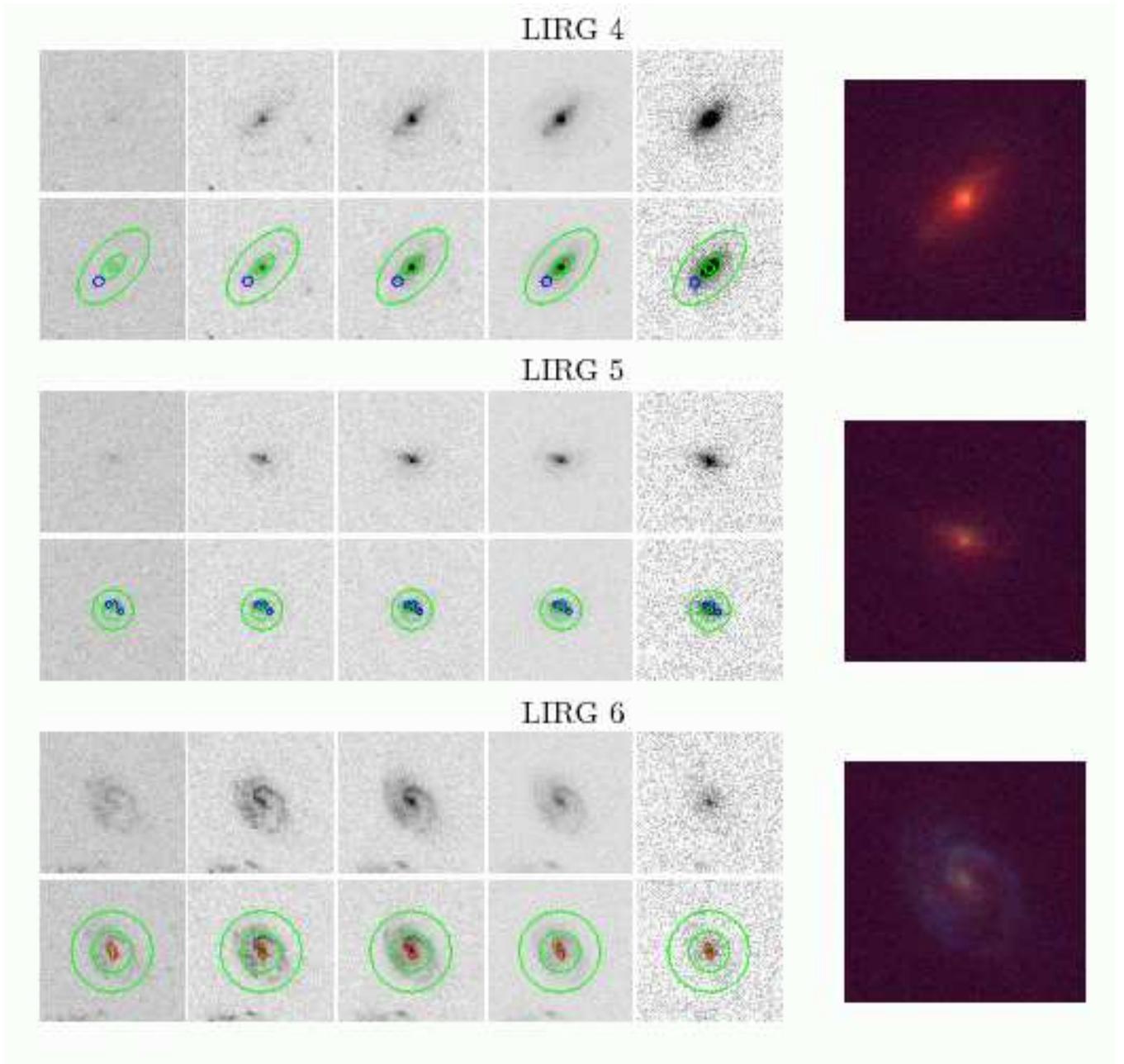}

\caption{\label{fig:LIRGsB} The same as Figure \ref{fig:LIRGsA} only for LIRGs 4 - 6.  LIRGs  4 and 6 are large spiral galaxies.  LIRG 4 also contains a prominent bar. LIRG 5 appears to be a compact spiral galaxy. }
\end{figure*}

\begin{figure*}
\center
\includegraphics[scale=1.4]{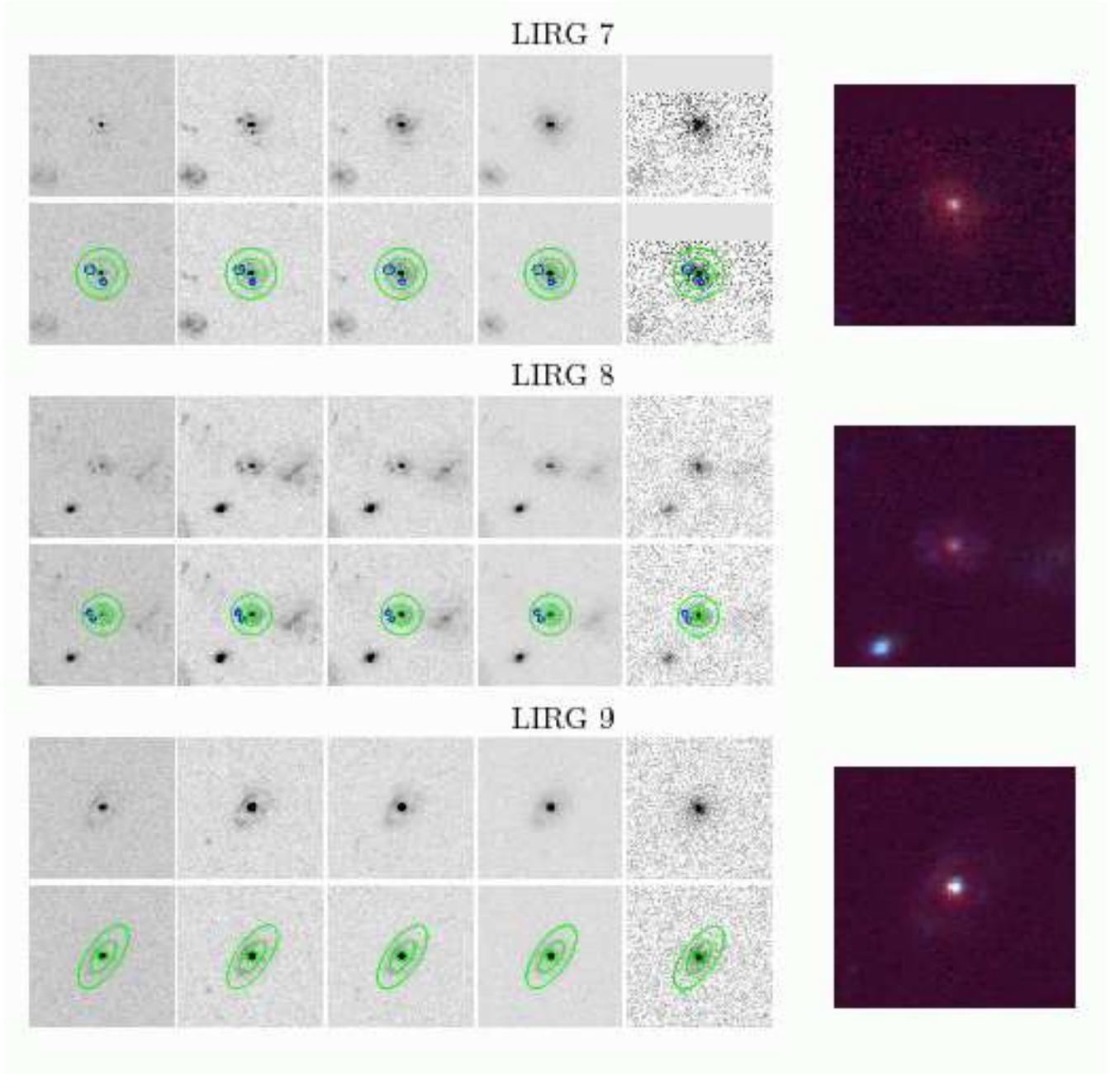}

\caption{\label{fig:LIRGsC} Images of LIRGs 7 - 9. LIRGs 7 and 8 are spiral galaxies with very red central cores.  LIRG 9 is a ring galaxy and a strong x-ray source.  Its core is probably dominated by an AGN.}
\end{figure*}

\begin{figure*}
\center
\includegraphics[scale=1.4]{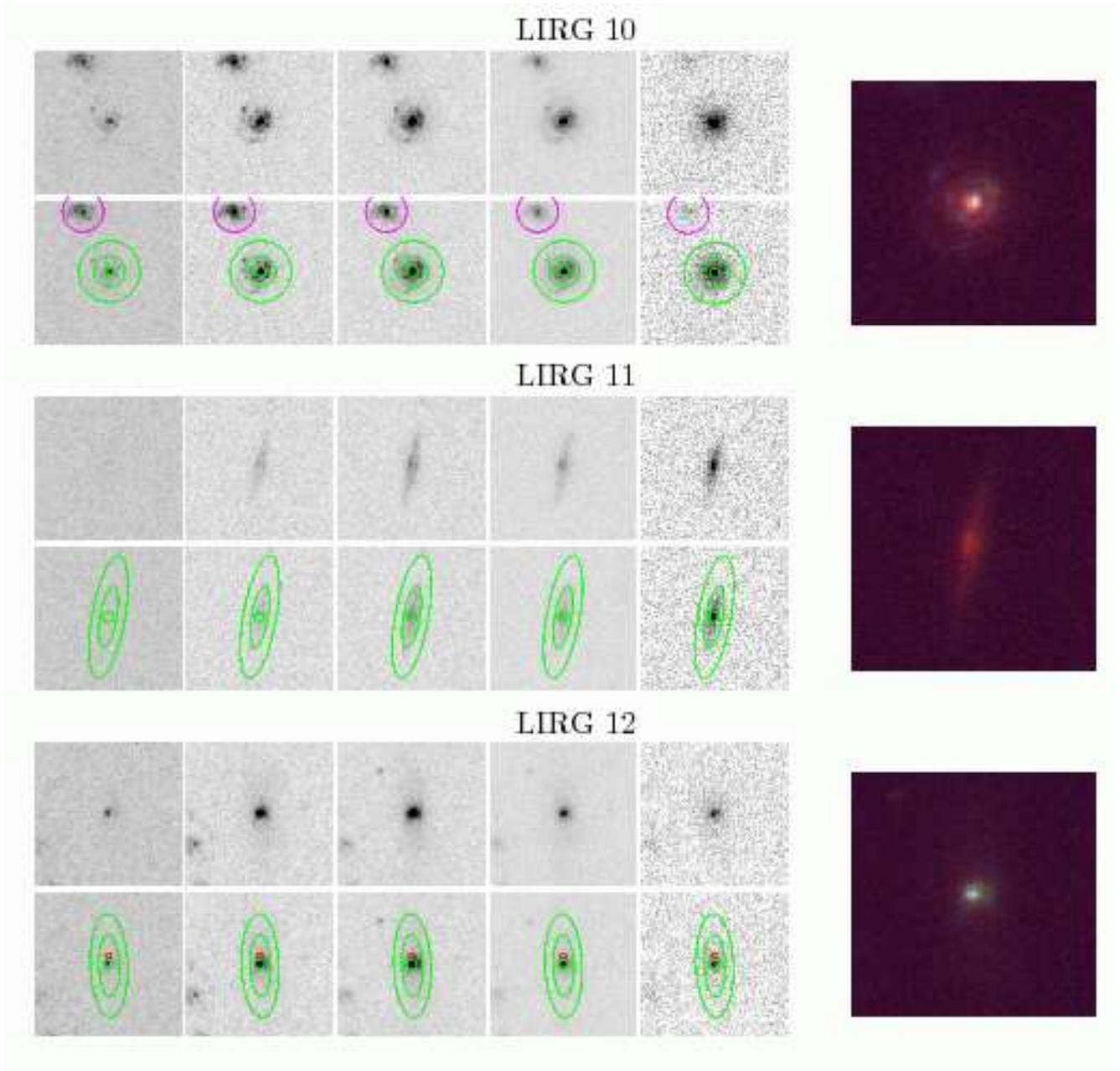}

\caption{\label{fig:LIRGsD} LIRGs 10 - 12.  LIRGs 10 and 11 are $z\sim1$ spiral galaxies. LIRG 11 is seen edge on.  LIRG 12 is at lower redshift and dominated by a central bulge.  LIRG 12  is also an x-ray source.  As there is not strong evidence for star formation in this galaxy, a central AGN may be the dominant contributer to the IR luminosity of LIRG 12.  }
\end{figure*}

\begin{figure*}
\center
\includegraphics[scale=1.4 ]{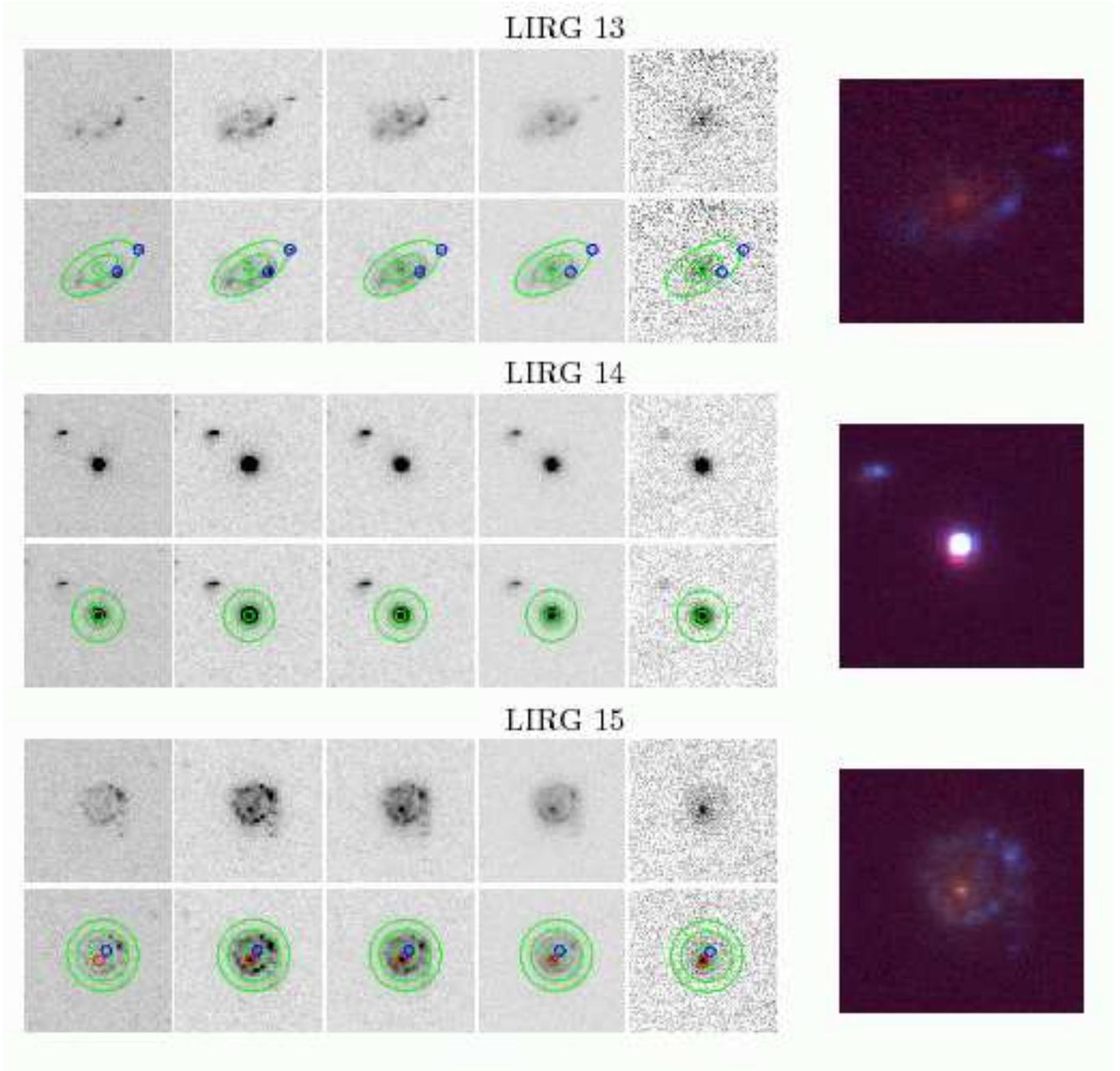}

\caption{\label{fig:LIRGsE} LIRGs 13 - 15.  LIRG 13 us a disk galaxy with a red core.  LIRG 14 is a QSO.  LIRG 15 has probably undergone a recent minor merger.
}
\end{figure*}

\begin{figure*}
\center
\includegraphics[scale=1.4 ]{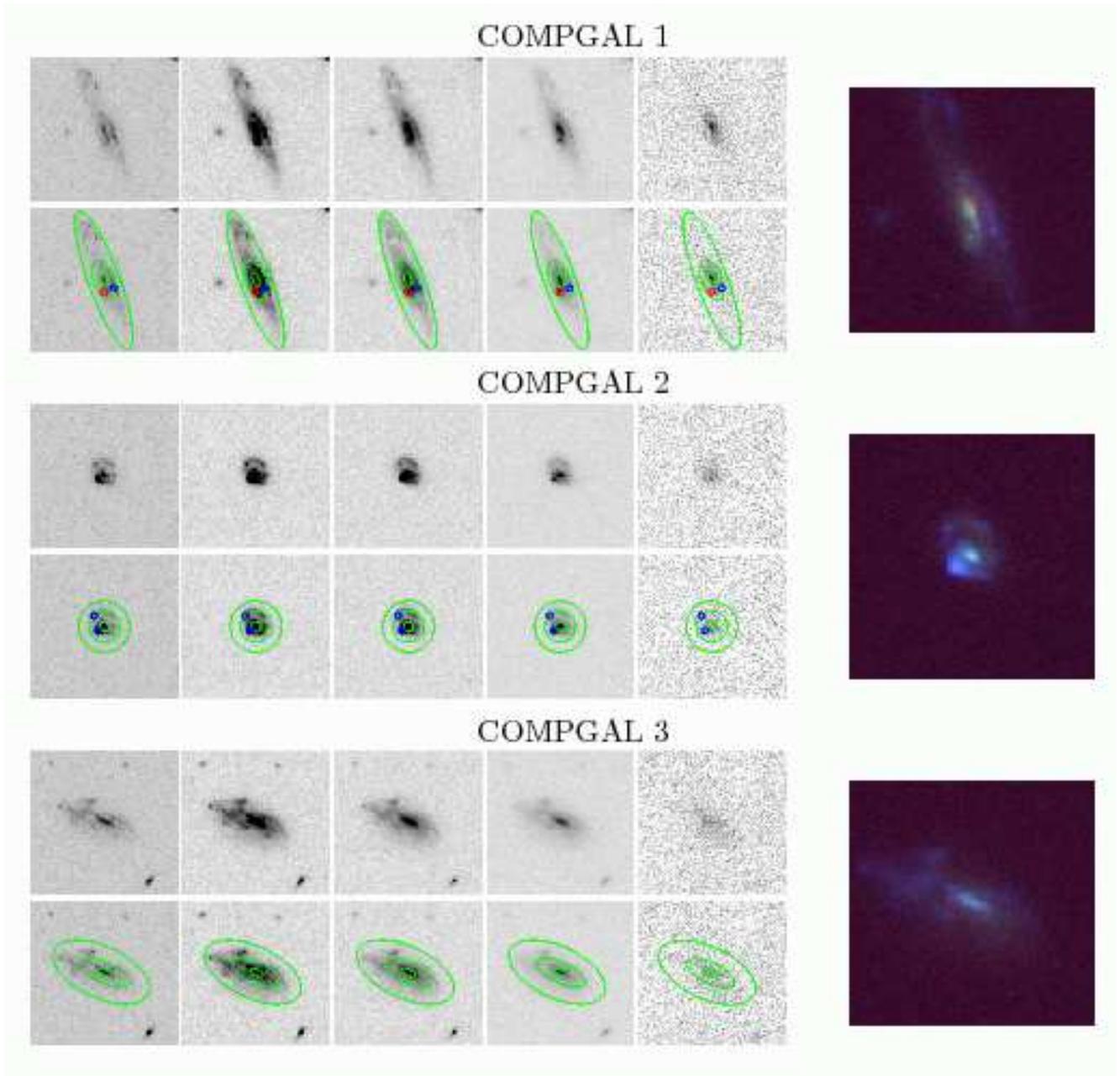}

\caption{\label{fig:COMPGALsA} COMPGALs 1 - 3.  Most of the Comparison galaxies have irregular morphologies in the bluer bands and disk-like morphology in the the redder bands.
}
\end{figure*}

\begin{figure*}
\center
\includegraphics[scale=1.4 ]{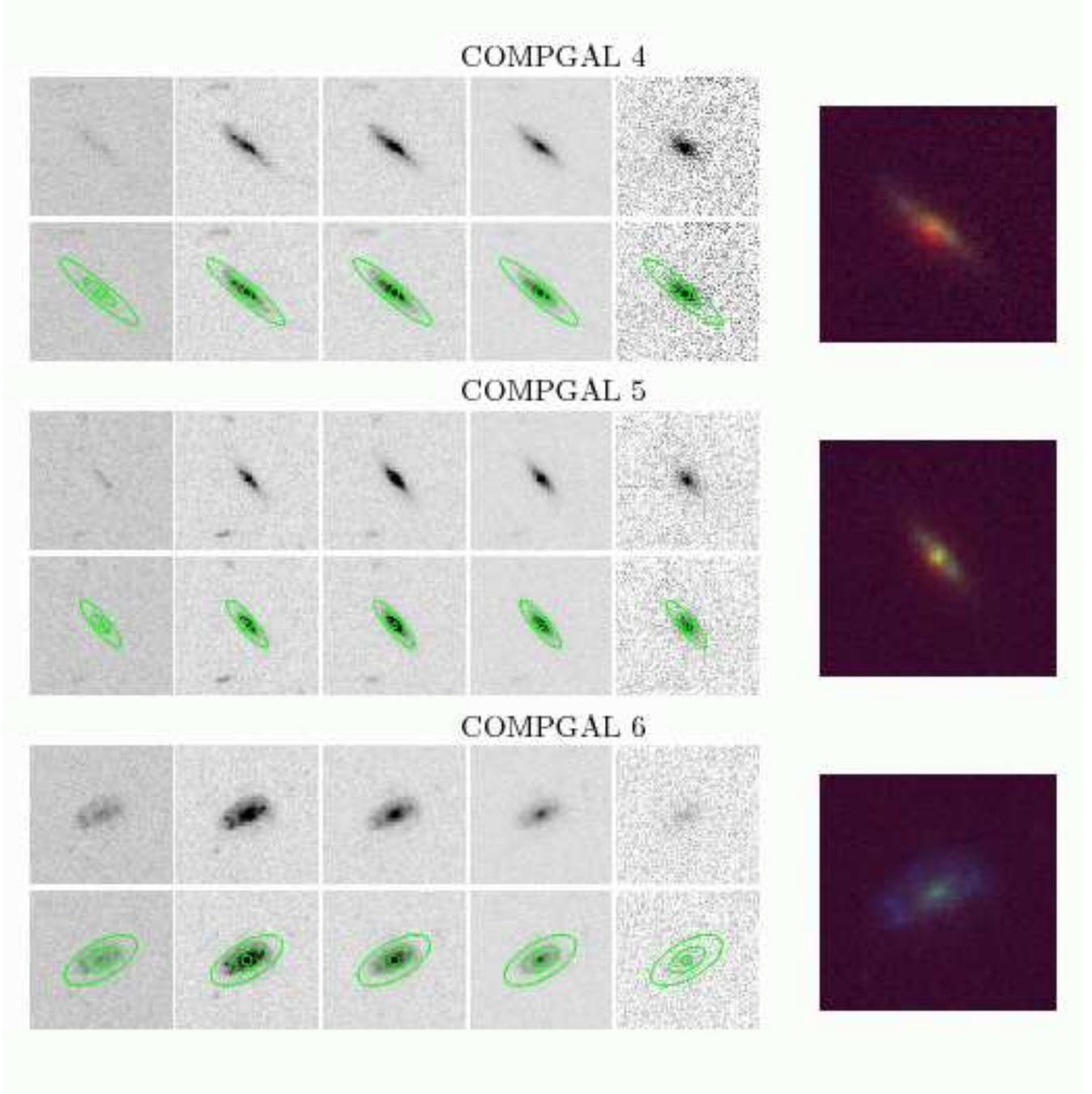}

\caption{\label{fig:COMPGALsB} COMPGALs 4 - 6.
}
\end{figure*}

\begin{figure*}
\center
\includegraphics[scale=1.4 ]{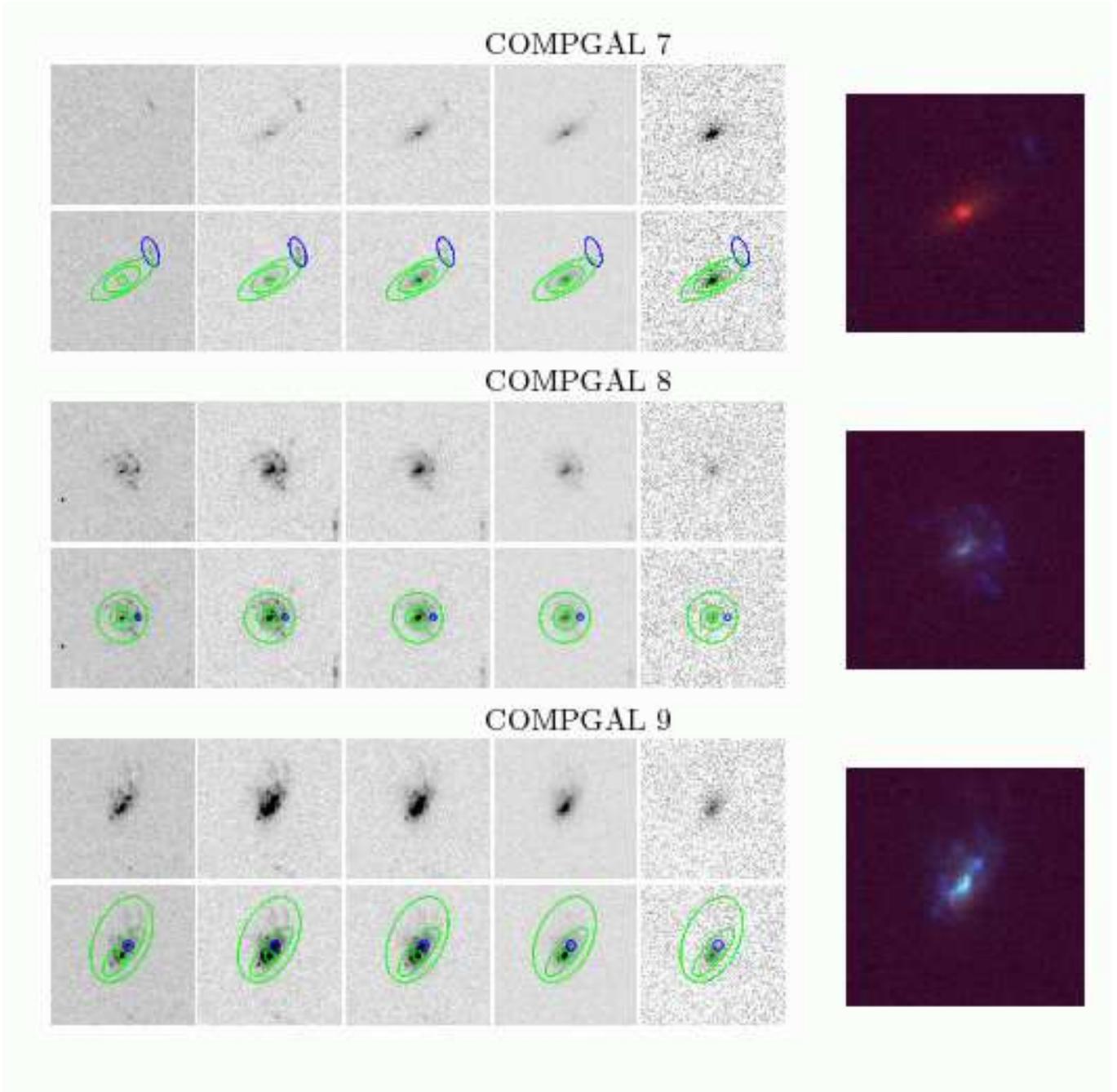}

\caption{\label{fig:COMPGALsC} COMPGALs 7 - 9.
}
\end{figure*}

\begin{figure*}
\center
\includegraphics[scale=1.4 ]{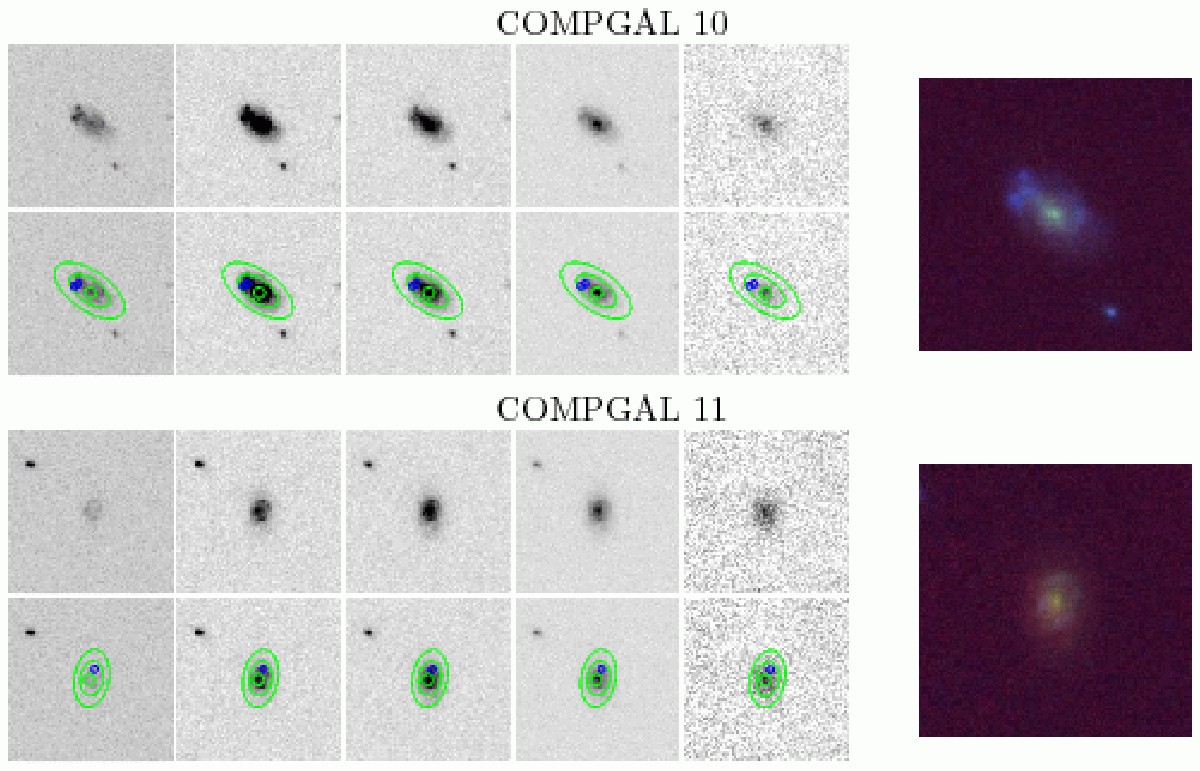}

\caption{\label{fig:COMPGALsD} COMPGALs 10 - 11.
}
\end{figure*}

In this section we present the 5 high resolution images of each galaxy in our study.  Figures \ref{fig:LIRGsA} - \ref{fig:LIRGsE} show the 5-band images of the LIRGs.  The images are 2.8$\arcsec$ on a side.   Color images of the LIRGs, created from the $B$, $i$, and $K$-band images (blue, green, and red respectively), are also presented.  Figures \ref{fig:COMPGALsA} - \ref{fig:COMPGALsD} show the comparison galaxies.  

In Sections 4 and 5 we will derive photometry from these images.  Section 4 will provide global photometry for each galaxy, from which we will derive physical properties such as absolute magnitude, rest-frame optical color, and stellar mass.  Section 5 will present photometry of galaxy subcomponents.   Figures \ref{fig:LIRGsA} - \ref{fig:COMPGALsD} show the apertures used to measure the SEDs of the galaxy subcomponents.  For each galaxy three apertures are centered on the nucleus (green apertures).  These are used to measure the SEDs of the inner, middle and outer regions of the galaxy.  Blue colored apertures are centered on blue knots.  Likewise, red colored apertures are placed around red knots.  Magenta colored apertures are placed around neighboring objects to see if they are consistent with minor merging or interacting objects. 

Detailed descriptions of the  the morphological characteristics of each galaxy are given in Appendix A (LIRGs) and B (comparison galaxies).  A summary of these characteristics and a discussion of the morphological differences between the LIRG and normal samples is provided here.  Visual morphological classifications are given in Table \ref{table:tot_phot}.
	 
\subsection{LIRG Morphology Summary}

	The largest morphological class in our LIRG sample is disk galaxies, 66\% (LIRGs 1, 3, 4, 6, 7, 8, 10, 11, 13, and 15) can be classified as disks.  The LIRG disks generally contain multiple, distributed blue knots suggestive of widespread active star formation.  They also tend to harbor very red central cores.  Several of the LIRG disks may be undergoing some sort of interaction or \emph{minor} merger.  The most obvious examples are LIRGs 1, 13, and 15, 33\% of LIRG disks and 20\% of the entire sample.  
	
	At least one of the LIRG disks, LIRG 4, is a barred spiral.  Bars can drive central star formation, and may be a trigger for LIRG events in disk galaxies.  While not obviously barred systems, the $K$-band images of LIRGs 3 and 6 and 13 show hints of linear structures in their cores.  These will be examined more closely in Paper 3.  
	
	While minor mergers and interactions may be contributing to the star formation of several LIRG disks, major mergers are not.   Only one galaxy in our sample is consistent with a major merger, LIRG 2.  In this case, the merger has destroyed any evidence of disk like morphology.  It has also produced both star formation and AGN activity.  LIRG 2 is a very strong x-ray source (see Table \ref{table:tot_phot}).  
	
	The second largest morphological class among the 15 LIRGs in our sample is AGN dominated objects.  These are galaxies with little evidence of major star formation, but with strong evidence for central AGN.  LIRGs 9, 12, and 14 (20\% of the entire LIRG sample) fall into this category.  All three of these systems are detected in x-rays (see Table \ref{table:tot_phot}), with LIRG 14 identified as a QSO by \citet{Szokoly04}.  Because these systems lack the multiple blue knots seen in the other LIRGs, we expect that most of the dust heating is occurring around the central AGN.  LIRGs 1 and 2 are also x-ray sources, and probably contain AGN, but their morphologies are dominated by  multiple blue knots suggesting strong star formation.
	
	The one additional LIRG in our sample,  LIRG 5, is a compact galaxy with several bright blue knots.  Some authors have suggested that the luminous blue compact galaxies seen at intermediate redshifts, are dwarf systems undergoing starbursts \citep[e.g.]{Koo95,Phillips97,Guzman97}.  Others have suggested that these galaxies are actually mergers of large galaxies seen in mid-merger \citep{Hammer05}.  In the later scenario, the high surface-brightness central cores are easy to observe, but the tidal features are low surface brightness and difficult to detect.  LIRG 5 does have a low surface brightness component surrounding the central core.  This low-surface brightness component is most easily seen in the $V$, $i$, and $z$- bands.

\begin{figure}
\includegraphics[scale=0.5]{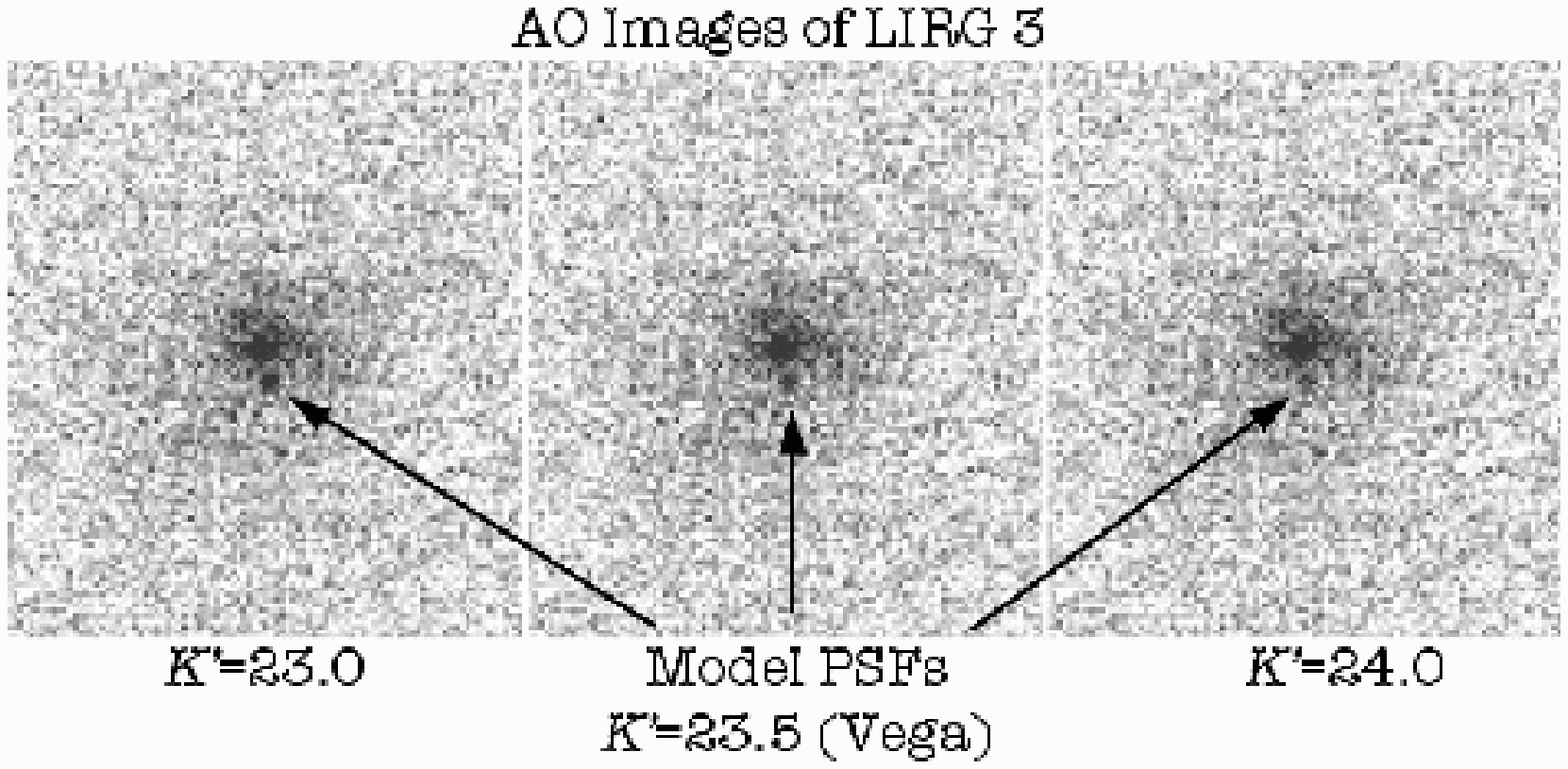}
\caption{\label{fig:model_knot} This figure shows the AO image of LIRG 3.  In each panel we have embedded a model PSF to mimic a star forming region. From left to right, the apparent magnitude of these model PSFs vary from $K'=23 - 24$ (Vega). While we can easily detect point-like structures that are brighter than $K'=23.5$, we do not find any structures like this in the actual AO images of the LIRGs.  This places a limit on the numbers of red super giant (RSG) stars within the knots that are seen at blue wavelengths.  These knots must contain fewer than $\sim$ 4000 RSGs (for a $z=0.8$ galaxy.)}
\end{figure}

\subsection{IR Morphologies of the LIRGs}		

	A unique aspect of our study over previous studies of intermediate redshift LIRGs is the inclusion of high spatial resolution IR imaging.  IR wavebands are preferentially less affected by dust in comparison with bluer wavebands.  In the bluer wavebands, the LIRGs in our sample contain significant substructure.  Most of the LIRGs contain one or more tight blue knots, presumably sites of recent star formation.  LIRG 11 shows evidence of a dust lane.  
	
	In the $K$-band, however,  these galaxies appear very smooth.  The bright blue knots generally are not bright enough in the $K$-band to be detected at our flux levels.  In addition, the $K$-band images do not reveal substructures missed in the bluer images.  Partly this is because the \HST\ images are significantly deeper than the AO images.  But for point like sources, as the blue knots appear to be, AO should greatly enhance sensitivity above the background \citep[e.g.][]{MelbourneAO07}.  For instance, Figure \ref{fig:model_knot} shows the AO image of LIRG 3 where we have placed model PSFs into the field.   We can visually detect point-like structures as faint as $K < 23.5$ (Vega).  Structures like these are not seen in our actual images.  
	
	The blue knots that are seen in the optical are presumably sites of significant and recent star formation, possibly containing multiple super star clusters (SSCs).  The most luminous red stars found in SSCs are red super giants (RSGs) with luminosities as bright a $M_J=-10$ to $-11$ \citep{Davidge06}.  Based on the lack of point-like sub-structure seen in the $K$-band, we can place an upper limit on the number of RSGs in these star forming knots.  Given the knot flux limit, $K<23.5$, and the typical LIRG redshift, $z\sim0.8$, these knots must contain fewer than 4000 RSGs.  For comparison, the core of 30 Doradus, a super star cluster in the Large Magellanic Cloud, contains $\sim 30$ RSGs \citep[e.g.][]{McGregor81}, so each of the  luminous blue knots probably contain fewer than 100 super star clusters.       

\subsection{Morphology of the Comparison Galaxies}
	
	The ``normal'' (non-LIRG) galaxies in our sample appear to be a mixture of irregulars and spirals.  If the classifications are made in the bluer bands one would conclude that this set of galaxies is primarily Irr.  But classifications in the redder bands suggest primarily disk like morphologies with red cores, similar to the majority of the LIRGs.  As a result, the differences in star formation between the ``normals'' and the LIRGs seem to be unrelated to morphology.  
	
	To drive this point home we can look at those systems that appear to be disturbed (possibly from the result of a minor merger) in the LIRG and normal samples. While a third of the LIRGs appear to be undergoing an obvious merger (major or minor), at least that fraction if not more of the ``normal'' systems have disturbed optical morphologies.  Examples of optically disturbed galaxies in the comparison sample include,  2, 3, 6, 8, 9, and 10, $\sim50$\%.  In addition COMPGALs 4 and 7 have very blue nearby companions that may be disrupting their morphology.  If we assume that the disturbed morphologies of the comparison sample are the result of minor mergers, then we must conclude minor mergers are not \emph{sufficient} for the production of a LIRG event.  
	

\begin{figure}
\includegraphics[scale=0.6]{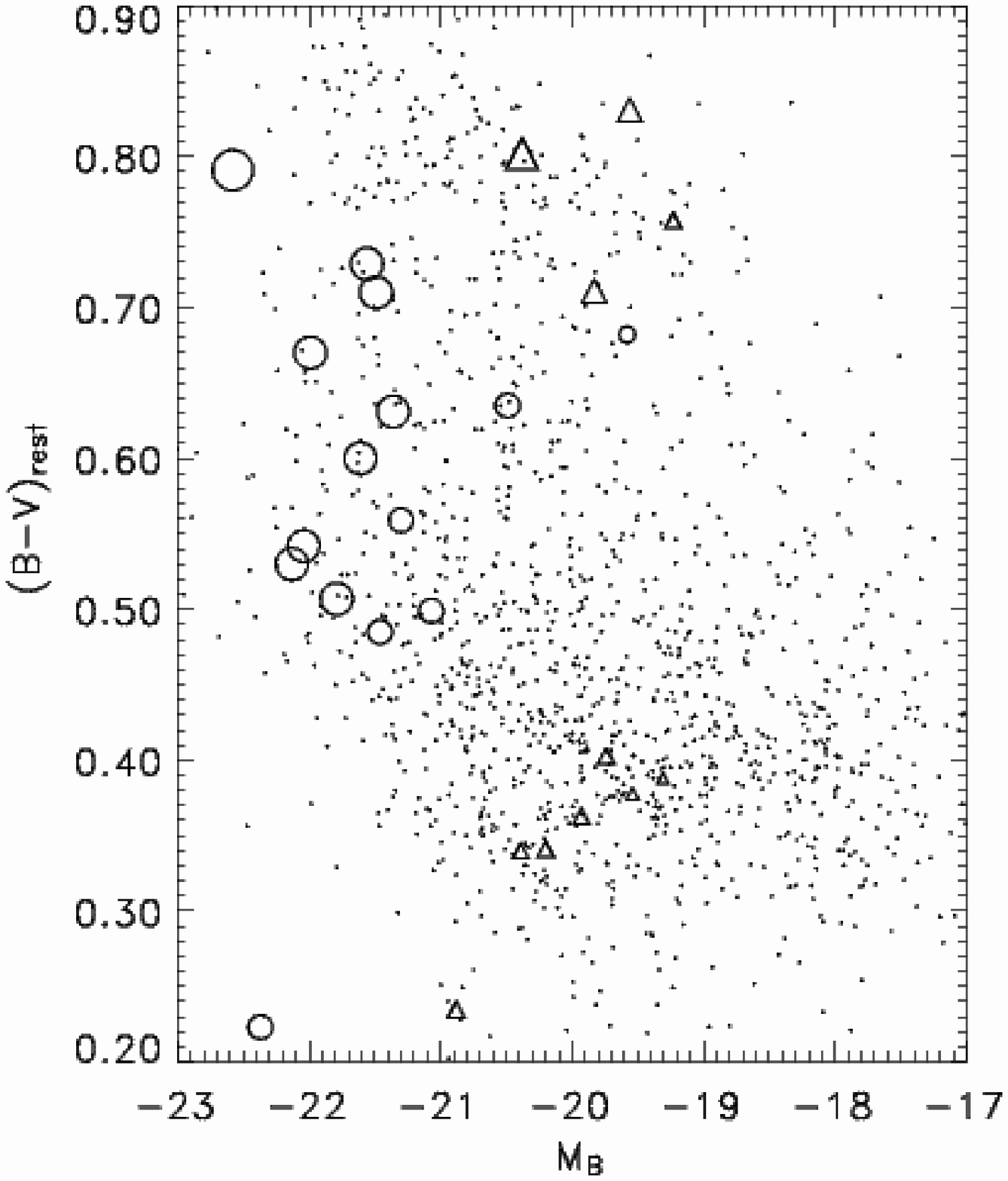}
\caption{\label{fig:cmd} The Color-Magnitude diagram (Vega mags) for the LIRGs (circles) and the ``normal'' (triangles) galaxies.  The size of the points is proportional to the stellar mass.  The small points are the photometry of 1000 galaxies in GOODS-N as measured by \citet{MelbourneGN07}.  We see that while the LIRGs tend to congregate in the ``green valley'' between the ``red sequence'' and ``blue cloud'', the comparison sample is bimodal in color.  The LIRGs also tend to higher luminosities and larger stellar masses.  Most of the LIRGs are brighter than $M_B < -21$, whereas none of the comparison galaxies are that bright.} 
\end{figure}

\begin{deluxetable*}{cccccccccccccc}
\tablewidth{0pt}
\tabletypesize{\tiny}
\setlength{\tabcolsep}{0.0in} 
\tablecaption{Photometry Summary \label{table:tot_phot}}
\tablehead{\colhead{Object} & \colhead{$z$} & \colhead{$B$} & \colhead{$V$} & \colhead{$i$} & \colhead{$z$}& \colhead{$K$} & \colhead{$M_B$ } & \colhead{$B-V\tablenotemark{a}$} & \colhead{log($M_\odot$)} & \colhead{$f_{24 \mu m}$ } & \colhead{$L_{IR}$  } & \colhead{$L_X$\tablenotemark{b}} & \colhead{Morphology} \\ & & \colhead{(AB)} &\colhead{(AB)} &\colhead{(AB)} &\colhead{(AB)} &\colhead{(Vega)} &\colhead{(Vega)} & \colhead{(Vega)} & & \colhead{($mJ$)} & \colhead{($L_\odot$)} & \colhead{}}
\startdata
                   LIRG 1&    0.74&   23.17&   22.18&   21.17&   20.75&   19.60&  -21.36&    0.63&   10.62&   0.293&  2.05e+11& 42.5 &               Disk\\
                   LIRG 2&    0.61&   21.74&   20.77&   19.90&   19.70&   18.58&  -22.04&    0.54&   10.74&   0.639&  3.12e+11& 43.5 &           Merger\\
                   LIRG 3&    1.02&   23.19&   22.75&   21.91&   21.44&   20.73&  -21.79&    0.51&   10.58&   0.176&  2.97e+11& -&              Disk?\\
                   LIRG 4&    1.10&   24.31&   23.17&   21.80&   20.92&   19.32&  -22.58&    0.79&   11.38&   0.136&  3.19e+11&  -&              Disk\\
                   LIRG 5&    1.08&   24.95&   23.71&   22.62&   21.92&   20.48&  -21.49&    0.71&   10.81&   0.223&  5.22e+11&  -&           Compact\\
                   LIRG 6&    0.84&   23.00&   22.38&   21.48&   21.20&   20.67&  -21.46&    0.49&   10.41&   0.172&  1.55e+11&     -&           Disk\\
                   LIRG 7&    1.04&   24.73&   23.53&   22.46&   21.70&   20.41&  -21.56&    0.73&   10.87&   0.167&  3.07e+11&   -&             Disk\\
                   LIRG 8&    1.26&   24.26&   23.86&   23.24&   22.52&   21.38&  -21.61&    0.60&   10.67&   0.227&  1.79e+11&   -&             Disk\\
                   LIRG 9&    1.22&   23.46&   22.85&   22.23&   21.72&   20.69&  -22.13&    0.53&   10.76&   0.159&  6.86e+11&    43.9&   Compact/Disk?\\
                  LIRG 10&    1.00&   23.92&   22.87&   21.79&   21.16&   19.92&  -21.99&    0.67&   10.94&   0.329&  5.79e+11&       -&         Disk\\
                  LIRG 11&    0.95&   25.86&   24.37&   22.96&   22.50&   20.82&  -20.49&    0.63&   10.28&   0.102&  1.14e+11&    -&            Disk\\
                  LIRG 12&    0.55&   24.50&   22.97&   21.91&   21.69&   20.79&  -19.58&    0.68&   10.00&   0.504&  1.89e+11&    42.2 &          Disk?\\
                  LIRG 13&    0.89&   23.39&   22.82&   21.93&   21.46&   20.70&  -21.30&    0.56&   10.48&   0.194&  1.96e+11&     -&           Disk\\
                  LIRG 14&    0.84&   20.88&   20.47&   20.57&   20.33&   19.82&  -22.37&    0.22&   10.33&   0.560&  5.67e+11&       44.1&          QSO\\
                  LIRG 15&    0.65&   22.70&   21.85&   21.08&   20.84&   20.27&  -21.07&    0.50&   10.28&   0.228&  1.44e+11&         -&       Disk\\
                COMP 1&    0.42&   22.65&   21.75&   21.28&   21.11&   20.56&  -19.74&    0.40&    9.59&  -     & -&    -&             Disk\\
                COMP 2&    0.67&   22.81&   22.44&   21.85&   21.70&   21.44&  -20.39&    0.34&    9.74&   0.045&  3.24e+10&      -&   Disk/Merger?\\
                COMP 3&    0.44&   22.48&   21.67&   21.26&   21.07&   20.72&  -19.93&    0.36&    9.59&  -& -&       -&          Disk\\
                COMP 4&    0.50&   24.72&   22.95&   21.98&   21.51&   20.23&  -19.23&    0.76&    9.99&   0.116&  3.33e+10&         -&        Disk\\
                COMP 5&    0.73&   25.54&   23.44&   22.12&   21.68&   20.26&  -20.38&    0.80&   10.52&  -& -&   -&      Disk/Compact\\
                COMP 6&    0.42&   23.06&   22.17&   21.72&   21.56&   21.25&  -19.31&    0.39&    9.39&  -& -&     -&            Disk\\
                COMP 7&    0.77&   26.69&   24.59&   23.17&   22.59&   20.91&  -19.56&    0.83&   10.24&   0.102&  7.18e+10&    -&             Disk\\
                COMP 8&    0.70&   23.11&   22.75&   22.20&   22.03&   21.45&  -20.20&    0.34&    9.66&  -& -&     -&            Disk\\
                COMP 9&    0.70&   22.64&   21.60&   21.53&   21.40&   21.21&  -20.88&    0.23&    9.76&   0.093&  5.22e+10&     -&       Irregular\\
               COMP 10&    0.46&   22.96&   22.16&   21.72&   21.54&   21.19&  -19.54&    0.38&    9.46&  -& -&   -& Disk/Minor Merger?\\
               COMP 11&    0.61&   24.52&   23.06&   22.04&   21.62&   20.83&  -19.83&    0.71&   10.15&  -& -&        -&         Disk\\
\enddata
\tablenotetext{a}{Restframe}
\tablenotetext{b}{[log ergs s$^{-1}$]}
\end{deluxetable*}

\section{Global Properties of the LIRG and Comparison Sample Galaxies\label{sec:tot_phot}}

In the previous Section we demonstrated that morphology alone does not differentiate between LIRGs and non-LIRGs, and minor mergers are not a ``sufficient'' condition for LIRG production.  In this section we calculate and compare the global properties of each galaxy, i.e. absolute magnitude, rest-frame color, stellar mass, and IR luminosity.  Global properties may shed more light on the differences between the LIRG and non-LIRG samples.  These derived quantities are based on photometry of the high resolution \HST\ and AO images,  published photometry of the lower resolution mid-IR and x-ray images, and published photometric and spectroscopic redshifts.

We measured the total magnitudes of the galaxies in the $B,\;V,\;i,\;z$ and $K$-band images in the following way.  We first extracted postage stamp images of each galaxy.  These images were $9\arcsec$ on a side, and all were resampled to the AO, 0.04$\arcsec$ pixel scale.  We created a white light image by summing the \HST\ images.  We ran the galaxy search algorithm  \sext\ \citep{Bertin96} on the white light images to create segmentation maps of the regions around each of our science targets.  These maps identified any neighbor galaxies.  We used the maps to mask out pixels from neighboring galaxies before measuring the magnitudes of the science targets.  

We performed curve of growth photometry on each LIRG and comparison galaxy.  The photometry was was done with circular apertures of increasing size out to a radius of $4\arcsec$.    All pixels associated with neighbors were masked out.  An initial estimate of the sky was made from the median of pixels unassociated with galaxies. The sky value was iterated so that the light in successively larger apertures remained constant beyond the edge of the galaxy.   

Table \ref{table:tot_phot} contains the results of the aperture photometry (columns 3 - 7).  We convert to absolute magnitude (column 8) and rest-frame color (column 9)  using the $K$-corrections developed by \citet{Willmer05} for the GOODS photometry.   We use the conversion to stellar mass given in  \citet{Bell05} and give the results in the column 10.  The Table \ref{table:tot_phot} also reports the mid-IR \tf\ photometry (column 11) as measured by \citet{LeFloch05} and the x-ray photometry (column 12) given by \citet{Rosati02}.
  
Figure \ref{fig:cmd} shows the color-magnitude diagram based on these measurements for the galaxies in our sample.  The LIRGs are shown as circles and the ``normal'' (non-LIRG) galaxies are shown as triangles. We scale the size of points in the figure by stellar mass.  For comparison we overplot (as small points) photometry of 1000 galaxies in the Team Keck Redshift Survey \citep{Wirth04} of GOODS-N as measured by \citet{MelbourneGN07}.  We see that that in general the LIRGs are more luminous and more massive than the comparison galaxies.  The typical LIRG has an absolute $B$-band magnitude brighter than $M_B < -21$ [Vega].  None of the comparison galaxies have a luminosity brighter than that limit. This luminosity division was anticipated based on the results of \citet{MKL05}, which found that 70\% of \emph{blue} galaxies brighter than $M_B=-21$ are LIRGs regardless of redshift.  The LIRGs are also typically higher stellar mass than the normal sample;  $M_{LIRG} \sim 10^{10.5} \;M_{\odot}$ vs. $M_{normal} \sim 10^{9.5} \;M_{\odot}$. 

We find that the comparison sample is made up of two distinct color classes; 1) Red galaxies with typical stellar mass of $10^{10} M_\odot$; and 2) Blue galaxies with typical masses of $10^{9.5} M_\odot$.  The LIRGs are more uniform in color, clustering around the green valley that separates the blue cloud from the red sequence, $(B-V)\sim0.7$ [Vega].    The one exception is the very blue LIRG 14 which is a QSO \citep{Szokoly04}, and not expected to have the colors of dusty star forming galaxies.  The ``green valley'' colors of the LIRGs might indicate that they are a transition population between the blue cloud of star forming galaxies and the red sequence. 

\begin {figure*}
\includegraphics[scale=0.8]{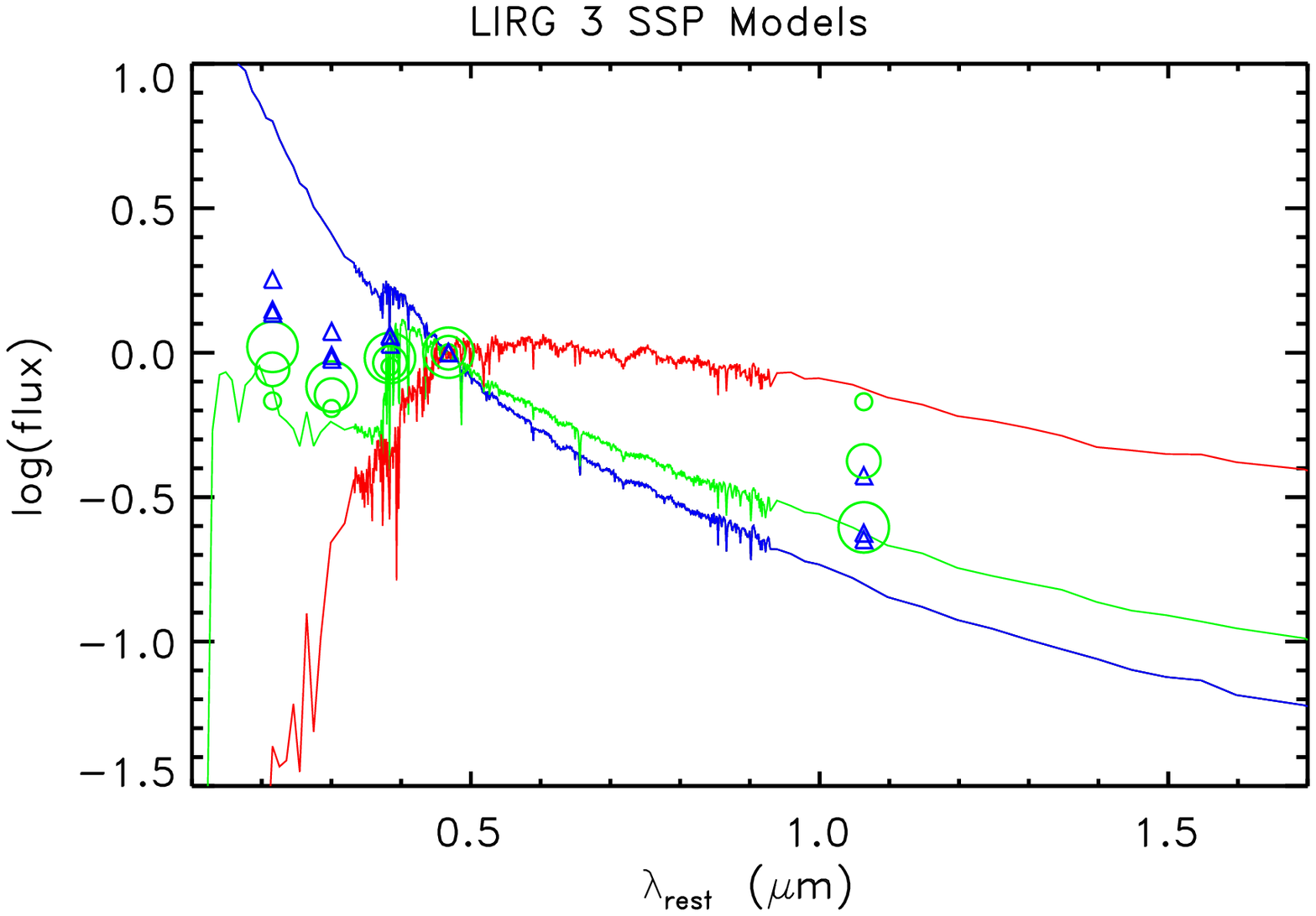}
\caption{ \label{fig:LIRGs_SED_A} SEDs of sub-components within LIRG 3, normalized in the observed $z$-band.  The photometry shown on these plots is from regions illustrated in Figure \ref{fig:LIRGsA}.  Green circles show the photometry of the main galaxy.  The smallest circle is the photometry of the galaxy core.  The intermediate sized circle shows the photometry from an annulus around the core.  The largest circle gives the photometry from the outer portions of each galaxy.  The triangles represent photometry of substructures, such as blue knots (blue triangles), red knots (red triangles), and neighboring objects (magenta triangles).  Over-plotted are galaxy stellar population synthesis models from \citet{BC03}.  All three models are single burst models (single stellar population, SSP, models) of differing ages.  The models have solar metallicity and a Chabrier initial mass function.  The Blue model has an age of 7 Myr, the green, an age of 300 Myr, and the red an age of 3 Gyr.  Figures \ref{fig:LIRGs_SED_B} - \ref{fig:COMPGALs_SED_C} will have the same format as this figure. }
\end{figure*}

\begin{figure}
\includegraphics[scale=0.4]{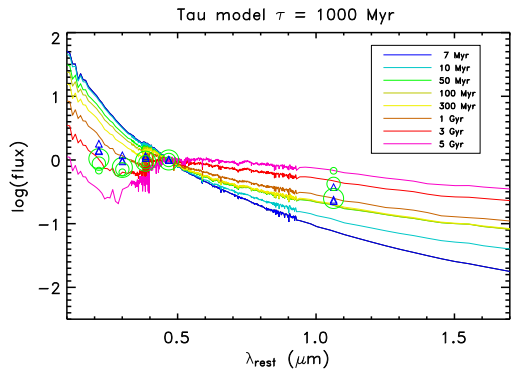}
\caption{ \label{fig:LIRG3_SED_tau} Similar to Figure \ref{fig:LIRGs_SED_A} only instead of SSP models, we plot models with exponentially declining star formation.  These tau models have an e-folding time of $\tau=1$ Gyr.  We plot multiple models each with a different age. While the SED of the LIRG 3 core does not fit one of these models, the model that is 3 Gyr old fits much better than an SSP model.}
\end{figure}

\begin{figure}
\includegraphics[scale=0.4]{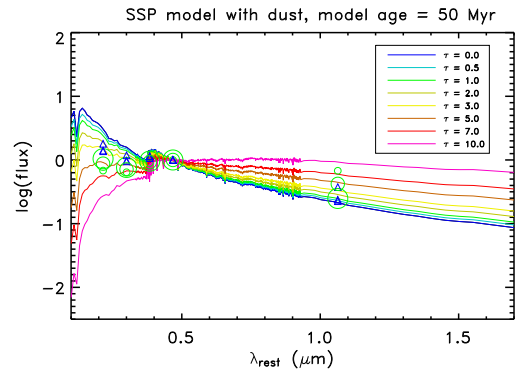}
\caption{ \label{fig:LIRG3_SED_dust} Similar to Figure \ref{fig:LIRGs_SED_A} only now the single burst models all have an age of 50 Myr, and the optical depth of dust changes.  Models plotted with bluer colors have less dust, while models with redder colors have more dust.  The core of the galaxy has an SED consistent with a single burst model of 50 Myr and a dust optical depth of $\tau=7$.  Because LIRGs are known to contain large amounts of dust, this model is reasonable for the galaxy core.}
\end{figure}

\begin{figure*}
\plottwo{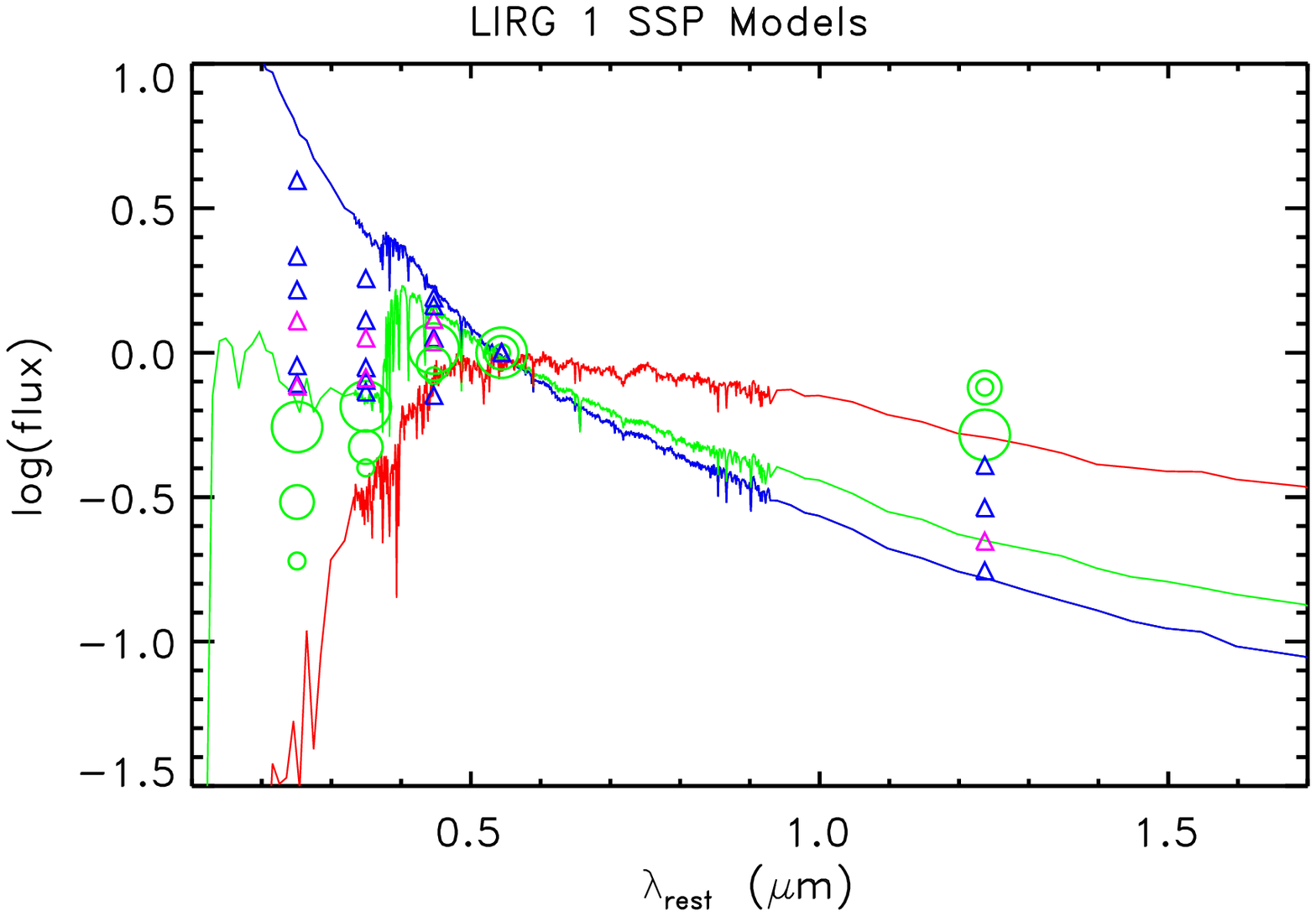}{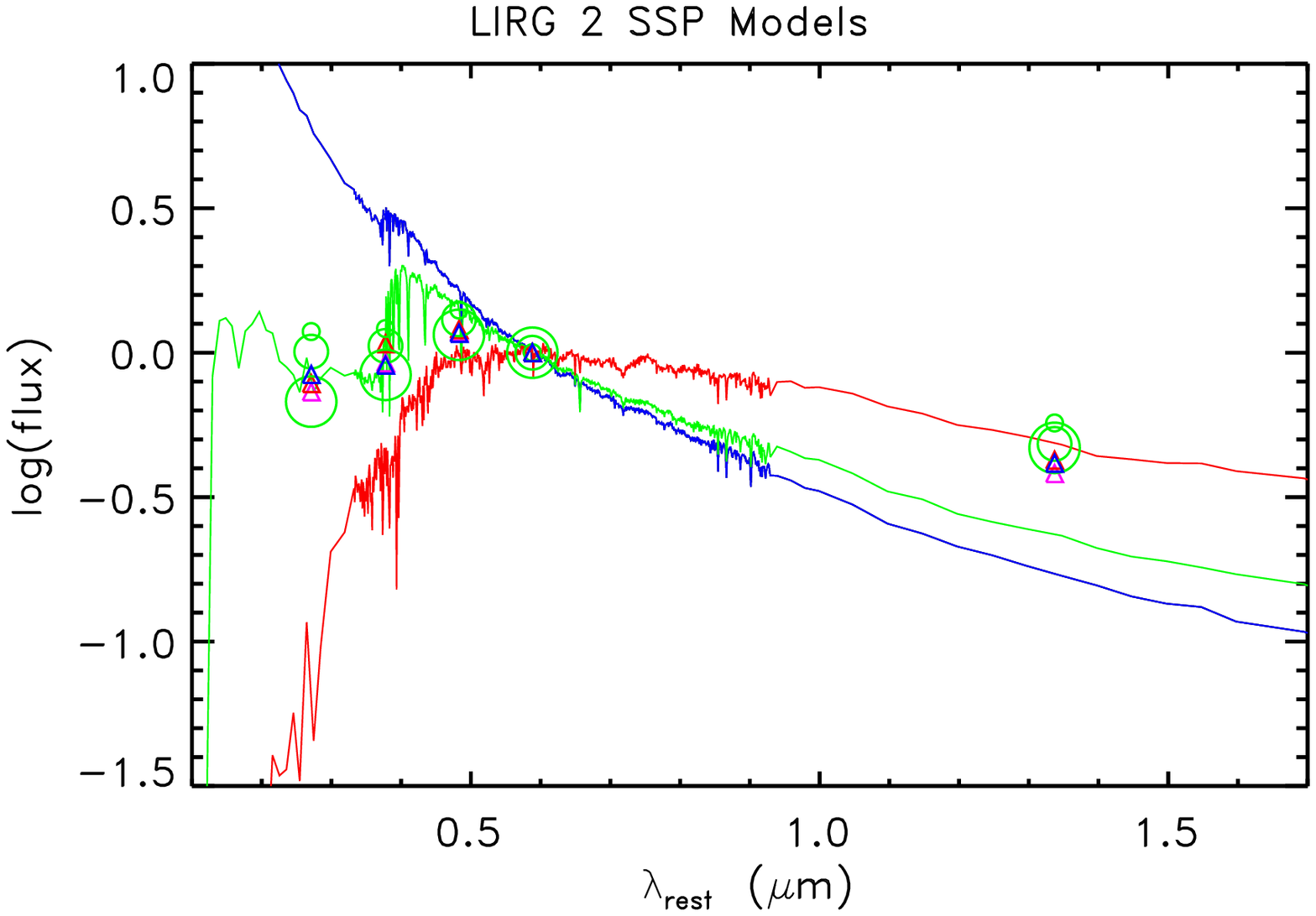}
\plottwo{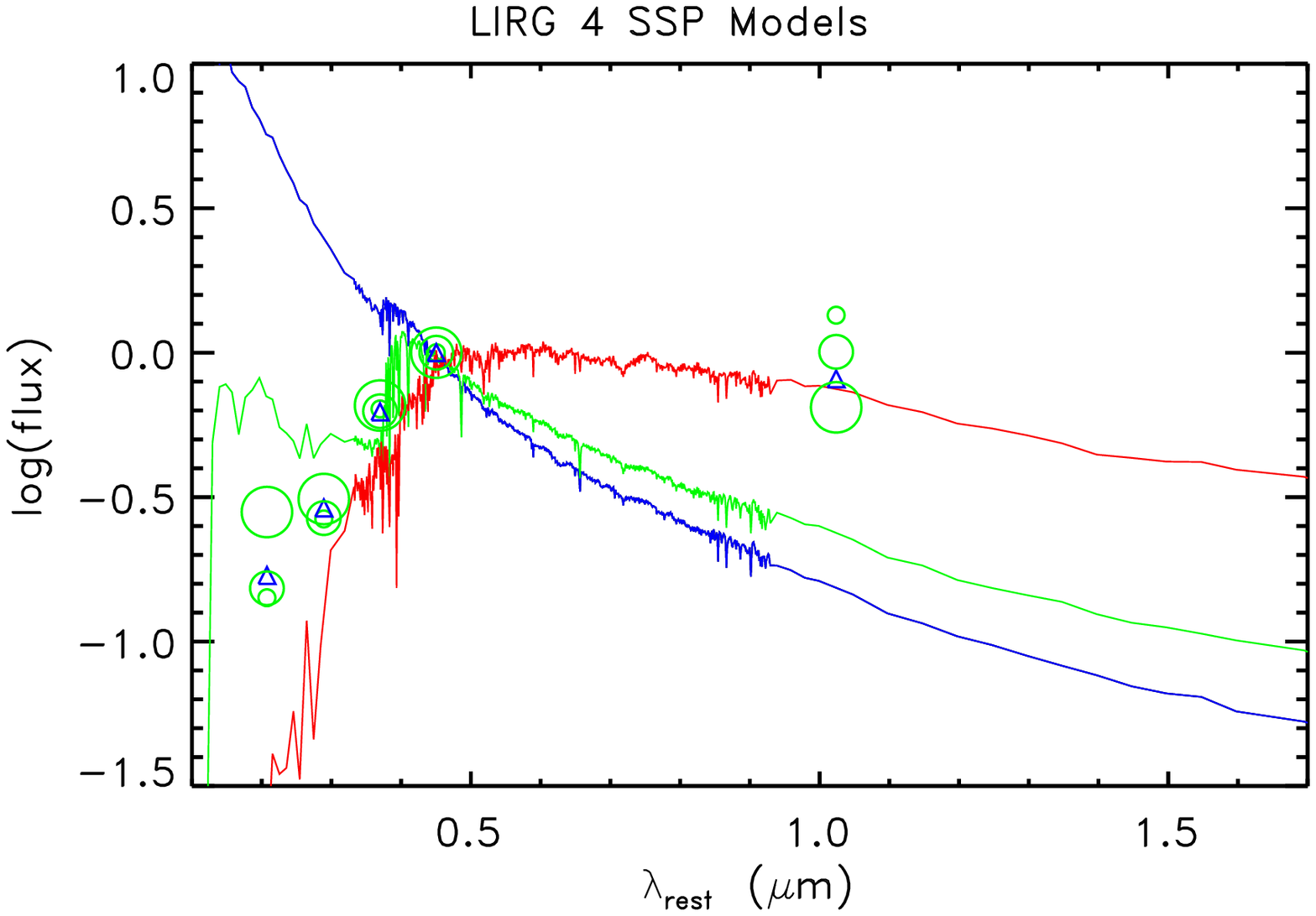}{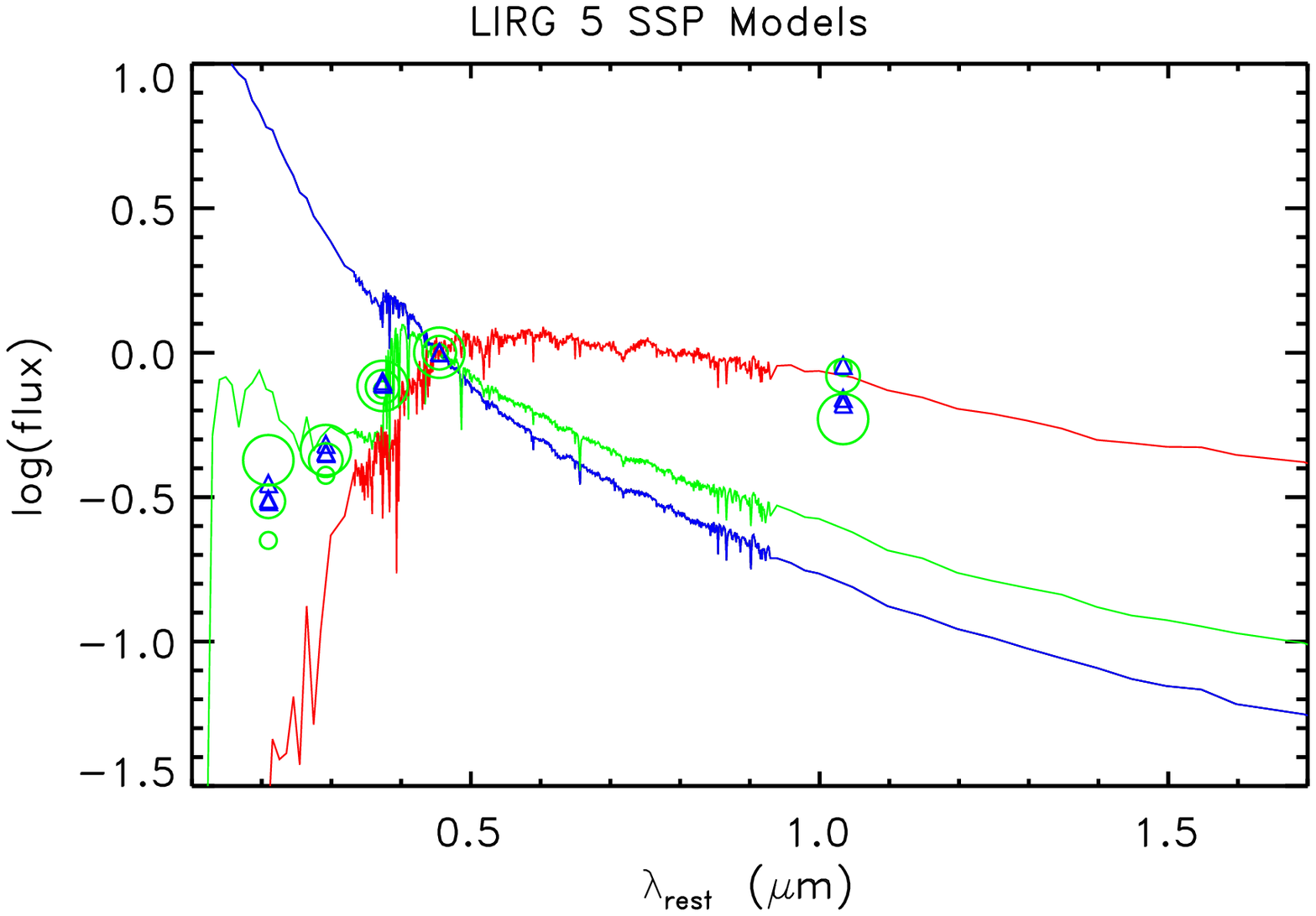}
\plottwo{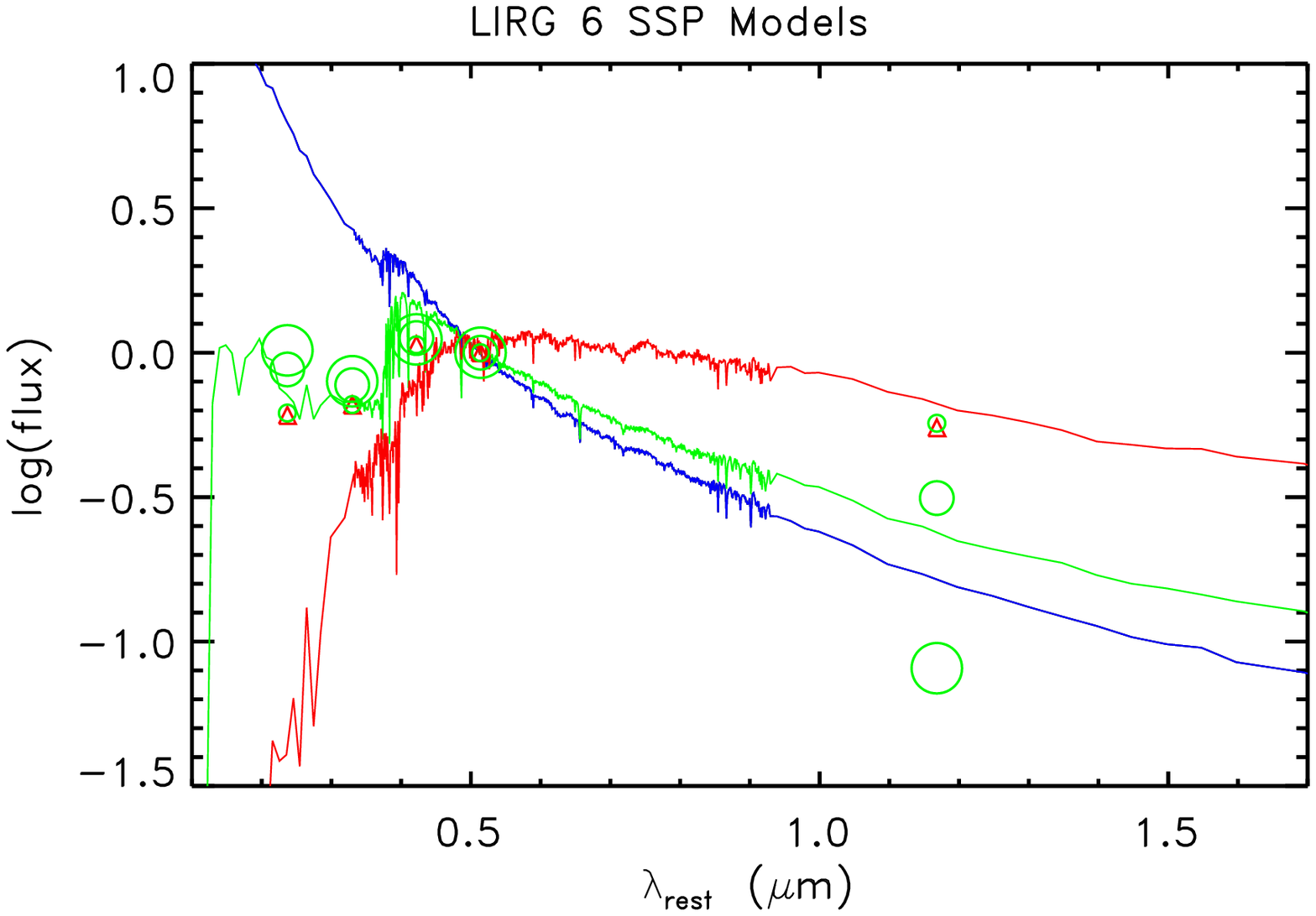}{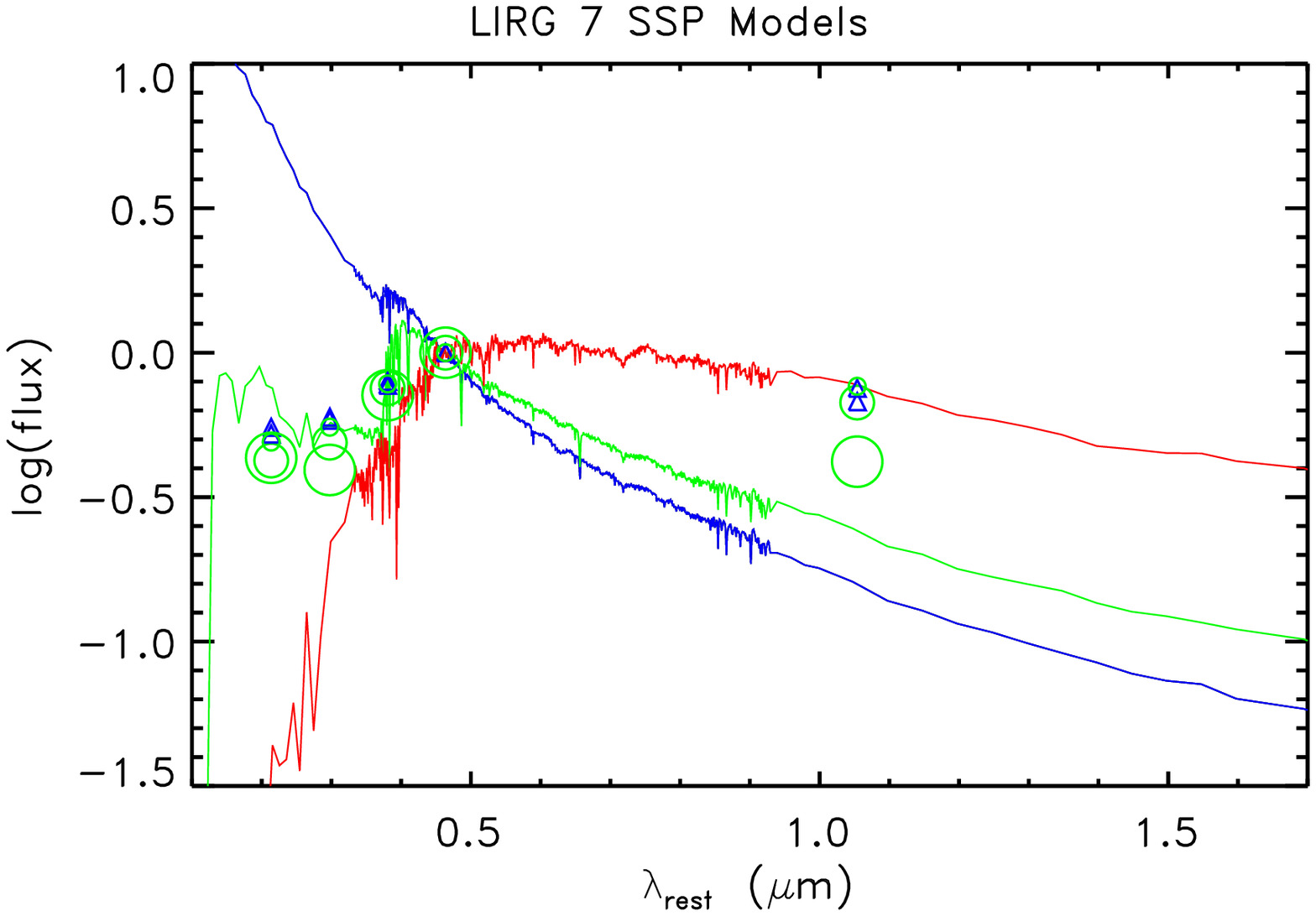}
\plottwo{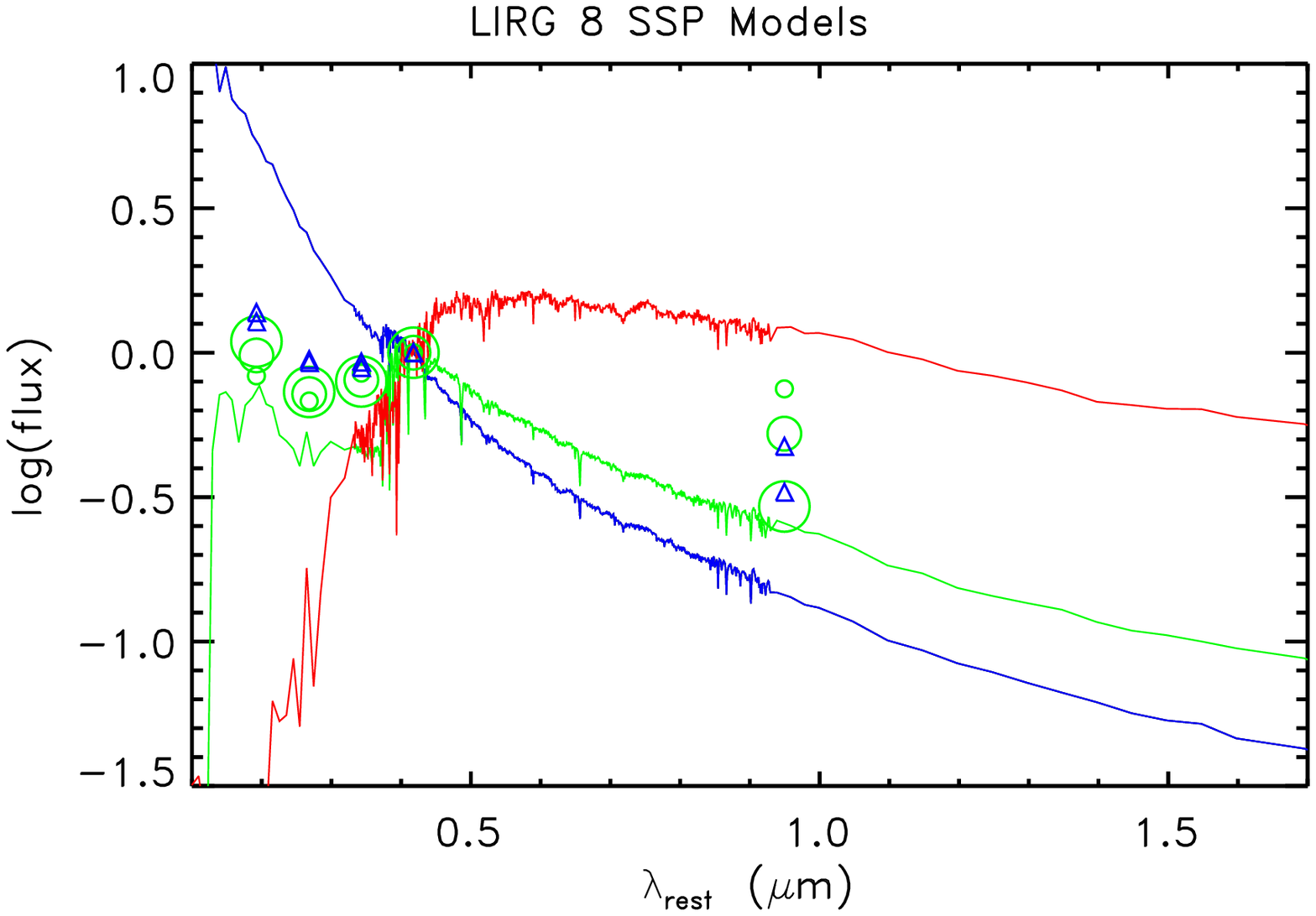}{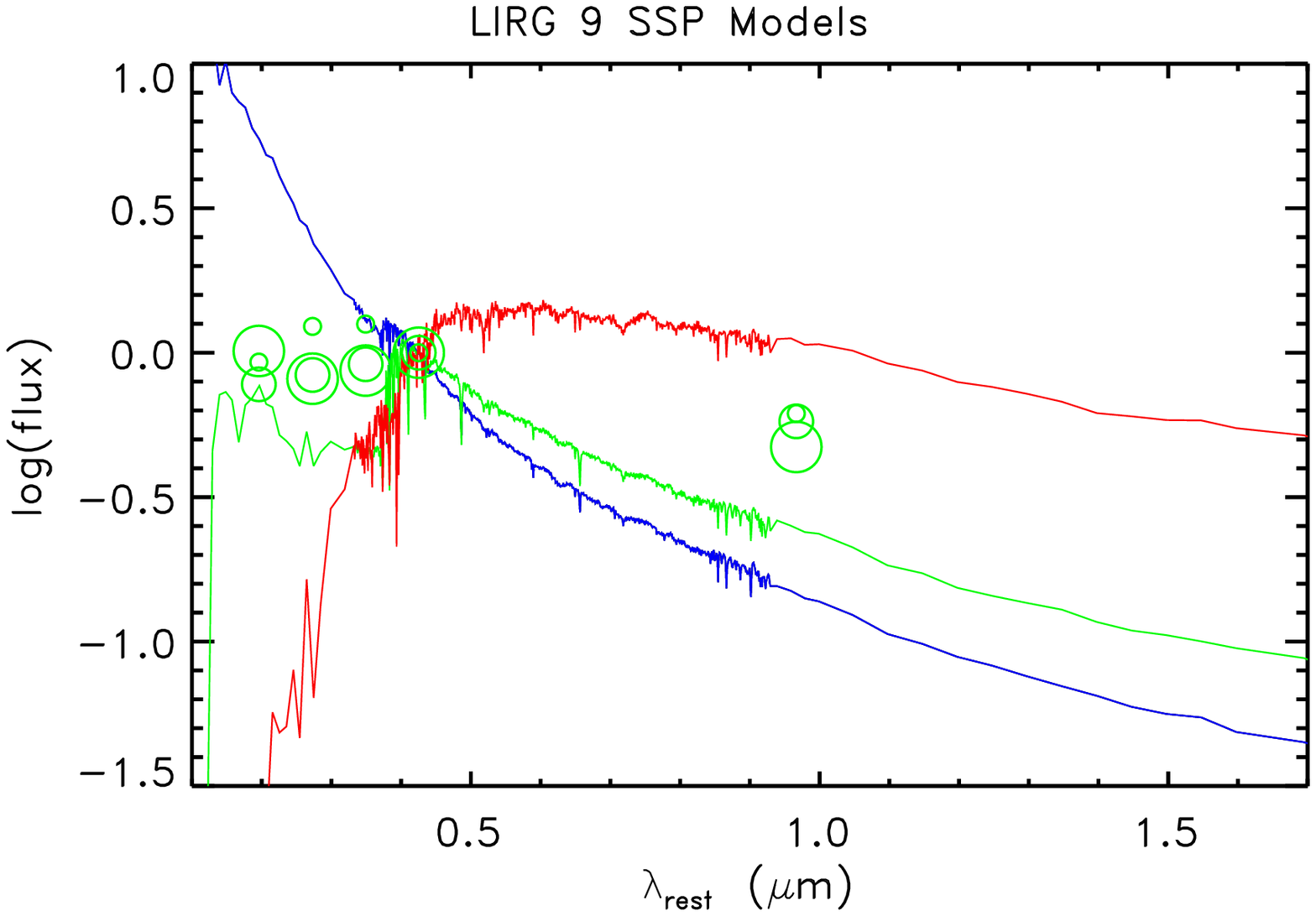}
\caption{\label{fig:LIRGs_SED_B} Same as Figure \ref{fig:LIRGs_SED_A} only now for SEDs of LIRGs 1 - 9.}
\end{figure*}

\begin{figure*}
\plottwo{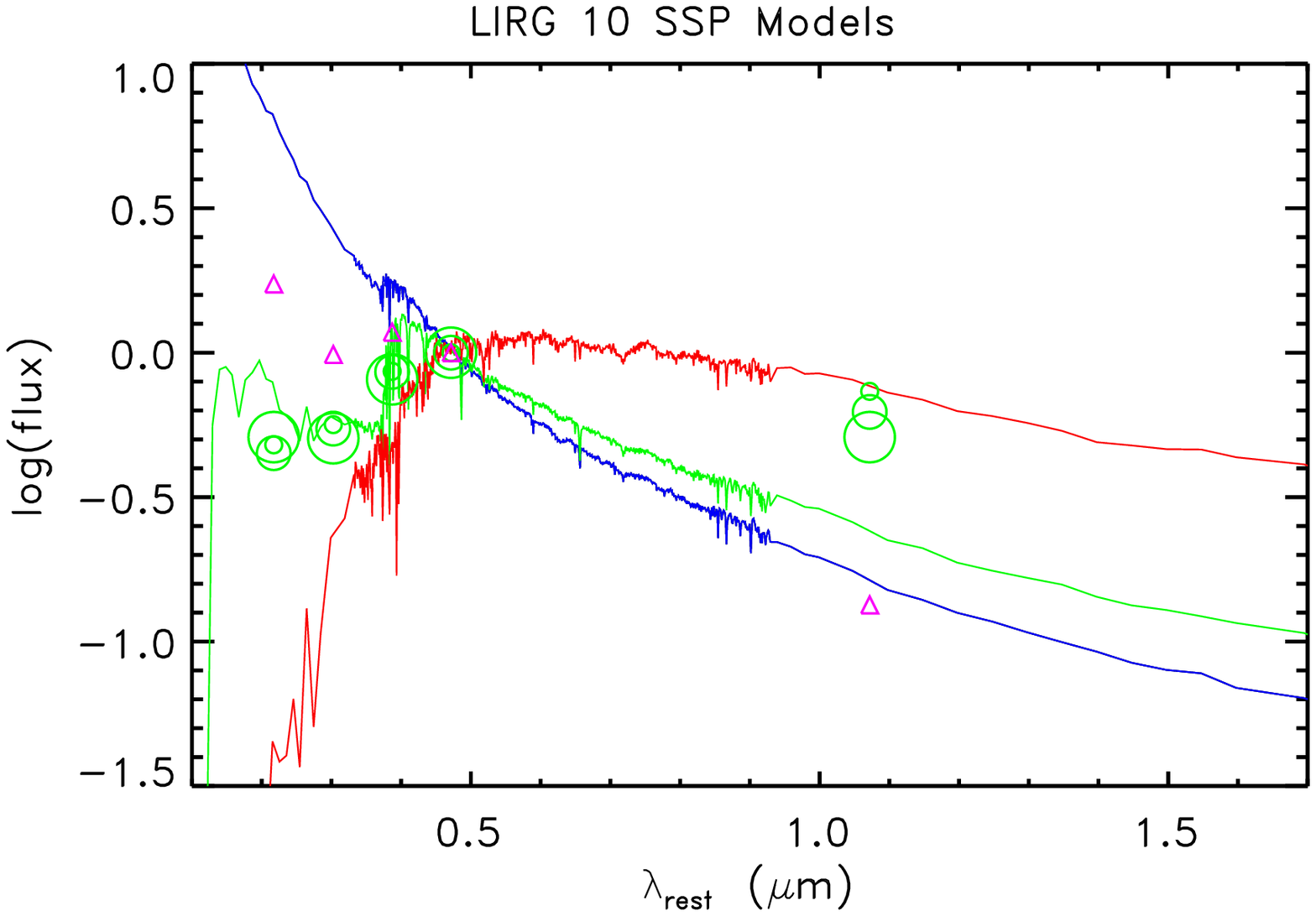}{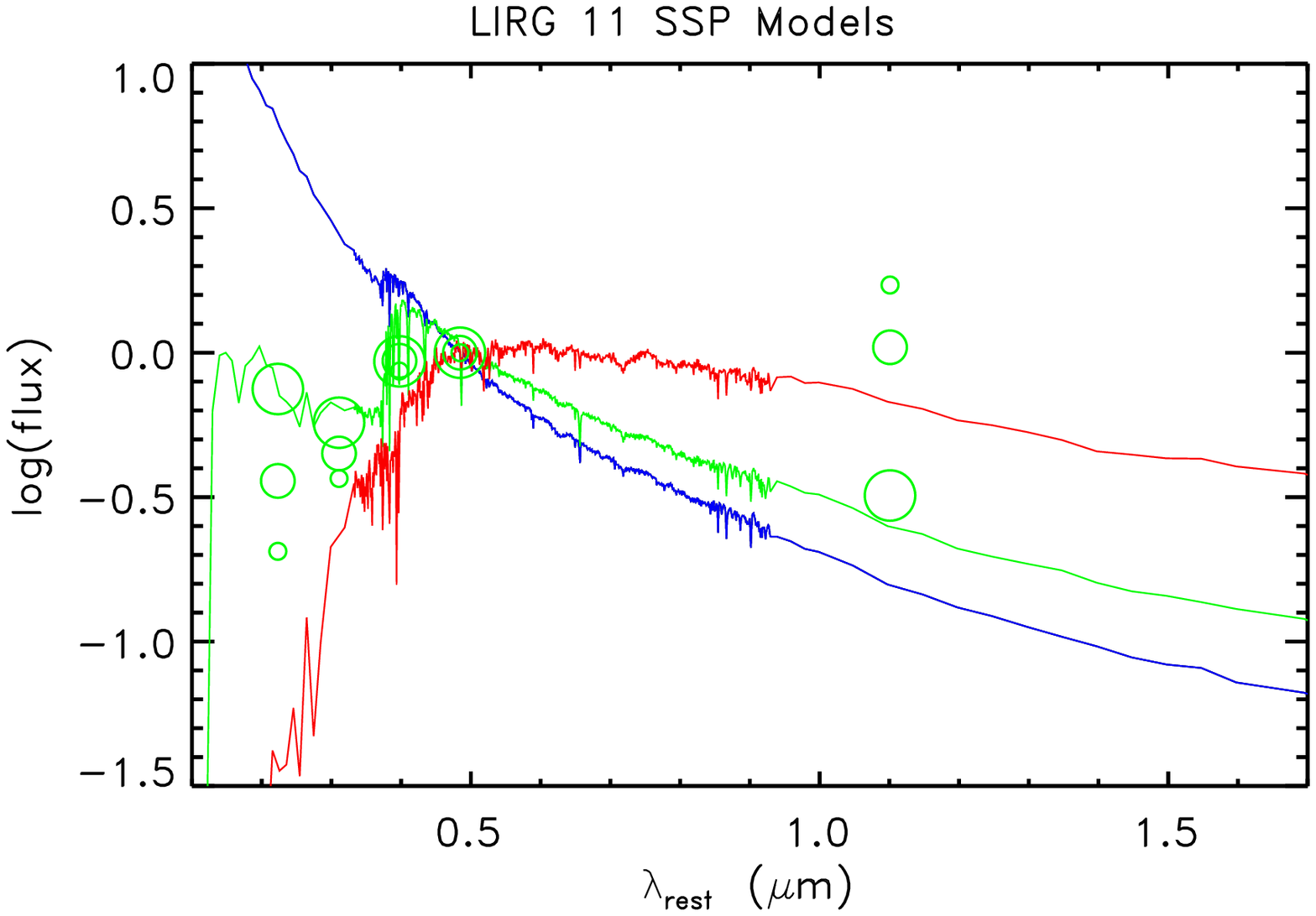}
\plottwo{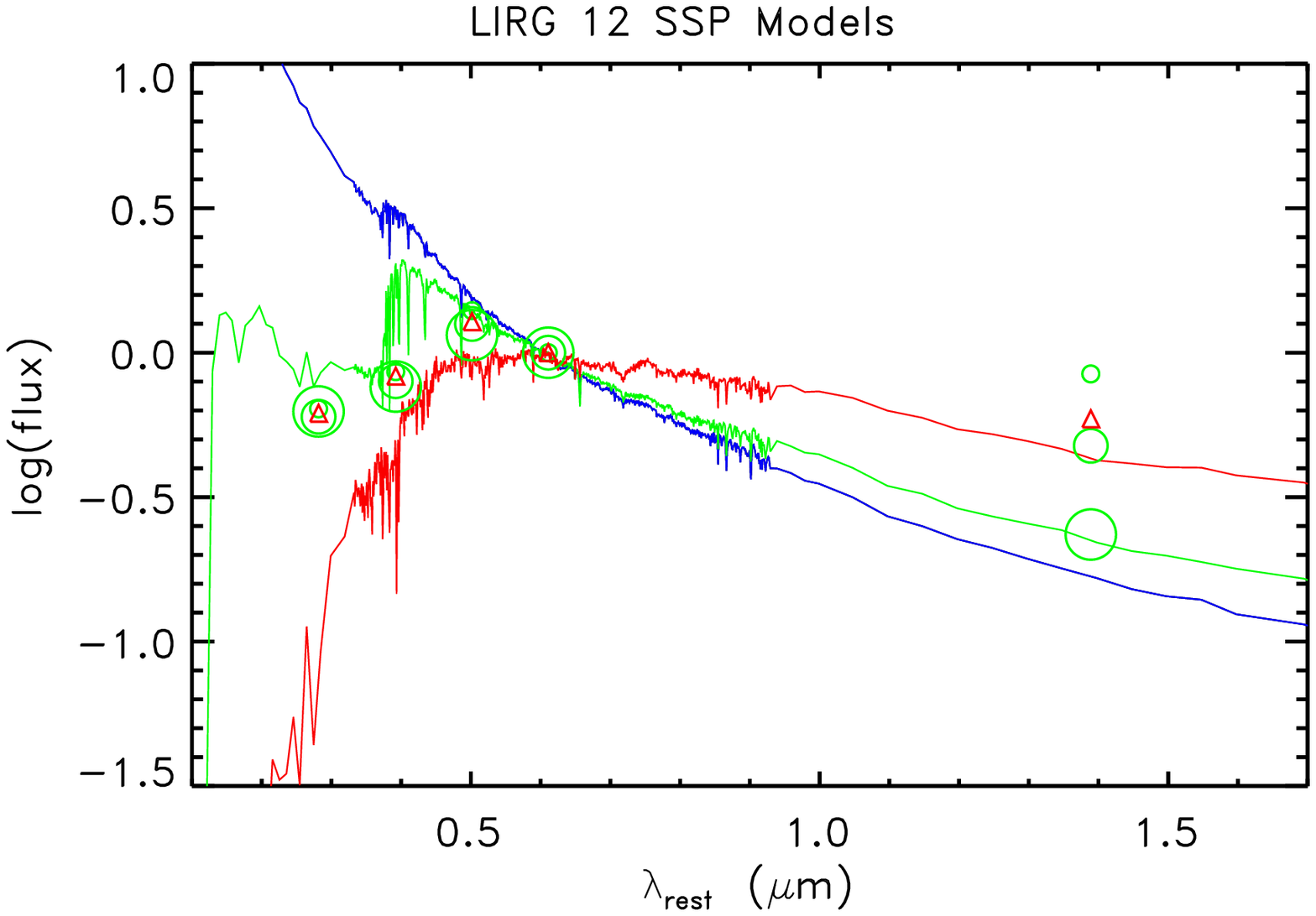}{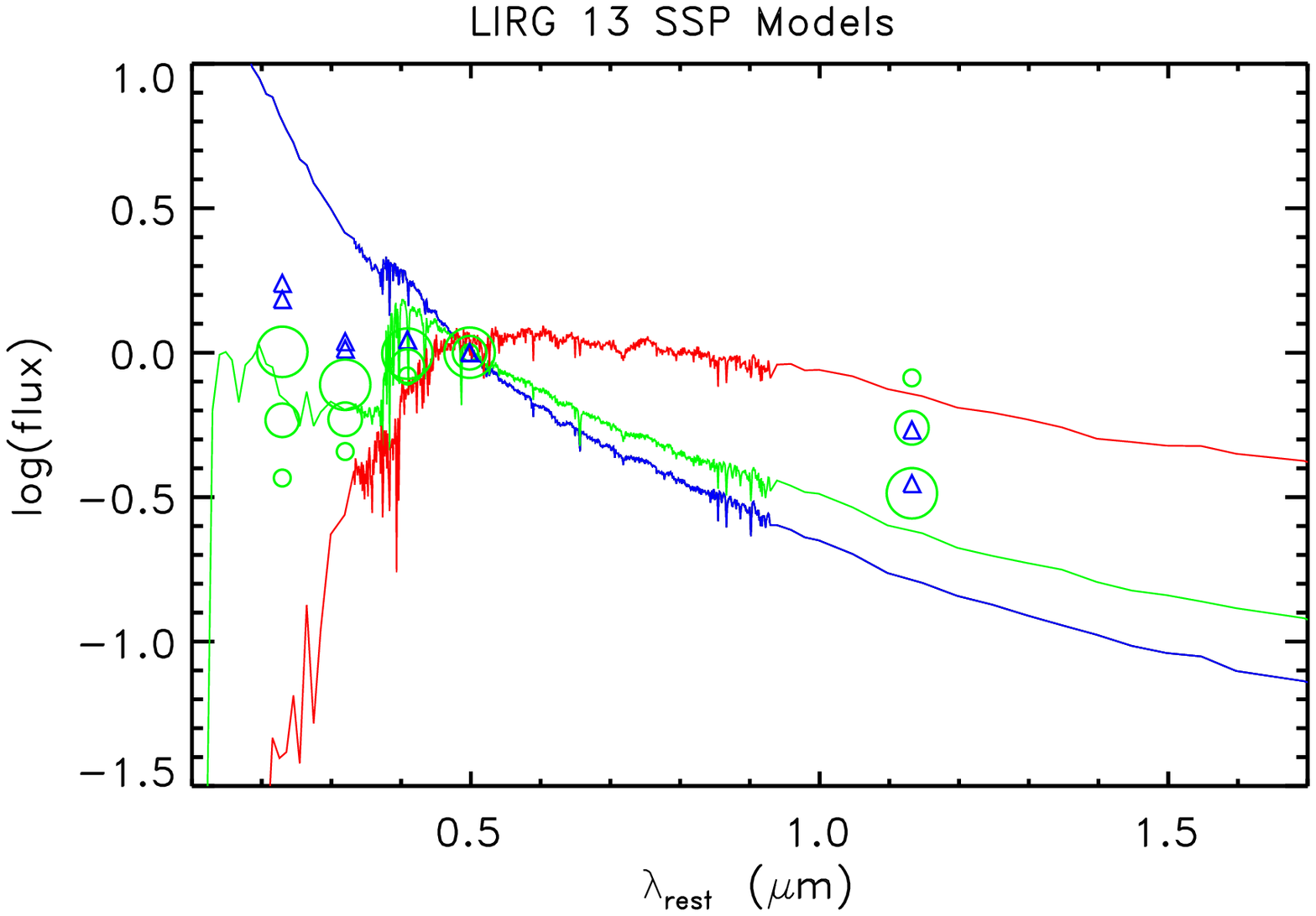}
\plottwo{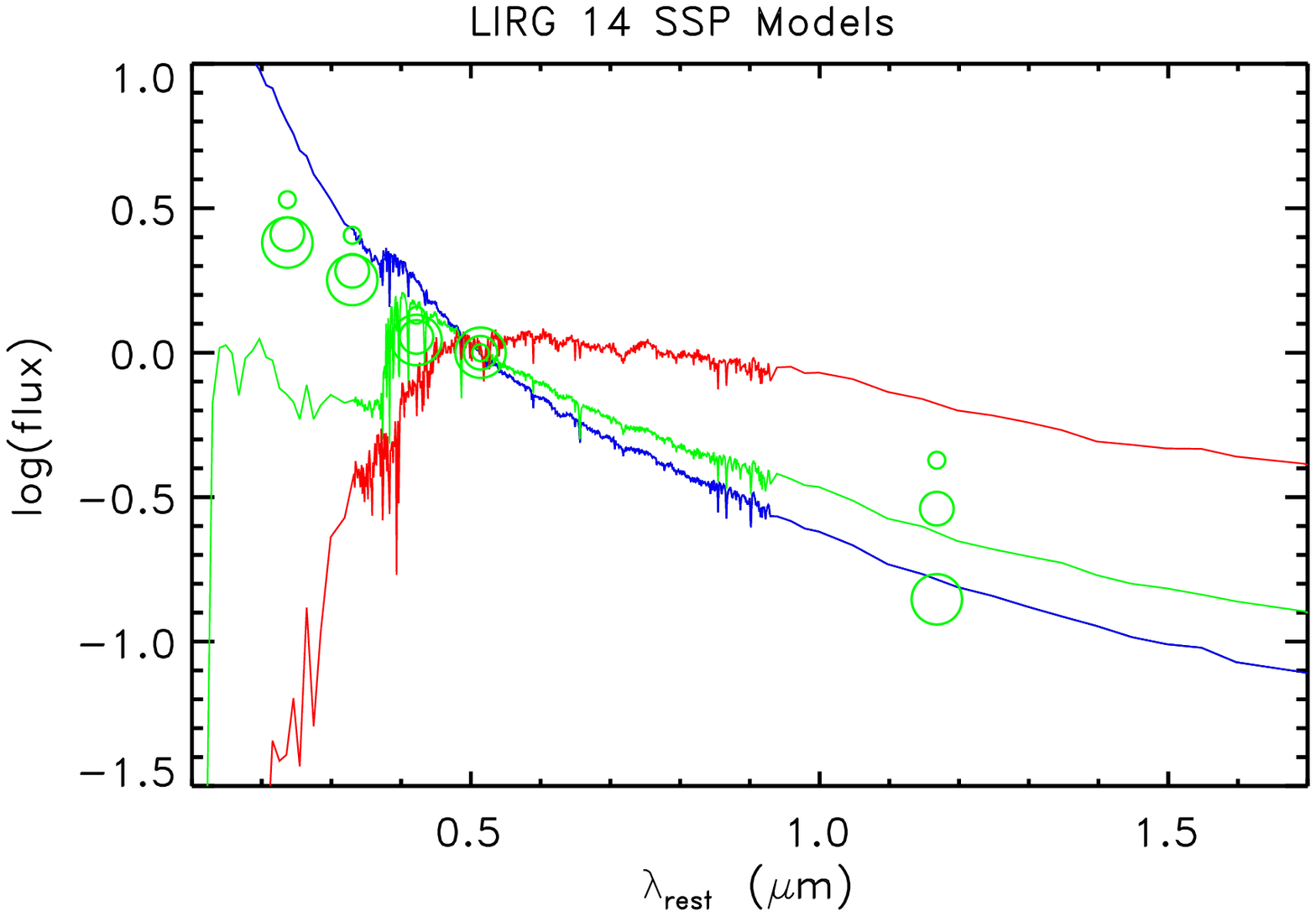}{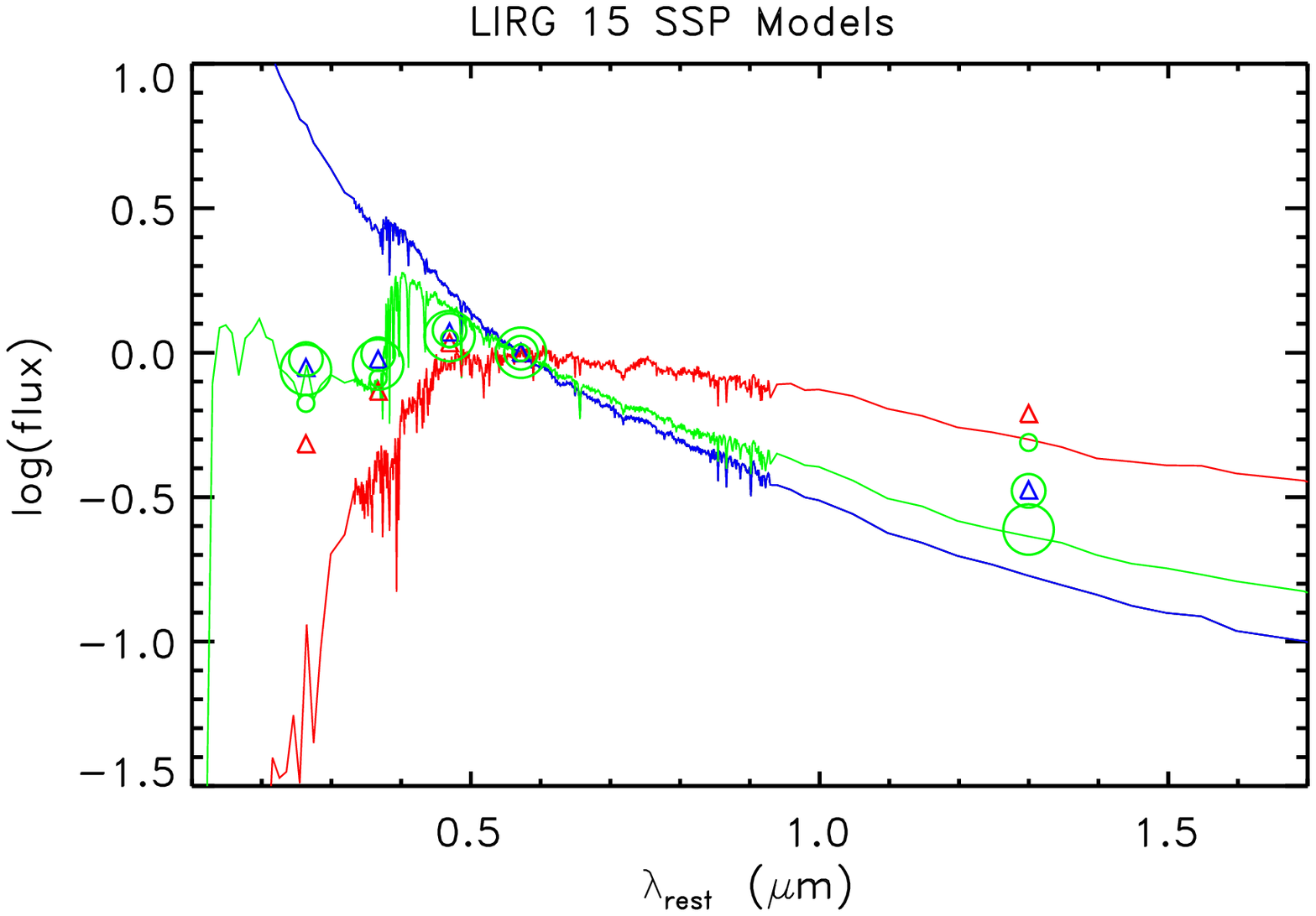}
\caption{\label{fig:LIRGs_SED_C} SEDs of LIRGs 10 - 15.}
\end{figure*}

\begin{figure*}
\plottwo{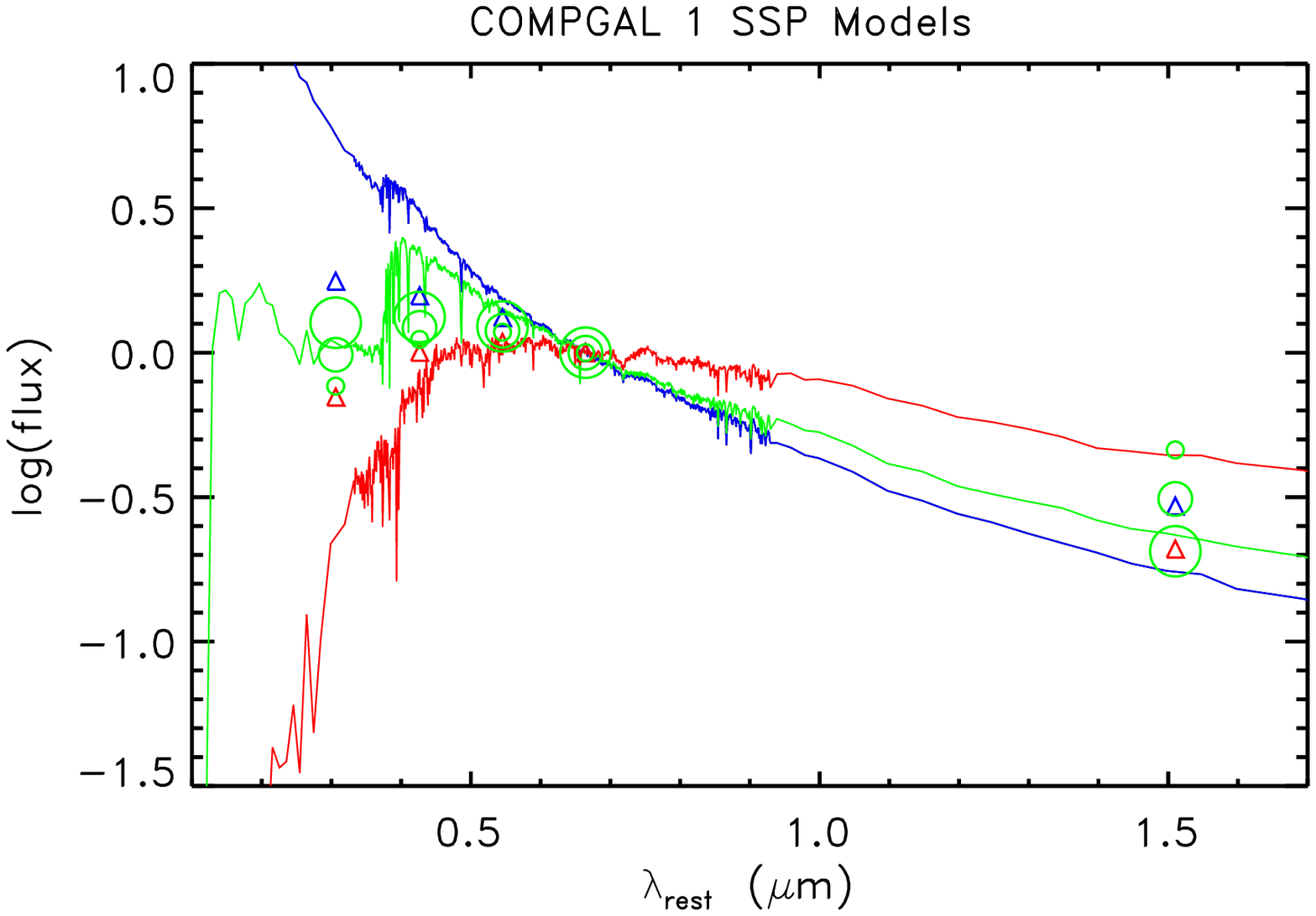}{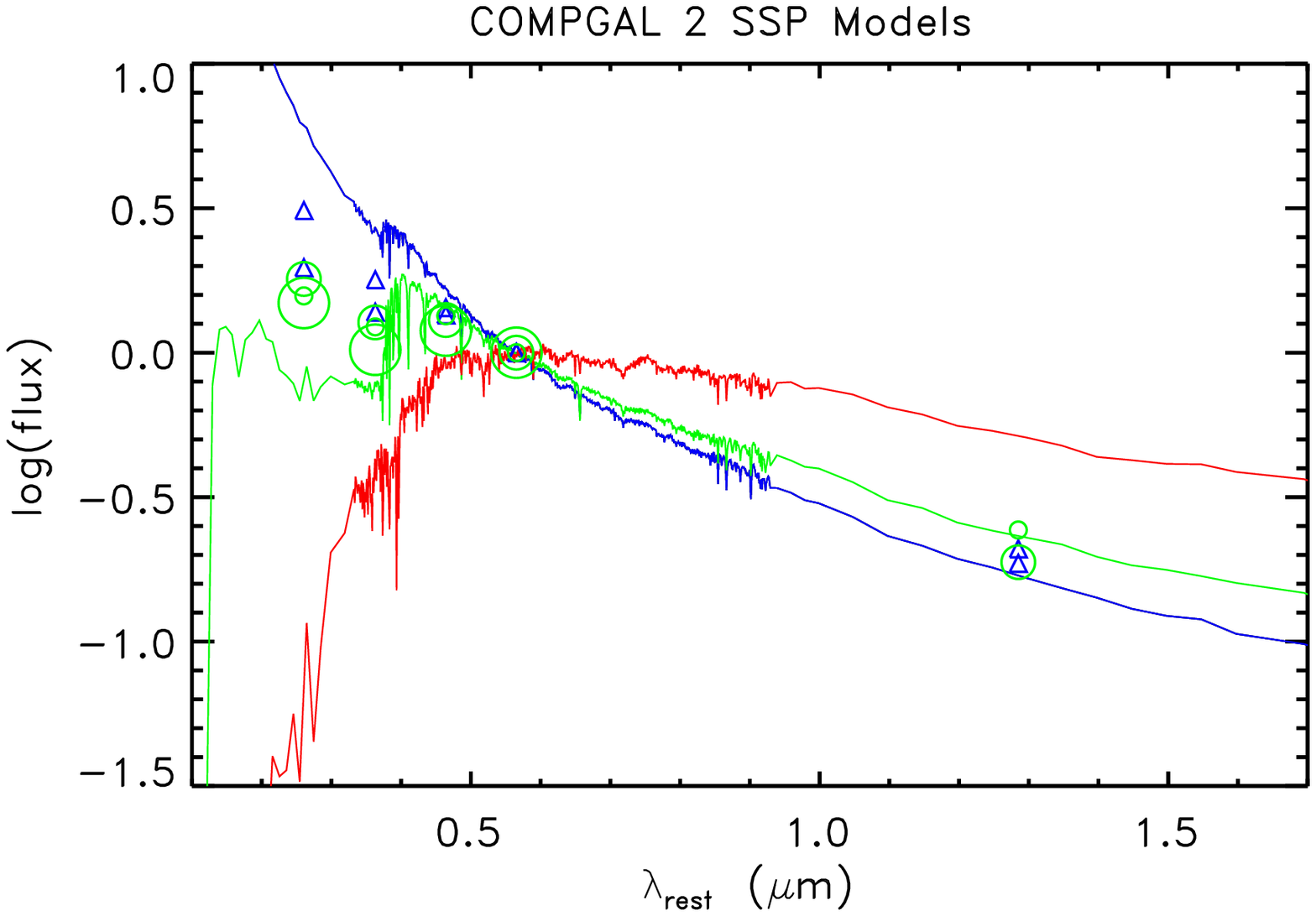}
\plottwo{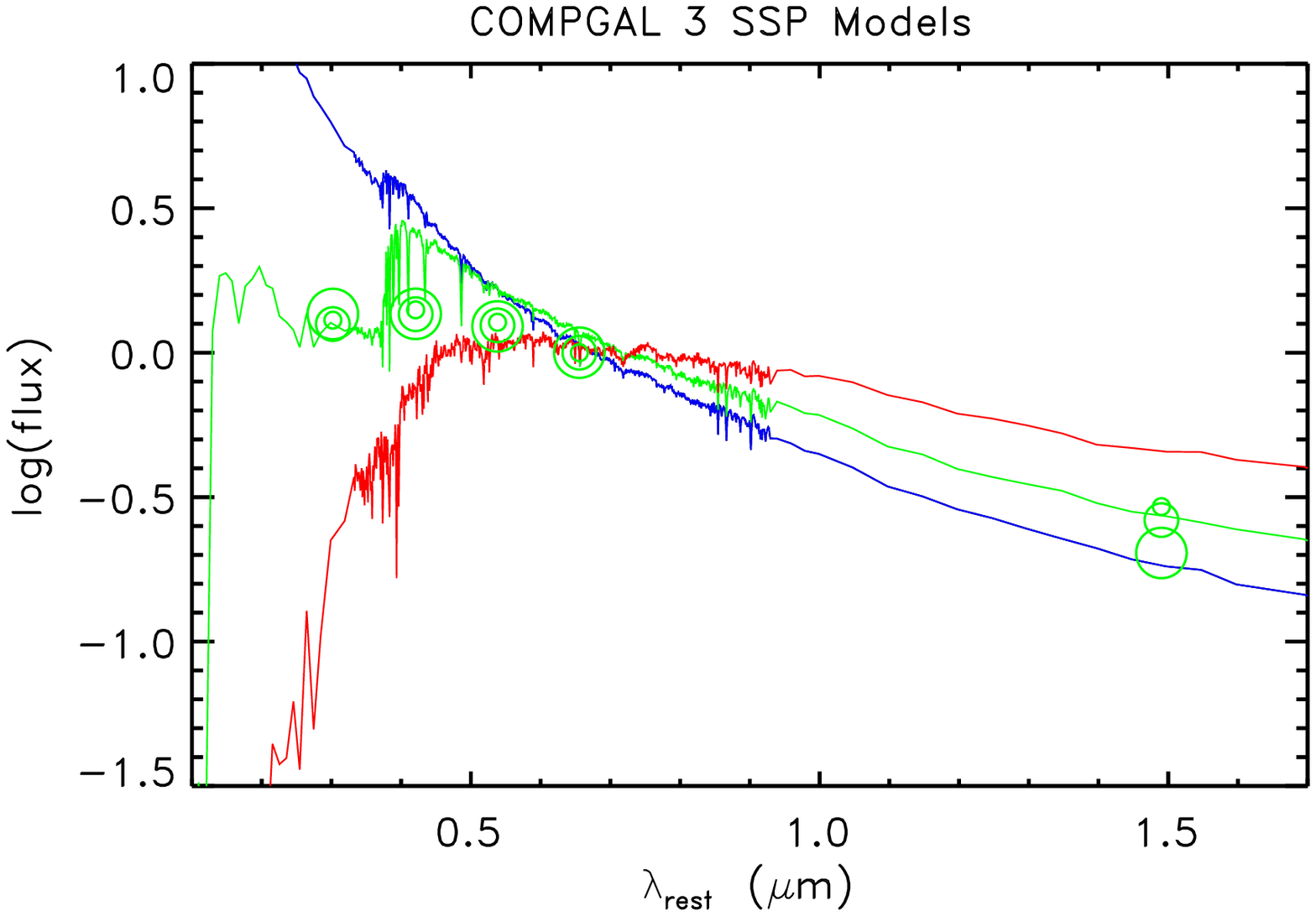}{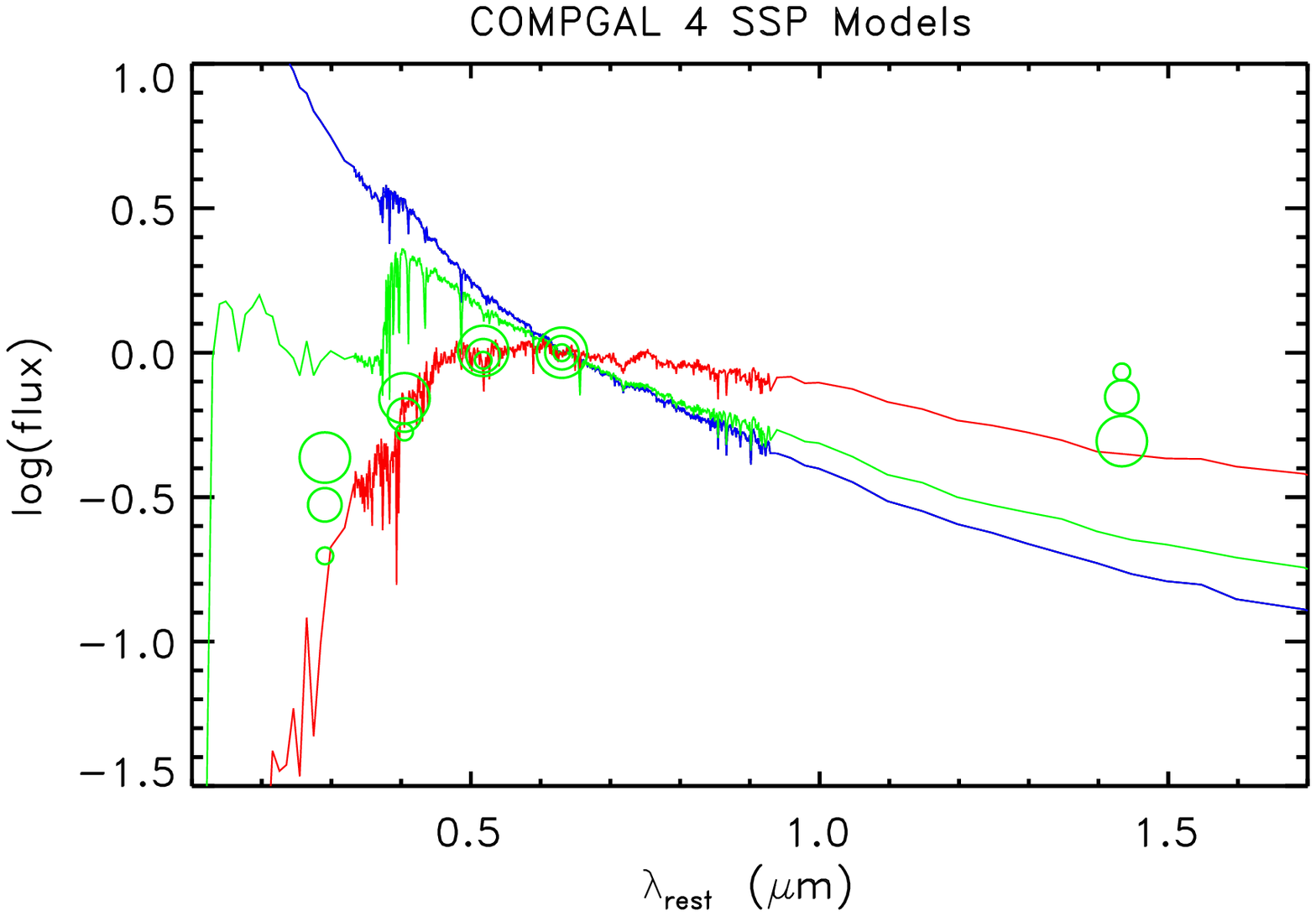}
\plottwo{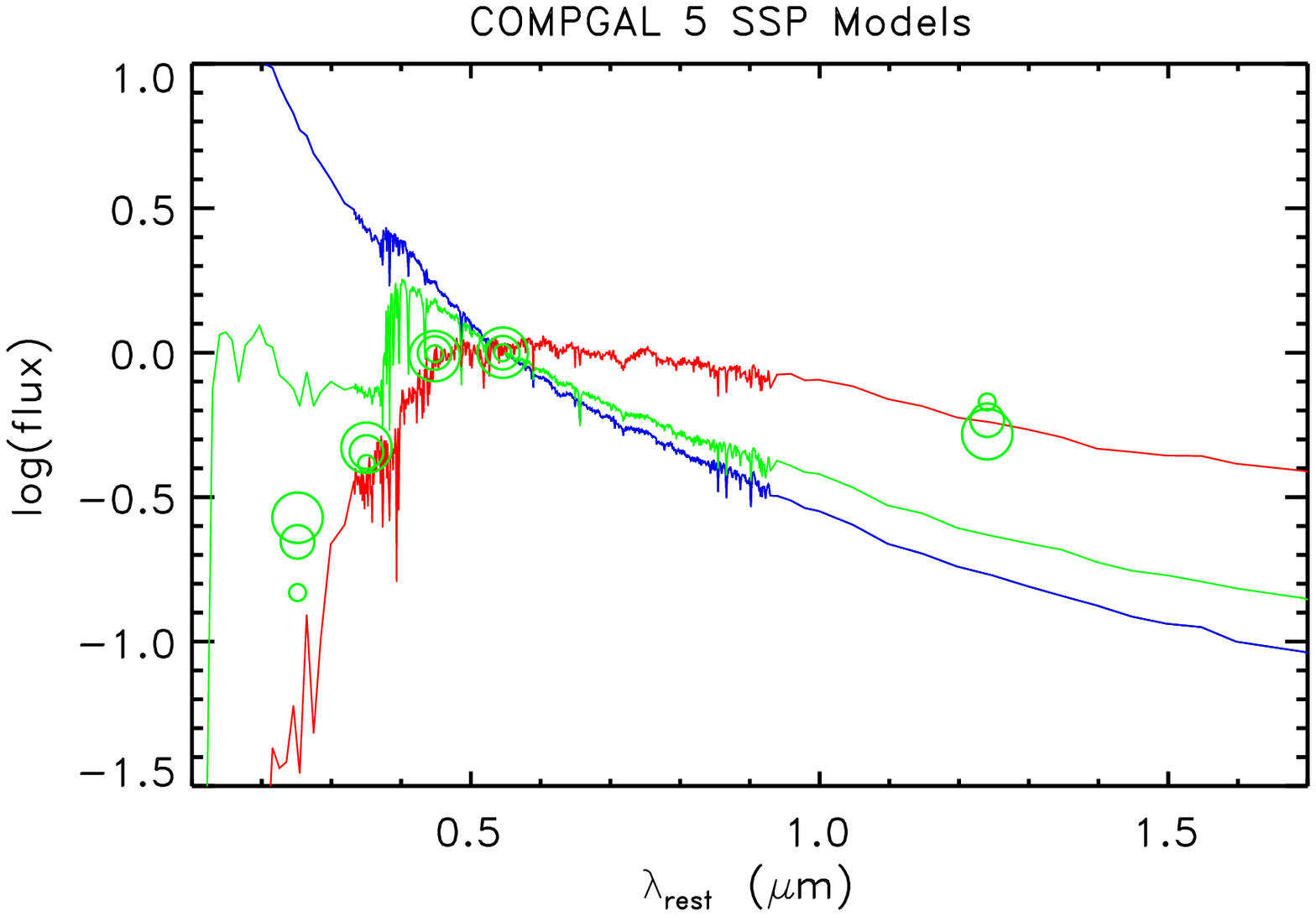}{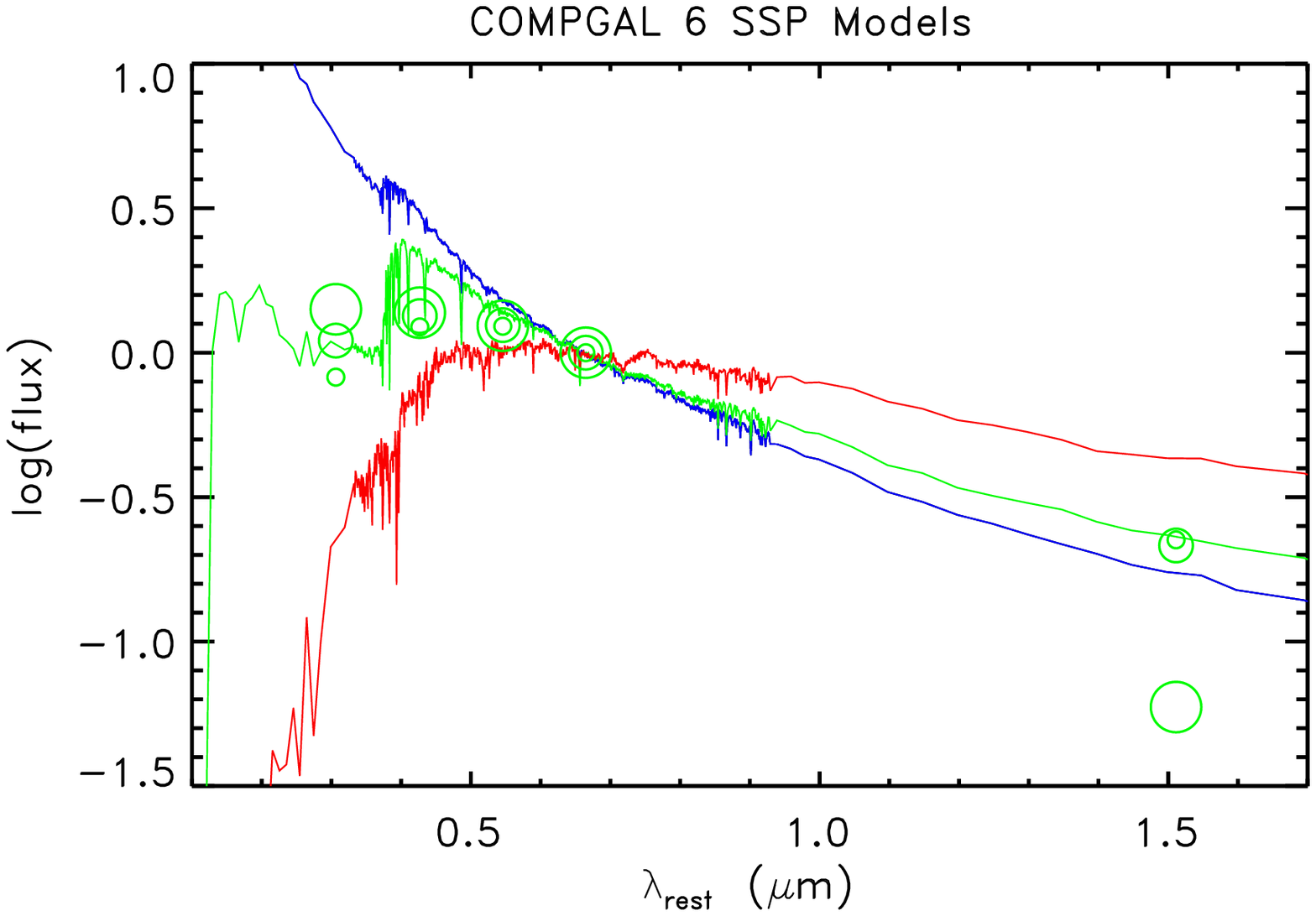}
\caption{\label{fig:COMPGALs_SED_A} Same as Figure \ref{fig:LIRGs_SED_A} only now for SEDs of COMPGALs 1 - 6.}
\end{figure*}

\begin{figure*}
\plottwo{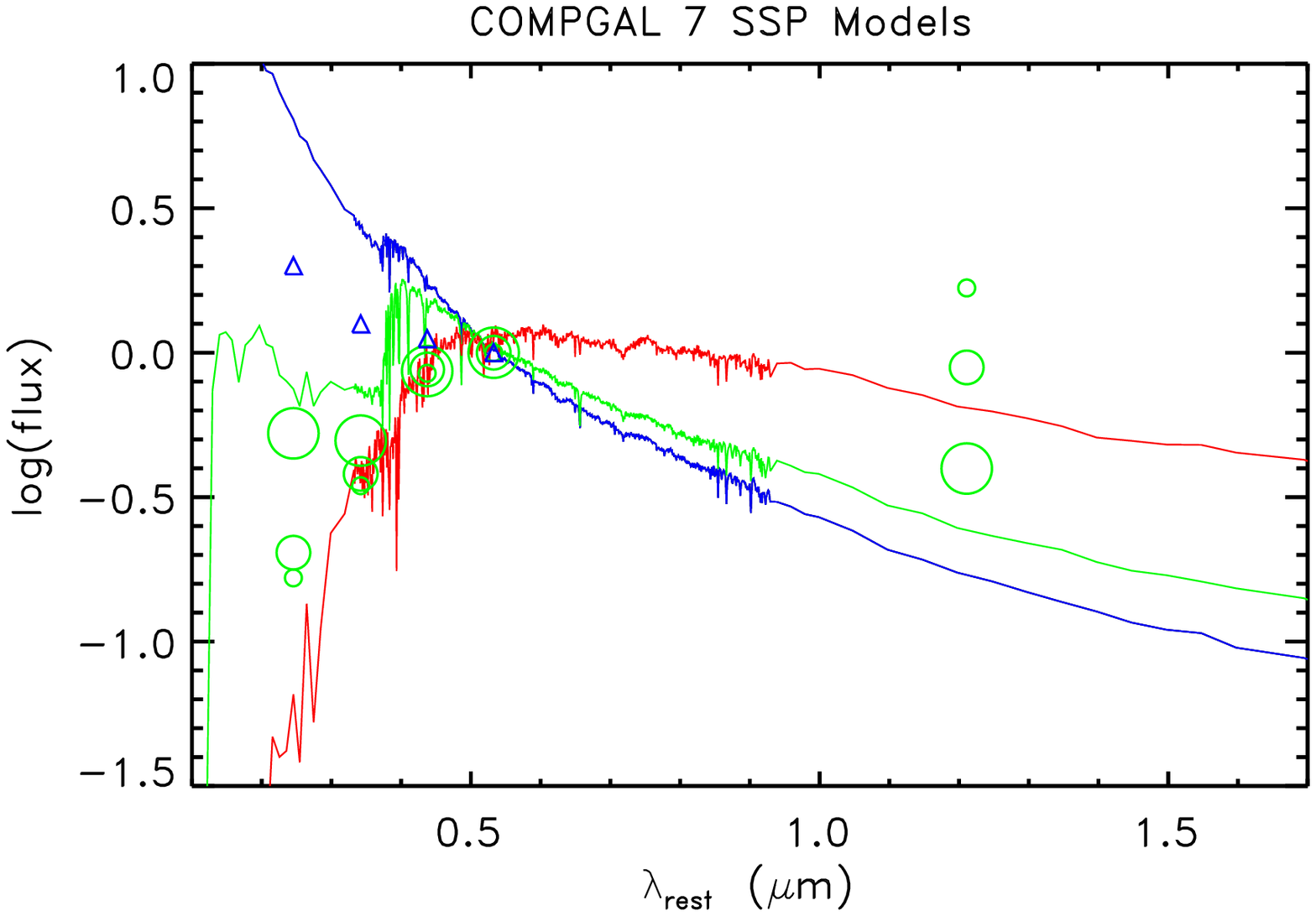}{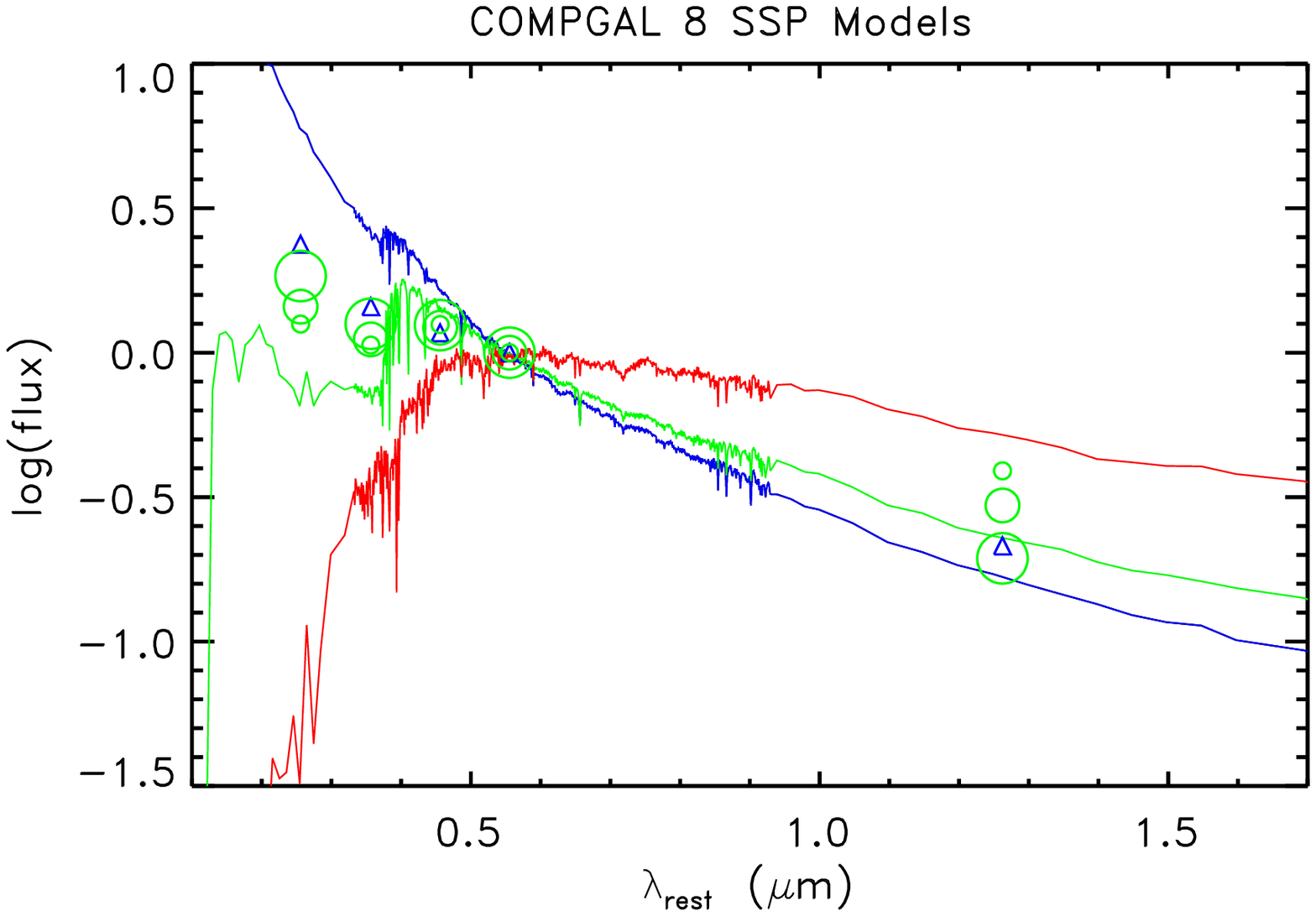}
\plottwo{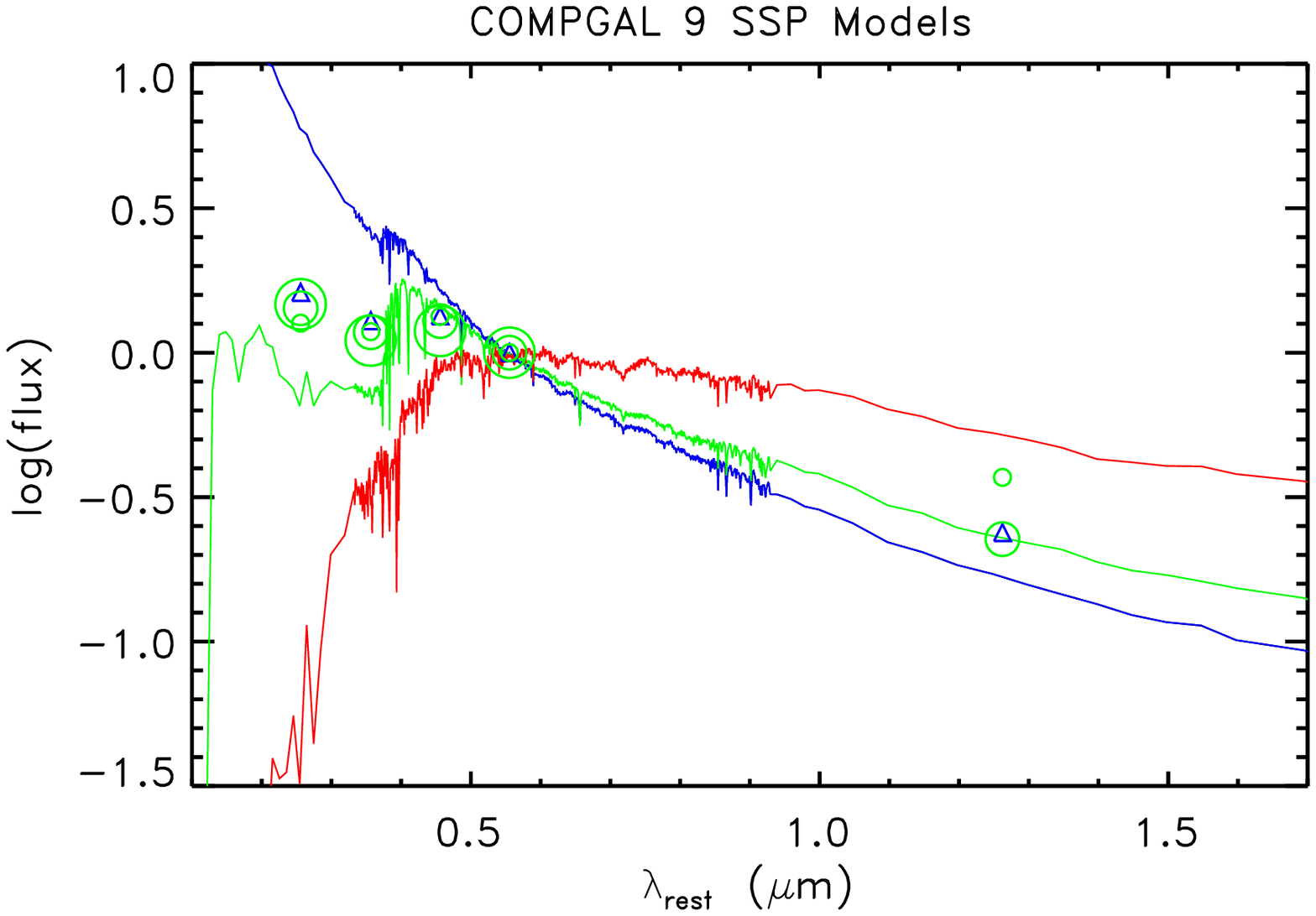}{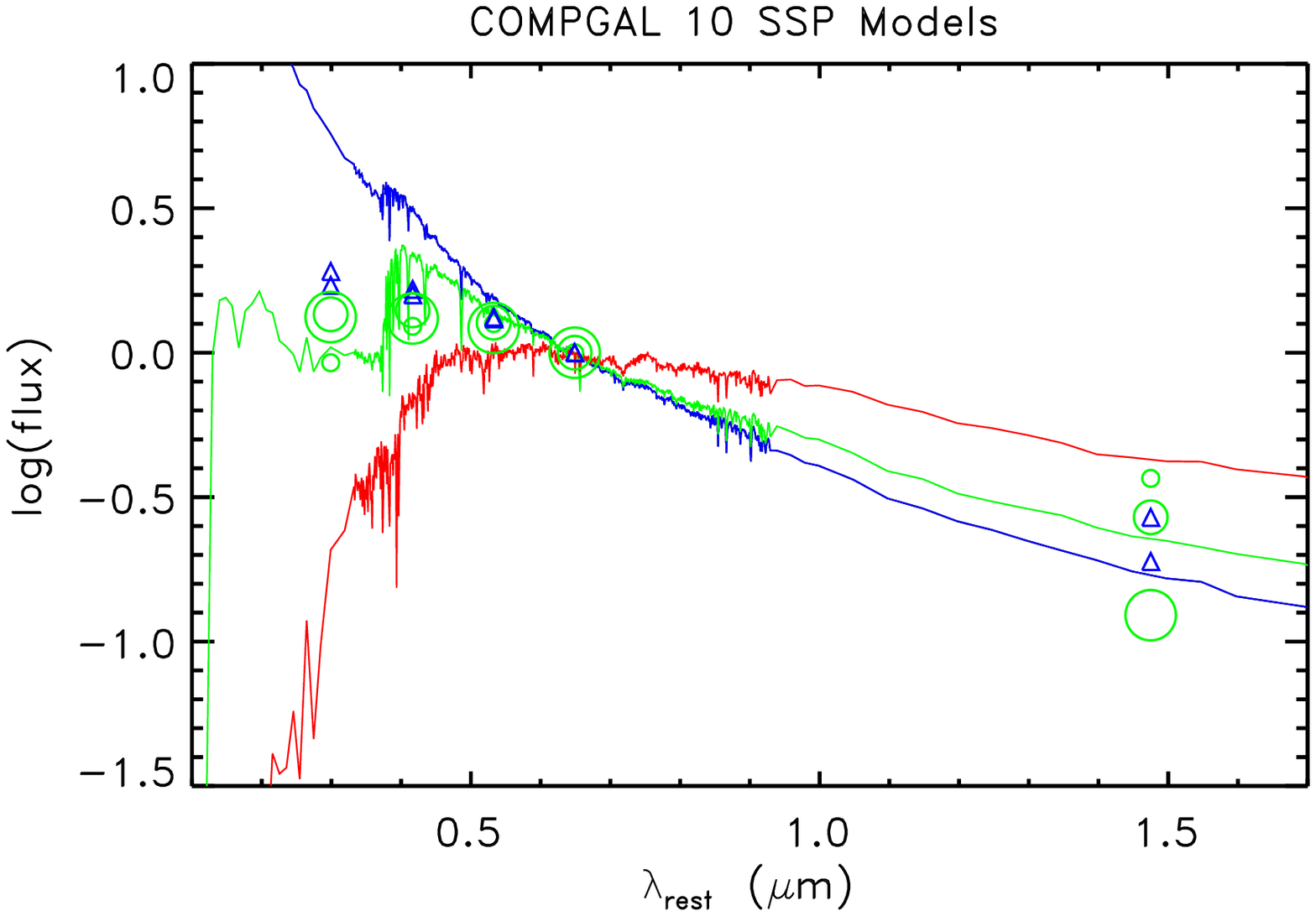}
\includegraphics[scale=0.4]{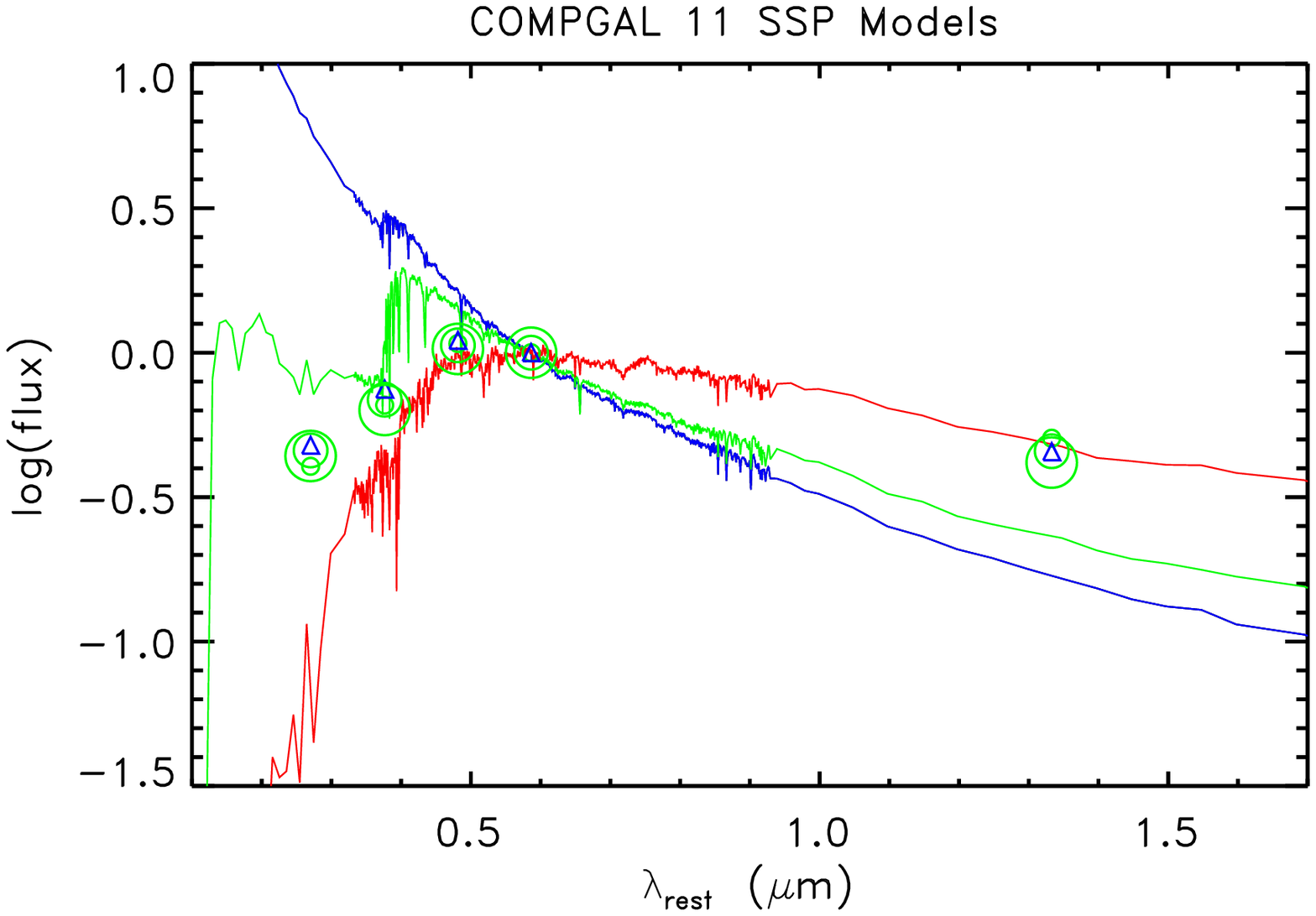}
\caption{\label{fig:COMPGALs_SED_C} SEDs of COMPGALs 7 - 11.}
\end{figure*}

\section{\label{sec:ap_phot} Photometry and SEDs of Sub-components }

Results from the previous section on global galaxy properties suggest that the intermediate redshift LIRGs are produced in high stellar mass systems, many of which are disk galaxies. Global properties, however, can not be used to identify where the star formation is occurring within those LIRGs.  Is star formation centrally concentrated as in the local LIRGs?  Or, is it more smoothly distributed, as in the normal disk galaxies of today?  In this section we discuss the photometry and SEDs of galaxy sub-components, which may begin to address these questions.  We compare the SEDs of the LIRGs to stellar populations synthesis models and to the SEDs of the comparison galaxies.  In Paper 2 we will \emph{fit} the SEDs of the bulge and disk components to stellar population synthesis models.  Fitting is a more complicated task than the qualitative comparison we perform here.

We measure the SEDs of galaxy sub-components with aperture photometry. For each galaxy, aperture positions and shapes were set by eye.  The apertures used are shown in Figures \ref{fig:LIRGsA} - \ref{fig:COMPGALsD}.  Green apertures are centered on the galaxy core and used to measure the SEDs of the inner, middle, and outer portions of the galaxy.  Blue apertures are centered on blue knots, red apertures on red knots, and magenta apertures on neighboring objects. 

Photometry was measured on the AO PSF convolved images (see Section 2.4).  Sky was set by the median of pixels unassociated with galaxies within a given frame.  A summary of the photometry of the inner, middle, and outer regions of each galaxy is given in Table \ref{tab:phot_reg}.

As an example, we first examine the SEDs of the sub-components within LIRG 3.  We chose LIRG 3 because it appears typical of the LIRG disks.  Then we discuss the SEDs of the larger sample of LIRG and Comparison galaxies.

\subsection{The SED of LIRG 3}

Figure \ref{fig:LIRGs_SED_A} shows the SEDs of the sub-components within LIRG 3.  The SEDs are normalized to the \HST\ $z$-band image (roughly the center of the 5 pass-bands).  The photometry of the inner, middle, and outer regions of LIRG 3 are shown as green circles of increasing size.  The photometry of three blue knots are also shown (blue triangles).  Over-plotted are Bruzual and Charlot (2003) stellar population synthesis, single-burst models (or single stellar population, SSP, models) of different ages (blue is 7 Myr, green is 300 Myr, and red is 3 Gyr).  The models contain no dust, have solar metallicity, a Chabrier initial mass function, and are normalized at the same point as the galaxy photometry.  Photometric uncertainties are about the size of the data points.  

This example is illustrative of the bulk of the LIRG photometry to be discussed later. First, lets examine the galaxy core (smallest green circle).  The four optical \HST\ bands are well matched to an intermediate aged population.  However, the Keck AO IR data point is significantly redder than expected for an intermediate aged population.  In contrast, the outer portion of the galaxy and the blue knots are consistent with a single model across all five bands.  For instance the outer portion could be ``fit'' by an intermediate age population with an age of 300 Myr (green line).  The middle annulus of this galaxy has an SED that lies between the SEDs of the inner and outer portions.

Why is the core of LIRG 3 significantly redder in the IR than might be expected from the optical?  We examine two scenarios, 1) multiple stellar populations, and 2) dust.  One way to increase the ratio of IR to optical light is to have a significant population of old stars.  The $z$ and $K$-band photometry of the LIRG 3 core can be fit by an SSP model with an old age (e.g. 3 Gyr).  But an old model does not fit the photometry in the bluer bands.   A younger model fits the bluer bands, while an older model fits the redder bands.   Therefore it is possible that some combination of old and young stars might produce the SED of the LIRG 3 core.  

One way to model a stellar population that contains both old and young stars is with ``tau'' models with exponentially declining star formation rates.  Figure \ref{fig:LIRG3_SED_tau} re-plots the photometry of LIRG 3 on top of tau models with an e-folding time $\tau = 1$ Gyr, seen at various different ages. While none of the plotted models exactly reproduce the photometry of the core, the model that is 3 Gyr old (reasonable age for the redshift of this galaxy, $z=1.0$) comes much closer than the SSP models shown in the previous figure.     

A second scenario for producing redder colors involves dust.  Dust preferentially obscures blue light over red light.  Therefore dust alone can produce an effect similar to that seen in the  SED of the core of LIRG 3.  Figure \ref{fig:LIRG3_SED_dust} shows the photometry of LIRG 3 over-layed with single burst models with an age of 50 Myr and different amounts of dust.  We use the \citet{CharlotFall00} dust model provided with the Bruzual and Charlot code.  If we assume that the light from the core of LIRG 3 is dominated by young stars (50 Myr in this case) then the core must be highly dust obscured.  A reasonable dust model for the core of LIRG 3 has an optical depth $\tau \sim7$. (Note: optical depth $\tau$, should not be confused with tau star formation models.) As LIRGs are expected to contain significant amounts of dust, this might be a reasonable model for the core.  If the core contains both dusty young stars and old stars, then the amount of dust necessary to produce the red colors within the core declines.
       
       Again the goal of this paper is not to ``fit'' for the best models in each case but rather to discuss the broader trends seen in the SEDs of the LIRGs, and compare them with models and the SEDs of the ``normal'' sample.  The next section will present the full set of SEDs and discuss these broader trends.
       
\subsection{SEDs of the LIRGs and ``Normal'' Galaxies}
 
SEDs of the additional LIRGs and ``normal'' galaxies are shown in Figures \ref{fig:LIRGs_SED_B} - \ref{fig:COMPGALs_SED_C}.  Just as with Figure \ref{fig:LIRGs_SED_A}, \citet{BC03} SSP models are over-plotted for reference.  The trend of redder cores and bluer outer regions seen in LIRG 3 is repeated in most of the LIRGs (LIRGs 1, 3, 4, 5, 6, 8, 10, 11, 13, and 15) 66\% of the sample, and 90\% of the LIRG disks.  In addition, just as with LIRG 3, the optical colors of the cores indicate one type of stellar population while the IR colors indicate a different model.  Absent dust, the IR photometry always would suggest the presence of older stars.  However, as with LIRG 3 this is not the most likely model in all cases.  In fact LIRGs 1, 4, 11, and 12 have IR colors that are redder than any SSP model would allow.  This comes as no surprise for LIRG 11 which is an edge on spiral with an obvious dust lane.

Because LIRGs are known to contain large amounts of dust we expect that the unusual combination of blue in the optical and red in the IR is at least partially the result of dust.  The sub-component that most exemplifies this phenomena is the core, but it is also often evident in the middle and outer annuli as well.  In Paper 2 we will attempt to quantify the amounts of dust in the cores in these systems.  

The SEDs of the AGN dominated LIRGs behave differently from the SEDs of the star formation dominated LIRGs.  For instance LIRG 14, a QSO, has a very blue SED across all five bands.  LIRG 9 another AGN dominated system has an almost flat SED across the five bands.  LIRG 2, a major merger, with a powerful AGN is more like the LIRGs discussed above, presumably because its photometry is dominated by star formation.  In fact \citet{Melbourneetal05} demonstrated that the AGN of this system is likely very dust obscured and much of the light we are receiving is likely to be from the dusty starburst.      

In contrast with the LIRGs, half of SEDs of the ``normal'' galaxies appear to be consistent with an SSP.  For instance COMPGALs 2, 3, and 6 are consistent with intermediate aged populations.  While COMPGALs 5 and 11 are consistent with older stellar populations.  The remaining 6 systems appear to be more complicated.  COMPGALs 1, 8, 9, and 10 are blue but likely contain some dusty star formation in their cores.  Based on the morphologies of these 4 systems, dusty cores are not surprising.  The outer regions of these systems tend to be consistent with intermediate aged populations.   COMPGALs 4 and 7 are red and also may have some dust.  These two systems are actually detected in the MIPS image below the LIRG level.  Interestingly COMPGALs 4 and 7 lie in a post starburst region on the red sequence of the CMD.   

In summary, the SEDs of the non-AGN dominated LIRGs are more likely to show evidence of recent dusty star formation, while  the comparison sample is more likely to show evidence for intermediate aged stellar populations and/or less dusty star formation. 

\section{Discussion}
One of the real surprises of the recent infrared space missions, \emph{ISO}\ and \emph{Spitzer}, has been the discovery of evolution in the morphology of the most actively star forming galaxies.  In the local universe LIRGs are dominated by interacting and merging systems \citep{Ishida04}, and the rare local LIRG disks are uniformly barred \citep{Wang06}.  At $z=1$, the LIRGs are predominantly spiral systems \citep{Bell05, MKL05, Lotz07}, with only small fraction showing evidence for major mergers or central bars \citep{Zheng05}.  This has raised the question of what is driving the elevated star formation in the LIRG disks.  Is it minor mergers?  Is it bars hidden in the \citet{Zheng05} optical images by dust?   Is it long distance interactions?  Or is it purely that disk galaxies have higher gas content at $z=1$ compared with today and therefore can sustain higher star formation rates?

By comparing high-resolution images of LIRG and non-LIRG systems at intermediate redshifts, our paper begins to address these questions.  Interestingly, it appears that the different star formation levels in these two  samples are unrelated to morphology.  For instance, while 30\% of our LIRG sample show some evidence for a major or minor merger, the ``normal'' sample has as high or higher fraction of disturbed optical morphologies suggestive of recent interactions.  This indicates that while minor mergers may contribute to the formation of LIRGs, they are not a sufficient condition for the production of a LIRG.  

Likewise, our data suggest the bars are neither a necessary nor sufficient condition for the production of intermediate redshift LIRG disks.  In the local universe, \citet{Wang06} found that LIRG disk galaxies are uniformly barred.  \citet{Sheth07}, however, showed that the bar fraction in disk galaxies declines by a factor of $\sim3$ out to $z\sim0.8$, and \citet{Zheng05} found bars to be rare in intermediate redshift LIRGs.  Several factors may be contributing to the Zheng et al. result.  First, the dusty nature of the LIRGs will tend to obscure the central regions of a galaxy where bars may reside.  Second, bars may contain significant numbers of older stars more easily seen in redder wavelengths.  Our paper has improved on the \citet{Zheng05} data set, by including the high resolution near-IR images, which are both less sensitive to dust, and more sensitive to older stars.  Despite this, only one of our systems  (10\% of the LIRG disks) contains a prominent bar, and that bar is visible in the optical wavebands as well.  In addition to this system, the $K$-band images reveal small ($<0.5\arcsec$) potential bars  in the cores of three more LIRGs.  These will be examined further in Paper 3.  Even if we include these systems, bared systems only make up about 40\% of our LIRG  disk sample.  Among the ``normal'' (non-LIRG) systems there are no obvious barred systems.  However, several, $\sim 30\%$, contain very small linear features in their cores similar to the potential bars seen in the LIRGs.  As with the minor mergers, this suggests that bars are not a sufficient condition to produce the excess star formation seen in the LIRGs.  Between the Zheng et al. result and our own, we are finding strong evidence for a lack of prominent bars in intermediate redshift LIRGs, a clear change from the  local universe.

While morphology alone can not explain the elevated star formation in intermediate redshift LIRGs, it appears that global properties, such as stellar mass, play a significant role.   The comparison sample galaxies are measured to have, on average, an order of magnitude  lower stellar mass.  They are also generally 1 - 2  magnitudes fainter than the LIRGs (in $M_B$).   This is true despite the fact that the selection criteria, proximity to a guide star, strong detection in the $K$-band, and non-elliptical morphology, should select for any massive non-LIRGs as well.  While we caution that our sample is small, and derived from a tiny area of the sky, it may be that massive \emph{non-LIRG} disk galaxies are less common than massive LIRG disks at these redshifts.  Interestingly the two most massive non-LIRG galaxies in our sample happen to be on the red sequence and presumably are no longer forming stars at very high rates.   It is possible the LIRGs fade into this portion of the CMD. While this is a possible evolutionary path, we stress that our study only encompasses a very small volume of the sky, so any conclusions dependent on number densities of systems will have to be verified with much larger samples.  While we demonstrated that minor mergers are not a sufficient condition for the production of a LIRG disk, it is possible that minor mergers in massive disks are important.  

It is also possible that interactions at large distance play a role.  For instance, \citet{Elbaz07} looked at the influence of environment on star formation rate density at $z=1$.  They  compared a local sample of galaxies from Sloan with a $z=1$ sample in GOODS, and found that while the star formation rate density declines with mass density in the local universe, it increases with mass density at $z=1$.  Even more telling \citet{Elbaz07} showed an example of a proto-cluster at $z=1$.  It was made up of three galaxy groups each with several LIRGs near their centers.  Their work is a strong indication that large scale structure and environment are playing a role in the elevated star formation of the LIRGs.  Despite the high density environments, \citet{Elbaz07} reiterate that major mergers are not the primary cause of the elevated star formation in the LIRGs.  Just as with the previous studies, the majority of their $z=1$ LIRGs, including those in the proto-cluster, are disk galaxies.  As with previous authors, they speculate that the high star formation rates may be related to minor mergers or higher gas densities within the disks.

Another approach to solving the nature of the LIRGs is to examine SEDs.  Much of the prior work on SEDs of  intermediate redshift LIRGs has focussed on global properties.  A very interesting result from \citet{Zheng07} showed that the mid-far IR SEDs of the intermediate redshift LIRGs are consistent with lower dust temperature than than for low redshift LIRGs.  They interpret this in terms of the morphology results now well established.  They suggest that because the intermediate redshift LIRGs are primarily disk galaxies, they are less likely to be undergoing central star bursts.  Rather they will have more distributed star formation with lower overall gas densities and temperatures.  In contrast the local LIRGs, dominated by mergers, show significant central star bursts with higher gas densities and temperatures.
 
These high gas densities have been shown in studies on low redshift LIRGs and ULIRGs, where high spatial resolution is easier to achieve even at longer wavelengths.   For instance \citet{Crawford96} showed that the radio emission from ULIRGs tends to be compact and best modeled by a nuclear star burst.  In contrast, less energetic galaxies with $L_{fir}<10^{11} L_{\odot}$ have diffuse radio emission.    At millimeter wavelengths, LIRGs are found to be extremely rich in molecular gas with central densities similar to giant molecular clouds \citep{Gao97}.  CO measurements of low redshift LIRGs show that 40-100\% of the CO luminosity is contained within the central kpc \citep{SandersMirabel96}.  In the optical, \citet{Hattori04} obtained $H\alpha$ imaging of 22 local LIRGs. They found a range of concentrations, with disturbed systems typically more concentrated. All of these studies indicate that star formation is centrally concentrated in local LIRGs, especially for the morphologically disturbed sample.  
 
Because they were measuring global SEDs \citet{Zheng07} were not able to verify the hypothesis of distributed star formation in the intermediate redshift LIRG disks directly.  We, however, have measured SEDs on a 1-2 kpc scale and can begin to address this problem.  The optical to IR SEDs in our study do show distributed, dusty star formation in all of the LIRG disks.  At the same time, the SEDs of the cores are also best explained by dusty star formation.  In fact, the SEDs suggest that the dust may be most concentrated in the cores.  In other words the LIRGs appear to be both forming stars in their disks and undergoing dusty central star formation.  We caution that we have yet to fit for the best stellar population model.  A combination of older stars and dusty star formation would reduce the amount of dust necessary to produce the red colors within the LIRG cores.  Paper two will attempt to quantify the general trends identified in this paper, and estimate the fraction of star formation occurring in the cores vs. the disks.  In contrast with the LIRGs, central dusty star formation is only about half as likely in the ``normal'' (non-LIRG) sample.      

The morphological evolution of the LIRGs appears to have had an effect on the sites of star formation.  In the local universe the LIRGs are dominated by mergers and star formation tends to be centrally concentrated.  At intermediate redshifts the LIRGs are dominated by disks and star formation is both distributed throughout the disks, and occurring within the cores.  It remains to be seen which component is contributing more to the overall star formation of the intermediate redshift LIRGs, the disk or the core.

\begin{deluxetable*}{ccccccccccccccccccc}
\tabletypesize{\small}
\setlength{\tabcolsep}{0.05in} 
\tablecolumns{19}
\tablewidth{0pt}
\tablecaption{\label{tab:phot_reg} Photometry (Apparent Magnitudes [AB]) of Galaxy Sub-components}
\tablehead{
\colhead{} & &\multicolumn{5}{c}{Inner Core} && \multicolumn{5}{c}{Middle Annulus} && \multicolumn{5}{c}{Outer Annulus} \\
\cline{3-7} \cline{9-13} \cline{15-19}
\colhead{Galaxy} && \colhead{$B$} & \colhead{$V$} & \colhead{$i$} & \colhead{$z$} & \colhead{$K$} &&
\colhead{$B$} &\colhead{$V$} & \colhead{$i$} & \colhead{$z$} & \colhead{$K$} &&
\colhead{$B$} &\colhead{$V$} & \colhead{$i$} & \colhead{$z$} & \colhead{$K$} 
}
\startdata
LIRG 1    &   &        26.68  &        25.15  &        23.82  &        23.19  &        21.71 &   &        25.72  &        24.52  &        23.27  &        22.74  &        21.26 &   &        23.96  &        23.06  &        22.03  &        21.64  &        20.56  \\
LIRG 2    &   &        23.19  &        22.44  &        21.75  &        21.69  &        20.51 &   &        23.09  &        22.32  &        21.56  &        21.42  &        20.42 &   &        23.24  &        22.29  &        21.41  &        21.13  &        20.17  \\
LIRG 3    &   &        26.80  &        26.14  &        25.25  &        24.70  &        23.34 &   &        24.72  &        24.22  &        23.41  &        22.89  &        22.05 &   &        23.56  &        23.19  &        22.40  &        21.93  &        21.66  \\
LIRG 4    &   &        27.46  &        26.06  &        24.57  &        23.66  &        21.55 &   &        26.58  &        25.25  &        23.79  &        22.86  &        21.06 &   &        24.62  &        23.78  &        22.44  &        21.56  &        20.25  \\
LIRG 5    &   &        28.23  &        26.95  &        25.67  &        24.92  &        23.27 &   &        26.26  &        25.18  &        24.01  &        23.29  &        21.70 &   &        25.62  &        24.81  &        23.72  &        23.01  &        21.80  \\
LIRG 6    &   &        26.49  &        25.70  &        24.65  &        24.29  &        23.12 &   &        23.83  &        23.24  &        22.30  &        22.00  &        21.47 &   &        23.60  &        23.15  &        22.25  &        21.94  &        22.88  \\
LIRG 7    &   &        26.40  &        25.55  &        24.65  &        23.95  &        22.45 &   &        25.29  &        24.41  &        23.41  &        22.67  &        21.32 &   &        25.58  &        24.97  &        23.79  &        22.99  &        22.15  \\
LIRG 8    &   &        26.33  &        25.83  &        25.05  &        24.45  &        22.98 &   &        25.60  &        25.22  &        24.56  &        23.90  &        22.81 &   &        24.87  &        24.58  &        23.96  &        23.28  &        22.83  \\
LIRG 9    &   &        25.33  &        24.30  &        23.75  &        23.57  &        22.31 &   &        25.17  &        24.37  &        23.75  &        23.22  &        22.03 &   &        24.43  &        23.95  &        23.34  &        22.76  &        21.79  \\
LIRG 10    &   &        25.96  &        25.07  &        24.07  &        23.48  &        22.03 &   &        24.71  &        23.78  &        22.75  &        22.16  &        20.89 &   &        24.60  &        23.89  &        22.84  &        22.18  &        21.13  \\
LIRG 11    &   &        28.34  &        26.99  &        25.53  &        24.94  &        22.57 &   &        26.13  &        25.17  &        23.84  &        23.34  &        21.51 &   &        25.37  &        24.94  &        23.88  &        23.38  &        22.83  \\
LIRG 12    &   &        26.48  &        25.44  &        24.38  &        24.31  &        22.71 &   &        25.03  &        24.00  &        22.97  &        22.79  &        21.81 &   &        24.99  &        24.05  &        23.07  &        22.80  &        22.59  \\
LIRG 13    &   &        27.34  &        26.39  &        25.20  &        24.58  &        23.01 &   &        25.23  &        24.50  &        23.51  &        22.97  &        21.83 &   &        23.89  &        23.46  &        22.66  &        22.22  &        21.65  \\
LIRG 14    &   &        21.74  &        21.33  &        21.48  &        21.38  &        20.53 &   &        22.34  &        21.93  &        21.97  &        21.68  &        21.25 &   &        22.70  &        22.30  &        22.29  &        21.97  &        22.32  \\
LIRG 15    &   &        24.89  &        23.96  &        23.08  &        22.77  &        21.77 &   &        23.17  &        22.41  &        21.67  &        21.43  &        20.85 &   &        24.40  &        23.63  &        22.85  &        22.57  &        22.31  \\
COMP 1    &   &        25.75  &        24.63  &        24.03  &        23.78  &        22.84 &   &        23.89  &        22.94  &        22.43  &        22.19  &        21.67 &   &        23.27  &        22.50  &        22.05  &        21.85  &        21.79  \\
COMP 2    &   &        24.91  &        24.46  &        23.83  &        23.72  &        23.47 &   &        23.44  &        23.10  &        22.55  &        22.40  &        22.43 &   &        24.41  &        24.09  &        23.39  &        23.16  &        25.60  \\
COMP 3    &   &        25.52  &        24.72  &        24.29  &        24.12  &        23.67 &   &        23.70  &        22.90  &        22.46  &        22.27  &        21.93 &   &        23.10  &        22.38  &        21.96  &        21.76  &        21.71  \\
COMP 4    &   &        27.19  &        25.40  &        24.25  &        23.75  &        22.14 &   &        25.83  &        24.34  &        23.29  &        22.83  &        21.43 &   &        25.23  &        23.99  &        23.05  &        22.64  &        21.62  \\
COMP 5    &   &        28.19  &        26.35  &        24.87  &        24.43  &        23.07 &   &        26.38  &        24.88  &        23.49  &        23.06  &        21.86 &   &        26.15  &        24.83  &        23.50  &        23.04  &        21.97  \\
COMP 6    &   &        26.11  &        24.95  &        24.41  &        24.21  &        24.05 &   &        24.15  &        23.21  &        22.76  &        22.57  &        22.45 &   &        23.86  &        23.17  &        22.75  &        22.56  &        23.84  \\
COMP 7    &   &        28.22  &        26.70  &        25.20  &        24.59  &        22.25 &   &        26.77  &        25.37  &        23.93  &        23.36  &        21.71 &   &        26.48  &        25.82  &        24.69  &        24.10  &        23.32  \\
COMP 8    &   &        25.83  &        25.30  &        24.58  &        24.40  &        23.64 &   &        24.70  &        24.27  &        23.63  &        23.42  &        22.96 &   &        23.92  &        23.61  &        23.09  &        22.90  &        22.90  \\
COMP 9    &   &        25.18  &        24.53  &        23.87  &        23.75  &        23.05 &   &        23.53  &        23.02  &        22.39  &        22.23  &        22.06 &   &        23.67  &        23.26  &        22.65  &        22.41  &         -  \\
COMP 10    &   &        25.56  &        24.52  &        23.97  &        23.79  &        23.09 &   &        23.78  &        23.02  &        22.60  &        22.43  &        22.07 &   &        24.25  &        23.55  &        23.09  &        22.88  &        23.37  \\
COMP 11    &   &        26.48  &        25.24  &        24.17  &        23.82  &        22.78 &   &        25.20  &        24.04  &        23.04  &        22.67  &        21.74 &   &        25.67  &        24.55  &        23.49  &        23.10  &        22.26  \\
\enddata
\end{deluxetable*}

\section{Summary}

We presented high spatial resolution near-infrared images of 15 intermediate redshift LIRGs and 11 comparison galaxies.  The comparison sample was chosen to be non-ellipticals with the same apparent $K$-band magnitude limit as the LIRG sample.  The images were obtained with the Keck Laser Guide Star Adaptive Optics Facility at the W.M. Keck Observatory, as part of the Center for Adaptive Optics Treasury Survey (CATS).  The images have a PSF FWHM of $\sim 0.1 \arcsec$, and resolve features on the 1-2 kpc scale.  The sample is located in the GOODS-S southern region, allowing us to measure SEDs of galaxy sub-components across the 4 optical HST bands from GOODS and the additional Keck AO $K'$ image from CATS.  

Morphologically the LIRG sample is dominated by large spiral galaxies with typical stellar masses of $M=10^{10} - 10^{11} \; M_{\odot}$.  While the optical (rest-frame blue to UV) images contain significant point-like substructures, presumably star forming regions, the near-IR AO images tend to be smooth.  In fact, the AO images place limits on the numbers of red super giant stars within these blue knots, at less then 4000 RSGs. 

The comparison sample tends to show irregular morphology in the optical bands, and typically disk-like morphology in the $K$-band.   The comparison sample also tends to be at lower stellar mass than the LIRG sample $M\sim10^{9.5} \; M_{\odot}$.  

About 30\% of the LIRG sample shows evidence for major or minor mergers.  The disturbed optical morphologies of the comparison sample suggest that they might be experiencing a similar or higher rate of minor mergers or interactions.  This indicates that minor mergers are not a sufficient condition for igniting a LIRG event.  Minor mergers within massive disk galaxies, however, may be important.

While the LIRG sample contains a high fraction of disk-like systems (66\%), only one (10\% of the disks) contains a obvious strong bar.  Three additional LIRG disks contain a small linear feature in their cores, visible in the $K$-band image.  Even if these structures are true bars, the fraction of disks with bars is only 40\%, far rarer than in low redshift LIRG disks where the fraction is close to 100\% \citep{Wang06}.

In addition to studying morphologies, we measured the SEDs of galaxy sub-components.  We find that the SEDs of 80\% of the LIRGs are consistent with dusty ongoing star formation. The remainder appear to be dominated by AGN light.  While we find evidence for spatially distributed star formation especially for the LIRG disk galaxies, we also show that the cores are likely to be sites of very dusty star formation.  More work needs to be done to quantify the fraction of star formation and dust in both core and disk components.  In contrast, only half of the comparison sample shows evidence for dusty star formation.  Even for the comparison galaxies that contain some dust, the outer regions usually appear to be relatively dust free.  

Locally LIRGs have been shown to have very concentrated central star formation \citep{SandersMirabel96} usually because of a recent merger or interaction.  The results of our work are preliminary but suggest that star formation is more smoothly distributed in intermediate redshift LIRG disks.  This type of star formation might be similar to the self regulated star formation seen in local disk galaxies (such as the Milky Way), but at 10 - 20 times higher rates.

\acknowledgments
We would like to thank Emeric Le Floc'h for contributing the \tf\ measurements of GOODS-S galaxies. 
This work has been supported in part by the NSF Science and Technology Center for Adaptive Optics, managed by the University of California (UC) at Santa Cruz under the cooperative agreement No. AST-9876783.  It was also supported by the \HST\ archival grant HST-AR-10965.
The laser guide star adaptive optics system was funded by the W. M. Keck
Foundation. The artificial laser guide star system was developed and
integrated in a partnership between the Lawrence Livermore National Labs
(LLNL) and the W. M. Keck Observatory. The laser was integrated at Keck with
the help of Curtis Brown and Pamela Danforth. The NIRC2
near-infrared camera was developed by CalTech, UCLA and Keck
(P.I. Keith Matthews). The data presented herein
were obtained at the Keck Observatory, which is operated as a
scientific partnership among the CalTech, UC and NASA.  
This work is supported in part under the auspices of the 
US Department of Energy, National Nuclear Security Administration and by 
the LLNL under contract W-7405-Eng-48.
The authors wish to recognize and acknowledge the very significant cultural
role and reverence that the summit of Mauna Kea has always had within the
indigenous Hawaiian community.  We are most fortunate to have the
opportunity to conduct observations from this superb mountain.

\appendix
\section{Morphologies of the LIRGs}
Figures \ref{fig:LIRGsA} - \ref{fig:LIRGsE} show the 5-band images of the LIRGs.  The images are 2.8$\arcsec$ on a side.   Color images of the LIRGs are also presented.  They were created from the $B$, $i$, and $K$-band images (blue, green, and red respectively.)

\subsection{LIRG 1}
	LIRG 1 is a $z=0.73$ disk galaxy.  The \HST\ images show well defined spiral structure.  A red core, not seen in the $B$-band image, is prominent at longer wavelengths and dominates the light in the $K$-band.  The $B$-image shows several compact knots that also appear in the $V,i,$ and $z$-band images.  Aside from the prominent core, the $K$-band image appears to be devoid of substructure. While this is a disk dominated galaxy, it may be undergoing  a minor merger.  A linear feature to the northwest may be the result of infalling material.  A similar structure is seen to the southeast.  This system is an x-ray source.  If there is an AGN in the nucleus, it may be contributing to the IR luminosity.

\subsection{LIRG 2}
	LIRG 2, first studied in Melbourne et al. (2005), is a $z=0.6$ major merger.  The morphology includes a concentrated central galaxy, as well as a prominent linear feature extending to the NE.  This linear feature is aligned with a secondary core to the SW of the main galaxy.  The secondary core is clearly seen across all wavebands, which may indicate the presence of an Active Galactic Nucleus.  The colors of the linear feature are consistent with star formation.

\subsection{LIRG 3}
	At first glance, LIRG 3 ($z=1.02$) appears to be a collection of star forming knots, with little underlying structure.  However, the $z$-band image reveals an underlying stellar disk, while the $K$-band, is dominated by a red central core that might contain a small (2.5 kpc diameter) horizontal bar.  The blue knots do not appear in the $K$-band image.

\subsection{LIRG 4}
	This $z=1.10$ LIRG shows tenuous spiral arms in the $V$, $i$, and $z$ band images.  This spiral structure does not appear in the lower SNR $B$ and $K$-band images.  The spiral arms emerge from a bar that is prominent in the redder wavebands.  The core is luminous in both the red and blue bands, possibly an indication of dusty star formation.
	
\subsection{LIRG 5}
	LIRG 5, $z=1.08$, is a compact source with a complicated nuclear structure composed of a central nucleus and 2-3 blue knots.  Diffuse emission appears to the south of the main galaxy, especially in the $V$-band image.  The $K$-band image is dominated by the nucleus.  The galaxy is elongated in the NE-SW direction in all bands.
	
\subsection{LIRG 6}
	LIRG 6 is a $z=0.78$ disk galaxy with strong blue spiral arms.  The $K$-band image shows a bar-like linear feature in the N-S direction.  The core appears under-luminous  in the $B$-band image suggesting dust.
	
\subsection{LIRG 7}
	LIRG 7 is a $z=1.01$ spiral galaxy.  The nucleus is very bright across all wavebands.  Several knots appear in the optical bands within the spiral arms.  The lower SNR $K$-band image does not show knots or spiral arms.

\subsection{LIRG 8}
	LIRG 8 is similar to LIRG 7.  It is a 1.2 spiral galaxy with a bright nucleus across all wavebands.  It contains several  blue knots in the disk. The $K$-band image is smooth and shows no sign of the blue knots.

\subsection{LIRG 9}
	LIRG 9, at $z=0.37$ is dominated by a luminous, possibly unresolved core, indicative of an AGN.  This galaxy is a strong x-ray source suggesting that the core is dominated by AGN light. The lack of significant blue star forming knots also suggests that the AGN may be the primary source of the IR luminosity in this system. The core is surrounded by a blue ring.  The $K$-band image appears relatively free of sub-structure aside from the core.
	
\subsection{LIRG 10}
	Another compact spiral galaxy with a red core, LIRG 10 has a redshift of $z=1.002$.  According to the photometric redshifts from COMBO-17, the irregular galaxy to the NE has a redshift of $z=0.91$.  Therefore it may or may not be associated with the LIRG.  Despite this the LIRG may still be undergoing a minor merger.  There is a blue arc to the SE that does not seem associated with the rest of the spiral structures but does look like it is merging with the system.   Other than the luminous core, the $K$-band image does not show signs of substructure.

\subsection{LIRG 11}
	LIRG 11 is an edge on spiral galaxy at $z=0.95$.  The $V$, $i$, and $z$-band images show a dust lane across the core.  The $K$-band image of the core does not appear to be affected by the dust lane.  The galaxy is almost invisible in the $B$-band image.
		 
\subsection{LIRG 12}
	At first glance this galaxy ($z=0.55$) looks like a very compact, centrally concentrated galaxy across all five bands. However, at fainter flux levels there appears to be a very extended low surface brightness disk component in the N-S direction.  This is the only LIRG in our sample that might contain $K$-band substructure not prominent in the other bands.  A small red knot appears just above the central core in the $K$-band image.  This galaxy is also an x-ray source.  The central AGN may be the main source of IR luminosity in this system as it also appears to be free of prominent blue star forming regions.

\subsection{LIRG 13}
		This $z=0.89$, disk galaxy has a very red core surrounded by a series of bright blue knots.  The $K$-band image of shows a warped linear feature running from the SE to the NW, along the semi major axis of the galaxy.  This galaxy may be experiencing a minor merger.  There is a very blue companion to the NW of the galaxy.
		
\subsection{LIRG 14} 
	LIRG 14 is a z=0.84 QSO and therefore is different from the other galaxies in the sample.  It appears as an unresolved point source in all 5 bands.  There is no evidence in these images for an underlying galaxy.    
	
\subsection{LIRG 15}
The morphology of LIRG 15 ($z=0.65$) is dominated by a ring of luminous blue knots.  The primary core is red but is off-center from the rest of the galaxy.  To the NW of the red core is a bluer knot.    Finally a blue tail trails off in the southward direction at the SW corner of the galaxy indicating recent merger activity.

\section{Morphologies of the Comparison Sample}	
Figures \ref{fig:COMPGALsA} - \ref{fig:COMPGALsD} show the 5-band images of the ``normal'' galaxies.  

\subsection{COMPGAL 1}
	COMPGAL 1 is an inclined disk galaxy at $z=0.42$.  It contains a very blue arc to the W of the nucleus.  At the southern tip of the blue arc is a very blue, luminous knot.  A red knot of similar size is located to the SE of the core.  The core of the galaxy is very red, and is the only structure seen in the $K$-band image.
	
\subsection{COMPGAL 2} 
	 At $z=0.67$, COMPGAL 2 is a compact disk galaxy with blue southern and northern arcs.  It also contains two blue knots, one associated with each arc.  The core of the galaxy is redder than the arcs, but the $K$-band image is not as nucleated as the other bands, implying the lack of a significant bulge population.
		
\subsection{COMPGAL 3}
	COMPGAL 3 is a $z=0.44$ irregular galaxy.  Its central region is dominated by a linear feature that runs NE to SW.  This feature is visible in all bands although is very faint in the $K$-band image.  A set of bright blue knots is seen to the NE of the core.
	
\subsection{COMPGAL 4}
	COMPGAL 4 ($z=0.50$) is an edge on disk galaxy with a well defined red bulge that is prominent in the $K$-band image.  The bulge is encircled by blue arcs in the other bands.  The disk may have a warp at large radii to the NE. The upturn in the disk is pointed towards a faint blue galaxy to the N.  It is not clear if this neighboring galaxy is a companion or just a superposition.
	
\subsection{COMPGAL 5}
	COMPGAL 5 is a compact disk galaxy at $z=0.73$.  It contains a prominent red bulge.  A blue arc is located to the NW of the galaxy core.

\subsection{COMPGAL 6} 
	COMPGAL 6 ($z=0.42$) appears to be either an inclined disk galaxy, or possibly an irregular galaxy.  It contains a fairly red core, which is diffuse in the $K$-band.  The core is surrounded by a series of blue knots. The south east corner of this galaxy shows a peculiar blue feature.
	
\subsection{COMPGAL 7}  
	At first glance COMPGAL 7 ($z=0.77$) appears to be another small edge on spiral with a prominent red core.  However, it is the reddest galaxy in our sample, so it might be a dwarf elliptical.  A faint blue galaxy lies to the NW and may be a companion.  

\subsection{COMPGAL 8} 
	COMPGAL 8 is a disk galaxy at $z=0.77$.  It has two prominent spiral arms to the N and another linear blue feature to the S.  The core of this galaxy is very blue.  The core is the only feature that appears in the $K$-band image, but even the core is faint in $K$.  There is a blue knot on the western spiral arm.  

\subsection{COMPGAL 9}
	COMPGAL 9 is an irregular galaxy at $z=0.70$.  It has a red core that is offset to the SE from much of the rest of the galaxy.  The galaxy also contains several blue knots and blue wisps that lie to the N of the galaxy core.  This galaxy is very luminous $M_B=-20.88$, but unlike the LIRGs, it is a very blue system $B-V=0.235$.

\subsection{COMPGAL 10}
	When viewed in the bluer bands, COMPGAL 10 ($z=0.46$) looks like another irregular system.  In the $z$ and $K$-band images it loots like a dwarf spiral or elliptical with a red central bulge.  The differences between the blue and red morphologies may be indicating a recent minor merger between a gas rich cloud and a dwarf elliptical or spiral galaxy. 

\subsection{COMPGAL 11}
	COMPGAL 11 ($z=0.61$) looks like a irregular galaxy in the blue bands and like a small disk galaxy in the redder bands.  It has a red core.  A blue arc runs through the core from the S to N, and may be a spiral structure.  In the bluest band, the galaxy looks like a ring.


\end{document}